\documentclass[rmp,aps,nofootinbib,twocolumn,floatfix]{revtex4}

\usepackage{graphicx,color}
\usepackage{url}

\def\inbar{\,\vrule height1.5ex width.4pt depth0pt}
\def\IR{\relax{\rm I\kern-.18em R}}
\def\IC{\relax\hbox{$\inbar\kern-.3em{\rm C}$}}

\newcommand{\la}{\langle}
\newcommand{\ra}{\rangle}

\newcommand{\beq}{\begin{eqnarray}}
\newcommand{\eeq}{\end{eqnarray}}

\newcommand{\mn}{{\mu\nu}}

\newcommand{\Dsl}{/ {\hskip-0.26cm{D}}}

\newcommand{\bx}{\mbox{\boldmath $x$}}

\newcommand{\by}{\mbox{\boldmath $y$}}

\newcommand{\bk}{\mbox{\boldmath $k$}}

\newcommand{\qL}{q_{_{\rm L}} }
\newcommand{\qR}{q_{_{\rm R}} }
\newcommand{\bqL}{\bar{q}_{_{\rm L}} }
\newcommand{\bqR}{\bar{q}_{_{\rm R}} }
\newcommand{\qLR}{q_{_{\rm L(R)}} }

\newcommand{\als}{\alpha_{\rm s}}

\newcommand{\Lamqcd}{ \Lambda_{_{\rm QCD}} }

\begin{document}

\title{Hadron properties in the nuclear medium}

\author{Ryugo S.~Hayano}
\email{hayano@phys.s.u-tokyo.ac.jp}
\affiliation{Department of Physics, The University of Tokyo,
Bunkyo-ku, Tokyo 113-0033}
\author{Tetsuo Hatsuda}
\email{hatsuda@phys.s.u-tokyo.ac.jp}
\affiliation{Department of Physics, The University of Tokyo,
Bunkyo-ku, Tokyo 113-0033}

\begin{abstract}
The QCD vacuum shows
the dynamical breaking of chiral symmetry.
In the hot/dense QCD medium,  the  chiral order parameter such as  $\left < \bar q q \right> $  is expected to  change as function of temperature $T$ and density $\rho$ of the  medium, and its experimental detection is one of the main challenges  in modern hadron physics. In this article, we discuss theoretical expectations  for the in-medium hadron spectra associated with partial restoration of chiral symmetry and the current status of  experiments with an emphasis on the measurements of properties of  mesons produced in near-ground-state nuclei.
\end{abstract}

\maketitle
\tableofcontents

\section{INTRODUCTION}
\label{sec:intro}

Quantum chromodynamics (QCD), which is the color SU(3) gauge theory
of quarks and gluons  \cite{Nambu:1966}, is now established as 
the fundamental theory of strong interactions.
The Lagrangian density of QCD reads
\beq
\label{eq:QCDaction-LR}
{\cal L} &=&  \sum_q 
 \left( \bqL i  \Dsl \qL + \bqR i  \Dsl \qR \right)
   - \frac{1}{4} G_{\mn}^{\alpha} G^{\mn}_{\alpha} \nonumber\\
& &   + \sum_q \left( \bqL m \qR + \bqR m \qL \right) 
\eeq 
where we focus  on three light flavors
 $q=(u,d,s)$ with the mass matrix
 $m= {\rm diag} (m_u,m_d,m_s)$ throughout this article. 
 The quark field $q$ belongs to the triplet representation of 
  the color gauge group SU(3)$_{\rm C}$.
  The right (left) handed quark $\qR=\frac{1}{2}( 1+\gamma_5)q$
  ($\qL=\frac{1}{2}(1-\gamma_5)q$) is  the eigenstate of the
  chirality operator $\gamma_5$ with the eigenvalue $+1 (-1)$. 
 The covariant derivative is defined as 
 $D_{\mu} \equiv \partial_{\mu} + 
 i g t_{_{\rm C}}^{\alpha} {\cal A}_{\mu}^{\alpha}$ with 
 $g$ being the strong coupling constant, $t_{_{\rm C}}^{\alpha}$ being the 
 SU(3)$_{\rm C}$ generator and  ${\cal A}_{\mu}^{\alpha}$ being the 
color-octet gluon field. The field strength tensor of the gluon is
 defined as $G_{\mn}^{\alpha} = \partial_{\mu}{\cal A}_{\nu}^{\alpha} -
\partial_{\nu}{\cal A}_{\mu}^{\alpha}
 - gf_{\alpha \beta \gamma} {\cal A}_{\mu}^{\beta} {\cal A}_{\nu}^{\gamma} $
  with $f_{\alpha \beta \gamma}$
 being the structure constant of SU(3)$_{\rm C}$.
The QCD Lagrangian Eq.(\ref{eq:QCDaction-LR}) is exactly invariant under the
 local SU(3)$_{\rm C}$ gauge transformation of quarks and gluons.  

The running coupling constant
$g(\kappa)$ is defined as an effective coupling strength
 among quarks and gluons at the energy scale $\kappa$.
Due to the asymptotic free nature of
QCD,   $g(\kappa)$ becomes small as $\kappa$ increases 
\cite{Gross:2005kv,Wilczek:2005az,Politzer:2005kc}.
This is explicitly seen in the two-loop perturbation theory as
\beq
\label{eq:run-g}
\als  (\kappa) \simeq
\frac{1}{4\pi \beta_0 \ln (\kappa^2 / \Lamqcd^2 ) }  \cdot
\left[ 1 -  \frac{\beta_1}{\beta_0^2 } 
\frac{\ln (\ln (\kappa^2/\Lamqcd^2))}
{\ln(\kappa^2/\Lamqcd^2)} \right] ,\nonumber \\
\eeq 
where $\als (\kappa)\equiv \frac{g^2(\kappa)}{4\pi}$,
$\beta_0= (11-\frac{2}{3}N_{\rm f})/(4\pi)^2$,
$\beta_1=(102-\frac{38}{3}N_{\rm f})/(4\pi)^4$, $N_{\rm f}$ is the number of flavors and 
$\Lamqcd$ is called the QCD scale parameter
to be determined from experiment. 
\begin{figure}
\includegraphics[width=0.6\columnwidth]{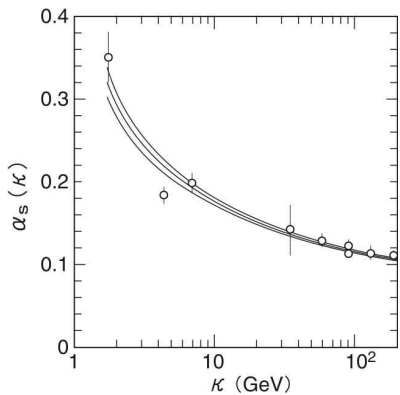}
\caption{The running coupling constant determined from 
 $\tau$ decay,    $\Upsilon$ decay,   deep inelastic scattering,
 ${\rm e}^+{\rm e}^-$ 
annihilation, and the $Z$-boson resonance shape and width \cite{Amsler:2008pj}. }
\label{fig:alpha_s} 
\end{figure}

Fig.\ref{fig:alpha_s} and Eq.(\ref{eq:run-g})
indicate that the running coupling constant increases and becomes strong
 at low energies $\kappa \sim \Lamqcd \sim 200 $ MeV.
This is the typical energy scale where  various non-perturbative effects
such as the confinement of quarks and gluons \cite{Wilson:2004de}
 and the dynamical breaking of
 chiral symmetry \cite{Nambu:1961tp,Nambu:1961fr}\cite{Hatsuda:1994pi}.
 Both effects are responsible for the formation of 
 composite hadrons and nuclei and for the origin of their masses.  
 In this article, we will focus on the dynamical breaking of chiral
  symmetry (DBCS) realized in the QCD vacuum and in the hot-dense QCD medium 
 by using in-medium hadrons as  useful probes of QCD matter.

\begin{figure}
\includegraphics[width=0.9\columnwidth]{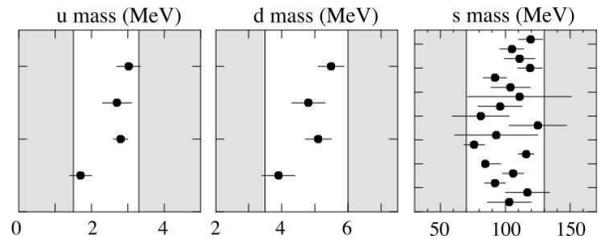}
\caption{The masses of $u$, $d$, and $s$ quarks at the scale $\kappa=2$ GeV
constrained by
various hadron masses using QCD sum rules and lattice QCD simulations
 \cite{Amsler:2008pj}. }
\label{fig:quark-mass_uds} 
\end{figure}

\subsection{QCD symmetries}
\label{sec:QCD-SYM}

 Similar to the running coupling constant  $\als (\kappa)$, the quark masses
 receive quantum corrections and become scale dependent, $m(\kappa)$.
 As seen from Fig.\ref{fig:quark-mass_uds}, the current determination of 
  the $u$ and  $d$ quark masses show that they are about 50 to 100 
  smaller than the QCD intrinsic scale $\Lamqcd$, while the $s$ quark mass
   is comparable to $\Lamqcd$.  
 Therefore, it is legitimate to treat $m_u/\Lamqcd$ and $m_d/\Lamqcd$
  as small expansion parameters.   

In the limiting case where $m_{u,d}=0$, which is called the SU$(2)$ chiral limit,
the QCD Lagrangian 
Eq.(\ref{eq:QCDaction-LR}) acquires an exact global symmetry called
chiral symmetry under  independent ${\rm SU(2)}$ rotations of the 
left handed and right handed quarks: 
$\qL \rightarrow U_{\rm L} \qL$ and $\qR \rightarrow U_{\rm R} \qR$ with $U_{\rm L,R}$ being
 the global  ${\rm SU(2)}$ matrices.  Thus we have the exact QCD symmetry for
 $m_{u,d}=0$,
\beq
\label{eq:G-symmetry}
{\cal G} = {\rm SU(3)}_{\rm C} 
\otimes {\rm SU(2)}_{\rm L} \otimes {\rm SU(2)}_{\rm R} 
\otimes {\rm U(1)}_{\rm B},
\eeq
where ${\rm U(1)_B}$ corresponds to the baryon number symmetry corresponding to the
global phase rotation, $\qLR \rightarrow e^{i \theta} \qLR$.
Although a similar phase rotation, $\qL \rightarrow e^{i \phi} \qL$ and
 $\qR \rightarrow e^{-i \phi} \qR$, looks like a symmetry of 
 Eq.(\ref{eq:QCDaction-LR}), it is broken explicitly by a quantum effect
  known as the axial anomaly. 
Currents associated with these ``symmetries" are defined as
 ${V}_{\mu}^a = \bar{q} \gamma_{\mu} t^a q$ (the triplet vector current),
 ${A}_{\mu}^a = \bar{q} \gamma_{\mu}\gamma_5 t^a q$ (triplet axial-vector current),
 ${V}_{\mu}^0 = \bar{q} \gamma_{\mu}t^0  q$ (baryon current), 
 ${A}_{\mu}^0  = \bar{q} \gamma_{\mu} \gamma_5  t^0 q$ (singlet axial current),
 where $t^{a=1,2,3}\equiv \tau^a/2$ with $\tau^a$ being
the Pauli matrices and  $t^0 \equiv \tau^0/2 \equiv 1/2$.
The divergences of these currents are 
\beq
\label{eq:V_a}
\partial^\mu {V}_{\mu}^a &=& i \bar{q} [m, t^a] q   ,\\
\label{eq:A_a}
\partial^\mu {A}_{\mu}^a &=& i \bar{q} \{ m, t^a \} \gamma_5 q   ,\\
\label{eq:V_0}
\partial^\mu {V}_{\mu}^0 &=& 0   ,\\ 
\label{eq:A_0}
\partial^\mu {A}_{\mu}^0 &=&  i \bar{q} m \gamma_5 q 
- 2 \frac{\als}{4\pi} G^{\mu \nu}_{\alpha} \tilde{G}^{\alpha}_{\mu \nu}  ,
\eeq
with $\tilde{G}^{\alpha}_{\mu \nu}=\frac{1}{2}\epsilon_{\mu \nu \lambda \rho}
{G}_{\alpha}^{\lambda \rho}$ being the dual field strength of the gluon.
 $[ \ , \ ]$ and $\{ \ , \ \} $ are the commutator and the anti-commutator
  in the flavor-space, respectively.
For later convenience, we define the scalar and pseudo-scalar density as
\beq
\label{eq:S_0a}
{S}^0 &=& \bar{q} t ^0 q, \ \ {S}^a = \bar{q} t ^a q, \\
\label{eq:PS_0a}
 {P}^0 &=& \bar{q} i \gamma_5 t^0 q, \ \ {P}^a = \bar{q} i \gamma_5 t^a q, 
\eeq 

From the time component of the currents, generators of the 
chiral transformation are defined as $Q^a(t) = \int V_0^a(t,\bx) d^3x$ and
$Q_5^a(t)=  \int A_0^a(t,\bx) d^3x$.  Then the bilinear quark operators defined
above obey the following  relations under the axial transformation $Q_5^a$ $(a=1,2,3)$;
\beq
\! \! \! \! \! \! \! \! \! \! \! \!   \! \! \! \! \! \! \! \!   
& & [Q_5^a(t), V_{\mu}^b(t,\bx)]= +i \epsilon_{abc} A_{\mu}^c(t,\bx),\\
\! \! \! \! \! \! \! \! \! \! \! \!  \! \! \! \! \! \! \! \!   
& & [Q_5^a(t), A_{\mu}^b(t,\bx)]=  +i \epsilon_{abc} V_{\mu}^c(t,\bx),\\
\! \! \! \! \! \! \! \! \! \! \! \!  \! \! \! \! \! \! \! \!   
& &[Q_5^a(t), S^0(t,\bx)]         =  +i  P^a(t,\bx), \\
\! \! \! \! \! \! \! \! \! \! \! \!  \! \! \! \! \! \! \! \!   
& &[Q_5^a(t), P^0(t,\bx)]         =  -i  S^a(t,\bx), \\
\! \! \! \! \! \! \! \! \! \! \! \!  \! \! \! \! \! \! \! \!   
& &[Q_5^a(t), S^b(t,\bx)]
= + i  \delta_{ab}  P^0(t,\bx) ,\\
\! \! \! \! \! \! \! \! \! \! \! \!  \! \! \! \! \! \! \! \!   
& &[Q_5^a(t), P^b(t,\bx)]
= -i  \delta_{ab} S^0(t,\bx) .
\eeq

\begin{figure}
\includegraphics[width=0.8\columnwidth]{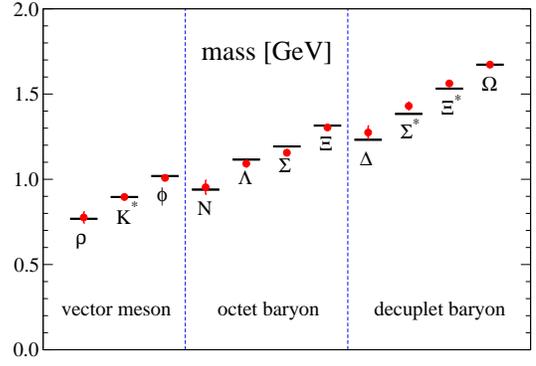}
\caption{Light hadron spectrum obtained from lattice QCD simulations
with dynamical $u, d, s$ quarks in the Wilson fermion formalism.
 The hadron masses are 
extrapolated to the physical quark masses
(determined by $m_\pi$, $m_K$ and $m_\Omega$) using the 
 data in the interval, 
 $\frac{1}{2}(m_u+m_d) (\kappa= 2 {\rm GeV})=3.5\ {\rm MeV} -67\ {\rm MeV}$. 
The spatial lattice volume $V$ and the lattice spacing $a$ are
$(2.9\ {\rm fm})^3$ and $0.09\ {\rm fm}$, respectively.
Horizontal bars denote the experimental values \cite{Aoki:2008sm}. }
\label{fig:lattice-spect} 
\end{figure}

In the past few years, remarkable progress was made in 
calculating the hadron spectra on the basis of lattice
QCD simulations with dynamical $u, d, s$ quarks.
This progress was achieved partly because
 the supercomputer speed is doubled every
 1.2 years  and partly because of 
  new simulation algorithms: The lattice QCD
  simulations for quark masses very close to the physical 
   point are now possible in the Wilson fermion formalism \cite{Aoki:2008sm,Durr:2008xx}.
 Shown in Fig.\ref{fig:lattice-spect}  is an example of 
the lattice results for  meson and baryon masses
extrapolated to the physical quark masses using the simulation data 
in the interval,
 $\frac{1}{2}(m_u+m_d) (\kappa= 2 {\rm GeV})=3.5\ {\rm MeV}-67\ {\rm MeV}$
(corresponding to $m_{\pi}=156\ {\rm MeV} -702\ {\rm MeV}$).
The experimental data are reproduced with 3\% accuracy.
The simulations right at the physical quark masses will be 
 performed in the very near future.

\subsection{Dynamical breaking of chiral symmetry in the vacuum}
\label{sec:DBCS}

Even if QCD in the SU$(2)$ chiral limit has the symmetries of Eq.(\ref{eq:G-symmetry}),
the ground state of the system may break some of the symmetries dynamically. 
Let us consider the QCD vacuum $|0 \ra$ at zero temperature and density.
 Assuming that the 
  vacuum is Lorentz invariant and taking into account the fact that
  QCD does not allow dynamical breaking of parity and vector  symmetry 
  in the vacuum \cite{Vafa:1983tf},
 we have the following possibility of the symmetry breaking pattern;
\beq
\label{eq:SB-pattern}
 {\rm SU(2)}_{\rm L} \times {\rm SU(2)}_{\rm R}
 \rightarrow {\rm SU(2)}_{\rm {L+R}} \equiv {\rm SU(2)}_{\rm V}.
\eeq
In terms of the generators of the vector and axial-vector rotations,
such a vacuum state is characterized as
\beq
\label{eq:Q-Q_5-vac}
Q^a |0 \ra = 0  ,\ \ \ Q_5^a |0 \ra \neq  0 .
\eeq
Strictly speaking, we need to take the SU$(2)$ chiral limit $m_{u,d}\rightarrow 0$
after taking the thermodynamic limit $V\rightarrow 0$ 
to make the matrix elements of $Q_5^a$ well defined. This is similar
to the case of the spin system where the external magnetic field plays
the role of $m_{u,d}$. 

At this point, it is in order to mention general definition of the order
parameter. 
Consider a symmetry group ${\cal G}$ and its generator $Q$. 
If there is an operator $\Phi$ such that
 $\la  [iQ, \Phi] \ra_0  (\equiv \la 0|  [iQ, \Phi] |0 \ra ) \neq 0$,
this expectation value is called the order parameter. If the vacuum is
symmetric under $Q$, the order parameter becomes zero. On the other hand,
 if 
the vacuum is not symmetric under $Q$,
there exists a Nambu-Goldstone boson having the same quantum number
as $\Phi$. Note that the 
order parameter is not unique for a given ${\cal G}$:
 one can introduce higher dimensional
 order parameters in principle to characterize the system
\cite{Kogan:1998zc,Watanabe:2003xt}.

\begin{figure}
\includegraphics[width=0.5\columnwidth,angle=90]{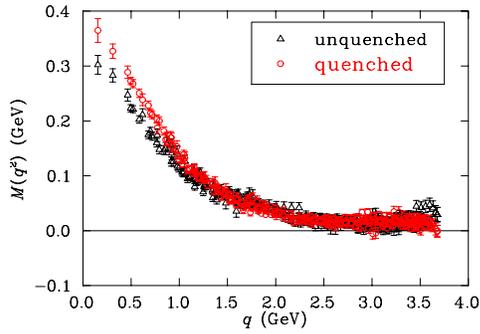}
\caption{Effective quark mass $M$ as a function of the 
Euclidean momentum $q$ obtained from lattice QCD simulations
 without dynamical quarks (quenched) and with three dynamical quarks (unquenched)
   \cite{Bowman:2005vx}. }
\label{fig:mass-func_lat} 
\end{figure}

For the symmetry breaking pattern as Eq.(\ref{eq:SB-pattern}),
$Q$ is identified as $Q_5^a$, and a simplest choice of $\Phi$ is
$P^a$.  Then, it leads to the order parameter $ \la S^0 \ra_0$.
 Recent lattice QCD simulation of the chiral condensate using
 overlap Dirac fermion
  with dynamical $u, d, s$ quarks  indicates  \cite{Fukaya:2009fh} 
\beq
\label{eq:quark-cond}
\la S^0 \ra_0 = \frac{1}{2}
\la \bar{\rm u}{\rm u}  + \bar{\rm d}{\rm d} \ra_0 =
- (242 (04)( ^{+19}_{-18}) \ {\rm MeV})^3, 
\eeq
where the renormalization scale is taken to be $\kappa = 2 \ {\rm GeV}$
with the statistical and systematic errors in parentheses.
This result implies that the QCD vacuum is the 
Bose-Einstein condensate of quark$-$anti-quark
pairs $\la \bar{q}q \ra_0=\la (\bqL \qR + \bqR \qL ) \ra_0$
and has the power to change  left handed quarks to right handed quarks  
and vice versa:  Namely the condensate induces a dynamical quark mass.
Since quarks are confined,
it is not possible to isolate a single quark to measure 
the dynamical quark mass.  Nevertheless, there is indirect evidence
 that the quarks inside hadrons have an effective mass (constituent quark mass) 
  $M \sim 350 $ MeV from the phenomenological
  description of hadrons. The effective quark mass
  near zero Euclidean momentum in lattice QCD
   simulations with Landau gauge fixing 
   leads to a similar value as shown in Fig.\ref{fig:mass-func_lat}.

\subsection{Chiral symmetry and hadron spectra}
\label{sec:SE-DBCS}

 The Nambu-Goldstone (NG) bosons associated with the DBCS of SU$(2)$ chiral
 symmetry  are nothing but the charged and neutral pions.
 Moreover, the partially conserved axial current (PCAC) relation,
  Eq.(\ref{eq:A_a}), leads to the 
  Gell-Mann$-$Oakes$-$Renner (GOR) relation  \cite{GellMann:1968rz} which
 relates the pion masses to the quark masses as
\beq
\label{eq:GOR-pi-pm}
f_{\pi}^2 m_{\pi^{\pm}}^2 &=&
- \hat{m} \la \bar{\rm u}{\rm u} + \bar{\rm d}{\rm d} \ra_0
+ O(\hat{m}^2) , \\
\label{eq:GOR-pi-0}
f_{\pi}^2 m_{\pi^{0}}^2 &=&
-  \la m_{\rm u} \bar{\rm u}{\rm u} + m_{\rm d} \bar{\rm d}{\rm d} \ra_0
+ O(\hat{m}^2) .
 \eeq
Here
 $\hat{m}$=$(m_{\rm u}+m_{\rm d})/2$
is the averaged mass of  u and d quarks, 
 $f_{\pi}$( = 92.4 MeV) is the pion decay constant,
and $m_{\pi^{\pm}} \simeq 140$ MeV ($m_{\pi^{0}} \simeq 135$ MeV)
 is the charged (neutral) pion mass.  
 Using these values together with the quark masses
  in Fig.\ref{fig:quark-mass_uds}, we obtain the 
  finite  chiral condensate comparable  to Eq.(\ref{eq:quark-cond}).

\begin{figure}
\includegraphics[width=0.7\columnwidth]{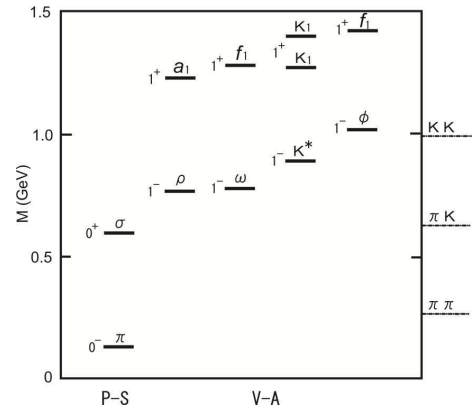}
\caption{Light scalar ($S$), pseudo-scalar ($P$), vector ($V$) and
axial-vector ($A$) mesons. Their spin and parity are denoted as
$J^p$. Threshold of the $\pi \pi$, $\pi K$ and $KK$ decays are also shown
\cite{Yagi:2005yb}. }
\label{fig:meson-spect} 
\end{figure}

Further experimental evidence of DBCS is obtained
 from the observed meson spectra.  
If  chiral symmetry is not broken, $Q_5^a |0 \ra = 0$, 
 vacuum expectation values of all the
the commutators, 
$\la  [Q_5^{a_n}, \cdots  [Q_5^{a_2}, [Q_5^{a_1}, \Phi ]] \cdots ] \ra_0 $,
should vanish for an arbitrary operator $\Phi$.  
The contraposition of this statement with $n=2$ 
and $\Phi = S^a(x)S^a(y), V_{\mu}^a(x) V_{\nu}^b(y)$
leads to a statement that DBCS must occur
if the correlation functions of the 
chiral partners are not degenerate. Namely,
\beq
\! \! \! \! \! \! \! \! \! \! \! \! \! \! \! \! 
& & \la S^a(x)S^a(y) - P^a(x)P^a(y) \ra_0 \neq 0 \rightarrow Q_5^a|0 \ra \neq 0,\\
\! \! \! \! \! \! \! \! \! \! \! \! \! \! \! \! 
& & \la V_{\mu}^a(x)V_{\nu}^a(y) - A_{\mu}^a(x)A_{\nu}^a(y) \ra_0 \neq 0 \rightarrow Q_5^a|0 \ra \neq 0.
\eeq
Experimentally,  the pion (the pseudo-scalar meson) does not have a 
  scalar partner at the same mass, and 
 the $\rho$-meson (the vector meson) does not 
 have an axial-vector partner at the same mass, which 
 are  the direct evidences of DBCS.
 Such  non-degeneracy is also seen in other channels, e.g. 
 $\omega$, $K^*$ and $\phi$
 as  illustrated in Fig.\ref{fig:meson-spect}. 


\section{CHIRAL SYMMETRY AND IN-MEDIUM HADRON SPECTRA}
\label{sec:CSHS}

Connections between properties of the QCD vacuum and 
 hadronic correlation functions as discussed in 
 Sec.\ref{sec:DBCS} and Sec.\ref{sec:SE-DBCS}
  can be generalized to QCD at finite temperature and density.
In this section, we will summarize such theoretical
 connections with special emphasis on the pion, the scalar meson, and
 vector mesons in the medium.

\subsection{Chiral condensate in the medium}
\label{sec:CCIM}

Let us now consider how the simplest chiral order parameter $\la \bar{q} q \ra $
changes its value inside the hot and/or dense medium.
Exact formula for the in-medium 
chiral condensate in terms of the QCD partition function $Z$ at finite
 temperature $T$ and the baryon chemical potential $\mu$ is given by
\beq
\label{eq:qqbar-m} 
\la \bar{q} q \ra_{T,\mu}  = \frac{1}{Z}{\rm Tr} \left[ \bar{q}q e^{-K_{\rm QCD}/T} \right] 
=  - \frac{\partial P(T,\mu)}{\partial m_q} ,
\eeq
where
\beq
\label{eq:Z-QCD}
\! \! \! \! \! \! \! \! \! \! 
Z(T,\mu) & = & {\rm Tr} \left[ e^{-K_{\rm QCD}/T} \right] = e^{P(T,\mu)V/T}, \\
\! \! \! \! \! \! \! \! \! \! 
K_{\rm QCD} & = & H_{\rm QCD}^{m=0} 
+ \int (\bar{q} m q  - q^{\dagger} \mu q ) d^3x ,
\eeq
with $H_{\rm QCD}^{m=0}$ being the QCD Hamiltonian without the quark mass term.

\subsubsection{Finite temperature}
\label{sec:CCFT}

\begin{figure}
\includegraphics[width=0.8\columnwidth]{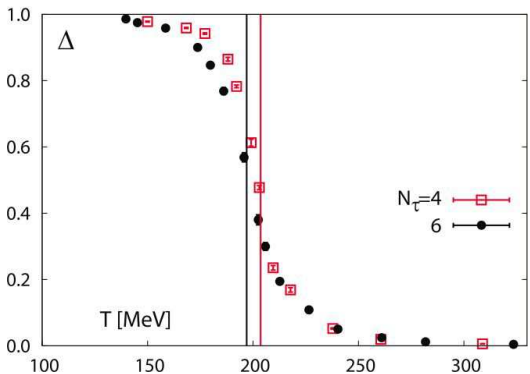}
\caption{Normalized chiral condensate 
$\Delta=[\la \bar{u}u \ra_T -(\hat{m}/m_s) \la \bar{s}s \ra_T]/
[\la \bar{u}u \ra_0 -(\hat{m}/m_s) \la \bar{s}s \ra_0]$
as a function of $T$ for two different lattice spacings,
$a= 0.24\ {\rm fm} \, (N_{\tau}=4)$ and 
$a= 0.17\ {\rm fm} \, (N_{\tau}=6)$, calculated by the 
 lattice QCD simulations with dynamical $u,d,s$ quarks in the 
 staggered fermion formalism \cite{Cheng:2007jq}.}
\label{fig:qqbar-T-lat} 
\end{figure}

Eq.(\ref{eq:qqbar-m})
can be evaluated analytically in some special cases.  For example,
in the two-flavor system ($N_{\rm f}=2$, i.e., $m_{u}=m_{d} < \infty, m_s=\infty$) 
at low $T$ with $\mu=0$,  
themal pions are the dominant contributor to the pressure. In the leading order of the 
virial expansion by the 
pion number density, we have \cite{Gerber:1988tt}
\beq
& &\left. \frac{\la \bar{q}q \ra_T}{\la \bar{q}q \ra_0} \right|_{N_f=2}
\simeq 1 - \frac{3}{4} \Theta(T),
\label{eq:qqbar-T0} \\
& &\Theta (T) = \frac{T^2}{6f_{\pi}^2}B_1(m_{\pi}/T),
\label{eq:qqbar-T}
\eeq
where $\int \frac{d^3k}{(2\pi)^3 2 \varepsilon_k} n_{_{\rm B}}(k;T)
= \frac{T^2}{24} B_1(m_{\pi}/T)$  with
$n_B$ being the Bose-Einstein distribution and 
 $\varepsilon_k=\sqrt{\bk^2+m_{\pi}^2}$. Eq.(\ref{eq:qqbar-T})
 shows a clear tendency that
 the magnitude of the chiral condensate decreases as  $T$ increases.  
 However, for $T > 150 $ MeV, the interaction among pions and the 
  contribution from other mesons become important and the estimate based on
  Eq.(\ref{eq:qqbar-T}) is not reliable. 
The two-flavor system at extremely high $T$ with $\mu=0$ can be also evaluated
 because the thermal quarks and gluons are the dominant contributor to the 
  pressure due to asymptotic freedom. In the leading order of $\als(\kappa\sim T)$, we have
$\la \bar{q}q \ra_T \simeq \frac{1}{2} m_q T^2$ ($q=u,d$), so that the chiral condensate
 vanishes in the chiral limit.   Lattice QCD simulations
 at finite $T$  in Fig.\ref{fig:qqbar-T-lat}
  indeed show a decrease of the normalized chiral condensate for $u,d$ quarks
 with a rapid crossover around $T \simeq 200 $ MeV \cite{Cheng:2007jq}.

\subsubsection{Finite baryon density}
\label{sec:CCFB}

A useful formula for the  chiral condensate at fixed baryon density $\rho$ with $T=0$
is obtained from  Eq.(\ref{eq:qqbar-m}) by using the thermodynamic relations;
$P+\varepsilon = \sum_q \mu_q \rho_q$,
$\rho_q =  \frac{\partial}{\partial \mu_q} P$
and $\mu_q =  \frac{\partial}{\partial \rho_q} \varepsilon$
 with $\varepsilon$ being the energy density of the system:
\beq
\label{eq:qqbar-rho}
\langle \bar{q} q \rangle_{\rho} 
= \langle \bar{q} q \rangle_{0} +
\rho {d \over dm_q} \left( \frac{E}{A} \right) , 
\eeq
where $\rho(=\sum_q \rho_q)$, $A$ and $E$ are the total baryon density, the 
total baryon number, and the total energy, respectively.
There is an alternative derivation of this formula
using the Hellmann-Feynman theorem  \cite{Cohen:1991nk}. 

\begin{figure}
\includegraphics[width=0.65\columnwidth]{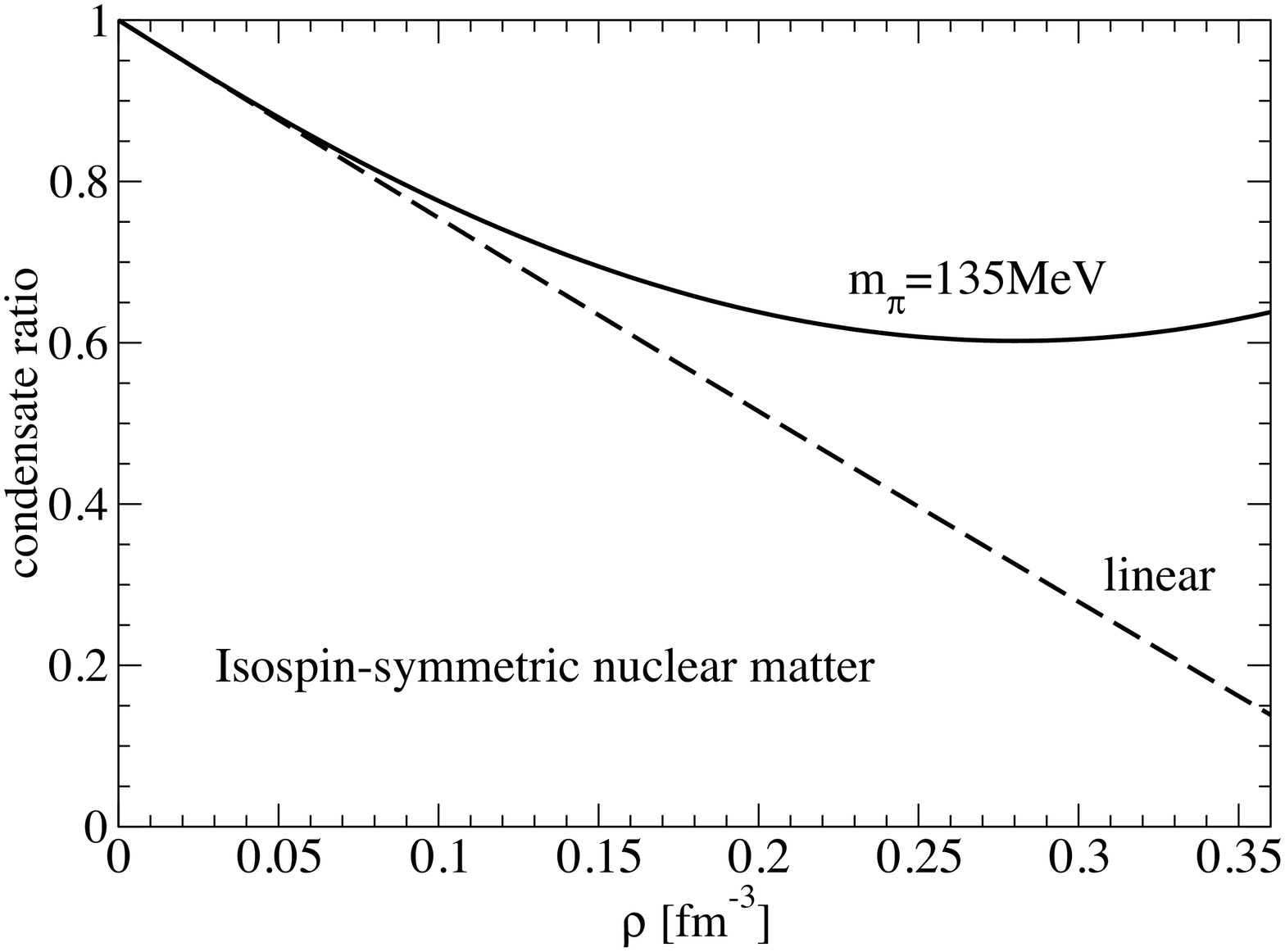}
\includegraphics[width=0.65\columnwidth]{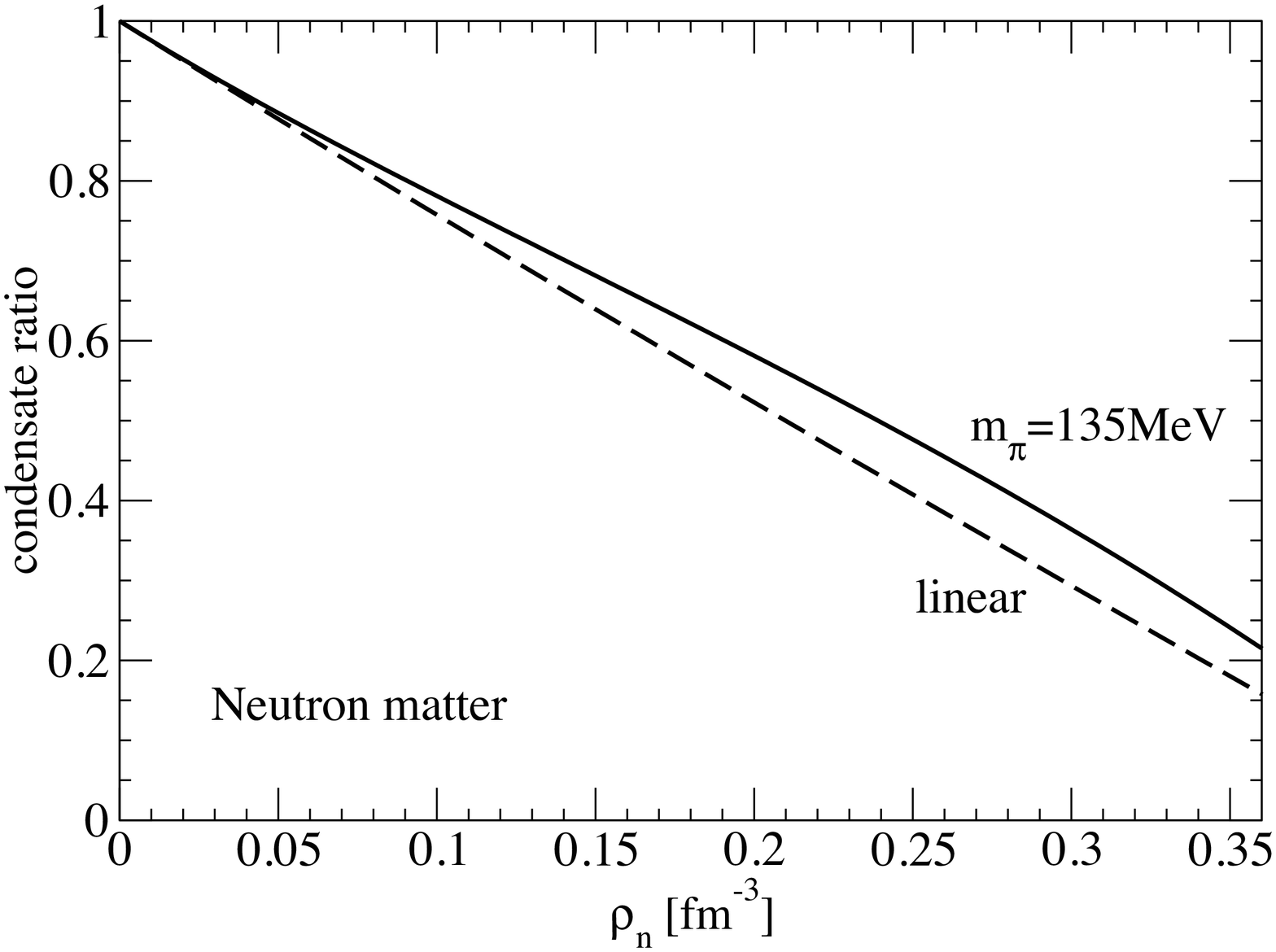}
\caption{Comparison of the linear density approximation of the 
chiral condensate (dashed lines) and the results with higher order terms
 originating from  nucleon-nucleon correlations calculated by the
  in-medium chiral perturbation theory (solid lines)  \cite{Kaiser:2008qu}.}
\label{fig:qqbar-rho} 
\end{figure}

For two-flavor nuclear matter 
with equal numbers of protons and neutrons,
the leading order of the virial expansion in terms of the 
baryon density reads \cite{Cohen:1991nk,Drukarev:1991fs,Hatsuda:1991ez}
\beq
\label{eq:uubar-rho}
\! \! \! \! 
& & \frac{\langle \bar{u} u + \bar{d} d \rangle_{\rho}}
{\langle \bar{u} u + \bar{d} d \rangle_{0}}
\simeq 1 - {\sigma_{\pi N} \over f_{\pi}^2 m_{\pi}^2 } \rho \ F_1(k_{\rm F}/m_{N}),  \\
\! \! \! \! 
& & \frac{\langle \bar{s} s \rangle_{\rho}}{\langle \bar{s} s \rangle_{0}}
\simeq 1 - y  { \sigma_{\pi N} \over f_{\pi}^2 m_{\pi}^2 } \rho \ F_1(k_{\rm F}/m_{N})    .
\eeq
Here 
$\sigma_{\pi N}= \hat{m} \la N | \bar{u}u + \bar{d}d | N \ra$
is the  $\pi N$ sigma-term, and 
 $y=2\la \bar{s}s \ra_N/\la \bar{u}u + \bar{d}d \ra_N$ is 
  the   strangeness content of the nucleon.
Also,  $\int \frac{d^3k}{(2\pi)^3 2 E_k} \theta(k_{\rm F}-|\bk|)
= \frac{k_{\rm F}^3}{6 \pi^2} F_1(k_{\rm F}/m_N)$  with
$k_{\rm F}$ being the Fermi momentum and 
 $E_k=\sqrt{\bk^2+m_{N}^2}$.  The low density expansion of $F_1$ reads
 $F_1(x) = 1- \frac{3}{10}x^2 + \frac{9}{56} x^4 + \cdots$
  \cite{Hatsuda:1995dy}. 
Using the empirical values,
$\sigma_{\pi N}= 45 \pm 10 $ MeV and $y = 0.12-0.22$ 
 \cite{Gasser:1990ce,Hatsuda:1991ez} the right hand side 
 of Eq.(\ref{eq:uubar-rho}) gives
 almost a 35 \% reduction of $\langle \bar{q} q \rangle_{\rho} $
 at  nuclear matter density $\rho_0 = 0.17\ {\rm fm}^{-3}$.
This leading order result together with a calculation of 
higher order corrections  on the basis of  in-medium chiral perturbation
 theory are shown in Fig.\ref{fig:qqbar-rho}
  for nuclear matter and neutron matter \cite{Kaiser:2008qu}.

The change of the chiral condensate induced by 
 the strong electric and magnetic fields
 and strong color electric and magnetic fields is 
also an interesting subject which may be relevant
to the physics of relativistic heavy ion collisions and
 the structure of compact stars with high magnetic field 
\cite{Klevansky:1992qe,Suganuma:1990nn,Miransky:2002eb}.

The experimental detection of the change of chiral condensate
in the hot and/or dense medium is   
one of the most interesting challenges in modern hadron physics.
Possible tools are hadron-nucleus and photon-nucleus reactions,
heavy-ion collisions, and deeply bound mesic atoms and mesic nuclei, 
which are summarized in the later chapters of this article.

\subsection{Spectral functions in the medium}
\label{sec:SP}

In the hot and/or dense environment,
all the hadrons including the pion undergo
 spectral changes  due to their strong interactions
with the medium. Therefore, it is not enough
 to talk about the ``mass" and ``width" of the 
 hadrons, but we need to study the 
hadronic  spectral functions.
For mesic resonances coupled to the 
composite operator ${\cal O}(t,\bx) = \bar{q}(t,\bx) \Gamma q (t, \bx)$
 with $\Gamma$ being an arbitrary combination of Dirac and flavor matrices, 
we have the spectral decomposition of the retarded correlation as
\beq
\label{eq:R-correlation-def}
G^{\rm R}(\omega,\bk) &=& i {\rm F.T.} 
\la {\rm R} \left( {\cal O}(t,\bx){\cal O}^{\dagger}(0) \right) \ra ,\\
&=& \int_0^{\infty} 
\frac{\rho_{_{\cal O}}(u,\bk)}{u^2 - (\omega+ i\delta)^2} du^2 , 
\eeq 
where F.T. stands for the fourier transform,
 ${\rm R}\left( A(t,\bx)B(t',\by) \right) =\theta(t-t')[A(t,\bx),B(t',\by)]$
is the retarded product of the operators $A$ and $B$, and $\la \cdot \ra$ implies
the expectation value at finite $T$ and $\mu$.
The spectral function, $\rho_{_{\cal O}}(\omega, \bk)$, has all the information of the 
states having the same quantum numbers with the operator ${\cal O}$.
In particular, $P^a$ and $V_{\mu}^a$ are the relevant operators
 for studying the in-medium pion and the $\rho$-meson, respectively.

\subsubsection{In-medium pion}
\label{sec:SP-pion}

The Nambu-Goldstone theorem which guarantees  massless pions 
in the QCD vacuum in the chiral limit holds also in the medium.
Indeed, by considering the correlation,
\beq
\label{eq:AP-correlation}
\Pi =  \int d^4x \partial^{\mu}  \la {\rm R}[A_{\mu}^a(t,\bx) P^b(0)] \ra, 
\eeq
the following sum rule can be derived \cite{Yagi:2005yb}:  
\beq
\label{eq:chiral-SR}
2m \int_0^{\infty} \frac{\rho_{_P}^{ab}(\omega,{\bf 0})}{\omega^2}d\omega^2 
= -\delta^{ab} \la S^0 \ra,
\eeq
where we considered a two-flavor system ($m=m_u=m_d < \infty, m_s=\infty$)
 for simplicity.  
$\rho_{_P}^{ab}$ corresponds to the spectral function associated with
$\Pi_{_P}^{ab}  (x) = \la {\rm R}(P^a(t,\bx) P^b(0)) \ra$.
To have a non-zero chiral condensate in the r.h.s.\ of
Eq.(\ref{eq:chiral-SR}) in the chiral limit $m \rightarrow 0$,
 the spectral function must have a pole,
$\rho_{_P}^{ab}(\omega,{\bf 0})|_{m \rightarrow 0} 
\sim \delta^{ab} [  C \delta(\omega^2 - a m) + \cdots ]$, so that
$m$ in the numerator is canceled by this pole in the chiral limit.
 This means nearly massless pions exist even in the medium as long
as DBCS takes place.

In the leading order of the virial expansion at $T\neq 0$ with $\mu=0$,
 the self-energy of the in-medium pion is dictated by the
  forward pion-pion scattering amplitude which vanishes 
   in the chiral limit. In this case, 
the pion is still a real pole with the mass $m_{\pi}(T)$
 and the decay constant $f_{\pi}^t(T)$
\cite{Goity:1989gs,Pisarski:1996mt,Toublan:1997rr}: 
\beq
\left( \frac{m_{\pi}(T)}{m_{\pi}} \right)^2 &=&  1+ \frac{1}{4} \Theta(T)   , \\
\left( \frac{f_{\pi}^t(T)}{f_{\pi}} \right)^2 &=&  1-  \Theta(T)   .
\label{eq:pion-T}
\eeq  
Here, the in-medium pion decay constant 
$f_{\pi}^t(T)$ is defined by the residue of the pion pole of the 
correlation function of $A_{\mu=0}^a$. Since 
 the Lorentz symmetry  does not hold in the rest frame of the 
medium, a difference arises between the  temporal residue
$f_{\pi}^t$ and the spatial residue $f_{\pi}^s$ in general, although
 they are equal in the leading order of the virial expansion at finite $T$.
Combining  Eq.(\ref{eq:pion-T}) and  $\la \bar{q}q \ra_T$ in
Eq.(\ref{eq:qqbar-T0}), the GOR relation turns out to hold
in the dilute pion gas at low $T$  \cite{Pisarski:1996mt,Toublan:1997rr}: 
\beq
\label{eq:GOR-T}
\frac{(f_{\pi}^t(T) m_{\pi}(T))^2}{\hat{m} \la \bar{u}u+\bar{d}d \ra_T}
\simeq -1 ,
\eeq
 which was originally noticed in the Nambu$-$Jona-Lasinio model
 at finite $T$ \cite{Hatsuda:1987kg}.

In the case $\rho \neq 0$ with $T=0$, the in-medium pion properties
at low baryon density is determined by the pion-nucleon forward
scattering amplitude which is dictated by several low energy constants
\cite{Thorsson:1995rj,Meissner:2001gz}. For the symmetric two-flavor nuclear matter, we have
\beq
\! \! \! \! \! \! \! \! \! \!  \! \! \! 
\left( \frac{m_{\pi}(\rho)}{m_{\pi}} \right)^2 &=& 
1+ \frac{2}{f_{\pi}^2} \left( 2c_1 -c_2 -c_3 + \frac{g_A^2}{8m_N} \right) \rho   , \\
\! \! \! \! \! \! \! \! \! \!  \! \! \! 
\left( \frac{f_{\pi}^t(\rho)}{f_{\pi}} \right)^2 &=& 
1+ \frac{2}{f_{\pi}^2} \left( c_2 +c_3 - \frac{g_A^2}{8m_N} \right) \rho   , \\
\! \! \! \! \! \! \! \! \! \!  \! \! \! 
\left( \frac{f_{\pi}^s(\rho)}{f_{\pi}} \right)^2 &=& 
1- \frac{2}{f_{\pi}^2} \left( c_2 -c_3 + \frac{g_A^2}{8m_N} \right) \rho   . 
\label{eq:pion-rho}
\eeq  
Using Eq.(\ref{eq:uubar-rho}) and a relation $\sigma_{\pi N} \simeq - 4c_1 m_{\pi}^2$, 
the GOR relation is shown to hold at low density:
\beq
\label{eq:GOR-rho}
\frac{(f_{\pi}^t(\rho) m_{\pi}(\rho))^2}
{\hat{m} \la \bar{u}u+\bar{d}d \ra_{\rho}} \simeq -1 .
\eeq
With the empirical values, $c_1=-0.81 \pm 0.12 \ {\rm GeV}^{-1}$, 
$c_2 = 3.2 \pm 0.25 \ {\rm GeV}^{-1}$, $c_3 = -4.70 \pm 1.16\  {\rm GeV}^{-1}$, 
$f_{\pi}=92.4(3) $ MeV and $g_A=1.2695(29)$ MeV, one finds
$f_{\pi}^t(\rho)/f_{\pi}= 1- (0.26\pm 0.04) (\rho/\rho_0)$,
$f_{\pi}^s(\rho)/f_{\pi}= 1- (1.23\pm 0.07) (\rho/\rho_0)$,
$\la \bar{q}q \ra_{\rho}/\la \bar{q}q \ra_0 = 1 - (0.35 \pm 0.09)(\rho/\rho_0)$. 
Note  that, in asymmetric nuclear matter, there is splitting between
$m_{\pi^-}$, $m_{\pi^+}$ and $m_{\pi^0}$: For example,
 $N/Z=1.5$ at $\rho=\rho_0=0.17 {\rm fm}^{-3}$,
  there is approximately a  $+18$ MeV shift for $\pi^-$,
  a $- 12$ MeV shift for $\pi^+$, and a $+2$ MeV shift for $\pi^0$
   \cite{Meissner:2001gz}. A  possible problem of the linear density formula
Eq.(\ref{eq:pion-rho}) is that $f_{\pi}^s$ vanishes even below  nuclear matter density.

In the leading order of baryon density and in the SU$(2)$ chiral limit,
 one can formulate  two different 
 representations of the in-medium 
change of the chiral condensate in terms of the 
 physical observables: 
\beq
\label{eq:GOR-ratio}
\frac{\la \bar{q}q \ra_{\rho}}{\la \bar{q}q \ra_0}
&\simeq & \left(  \frac{f_{\pi}^t(\rho)}{f_{\pi}} \right)^2
    \left(  \frac{m_{\pi}(\rho)}{m_{\pi}} \right)^2 , \\
\label{eq:TW-ratio}
&\simeq & Z_{\pi}^{1/2}(\rho)  \left(  \frac{b_1}{b_1(\rho)} \right)^{1/2}  .
\eeq
The first one is the in-medium GOR relation Eq.(\ref{eq:GOR-rho}), while
the second one is a combination of the in-medium
Tomozawa-Weinberg relation  \cite{Kolomeitsev:2002gc}, 
$b_1/b_1(\rho) \simeq (f_{\pi}^t(\rho)/f_{\pi})^2$, and 
the in-medium  Glashow-Weinberg relation  \cite{Jido:2008bk},
$\la \bar{q}q \ra_{\rho}/\la \bar{q} q \ra_0 \simeq 
Z_{\pi}^{1/2}(\rho) \times (f_{\pi}^t(\rho)/f_{\pi})$.
Here   $b_1(\rho)$ ($b_1$) is the isovector pion-nucleus (pion-nucleon)
scattering length in the chiral limit and 
 $Z_{\pi}(\rho)$ is the in-medium change of the pion pole residue
 of the correlation function of the pseudo-scalar operator $P^a(x)$.
  The slope of $Z_{\pi}(\rho)$ as a function of $\rho$
  at low density is related to the isoscalar pion-nucleon scattering
  amplitude, while   $b_1(\rho)$ is related to  the 
  energy levels of the deeply bound
   $\pi^-$-atom. Experimental data at $\rho < \rho_0$  indicate
$Z_{\pi}(\rho)<1$ and $b_1/b_1(\rho) < 1$, so that the chiral 
 condensate indeed decreases at finite baryon density.   

\subsubsection{In-medium scalar meson} 
\label{sec:SP-scalar}

The light scalar-isoscalar meson 
has been customarily  called the $\sigma$. 
Since it has the same quantum number as the vacuum,
$\sigma$ is analogous to the Higgs boson $H$ in the electro-weak (EW) theory.
The  $\sigma$ may be interpreted as the excitation associated with the 
  amplitude fluctuation of the chiral condensate $\langle \bar{q}q \rangle$
\cite{Nambu:1960xd,Nambu:1961tp,Nambu:1961fr} 
\cite{GellMann:1960np}
\cite{Delbourgo:1982tv,Weinberg:1990xn}.
  However, there is  a marked difference between $H$ and $\sigma$:  
  In the EW theory,   the NG bosons associated with  spontaneous symmetry breaking 
   ${\rm SU(2) \times U(1)_{\rm Y} \rightarrow U(1)_{\rm em}}$ are 
  absorbed into the gauge bosons, while the NG bosons in QCD 
  (the phase fluctuation of  $\langle \bar{q}q \rangle$) are nothing but physical
  pions.  Therefore, $\sigma$ is allowed to have $s$-wave decay into two pions as long as 
  $m_{\sigma} > 2 m_{\pi}$. Thus $\sigma$ should be a very broad resonance even if it 
  exists.

 Because of the above reason,
  it has been long debated whether there is unambiguous experimental 
  evidence of such a light and broad resonance in  $\pi-\pi$ scattering,
   $\gamma-\gamma$ collision, heavy meson decays, and so on
  \cite{Pennington:2007yt}.
 Recently,  an analysis based on the model independent Roy equation for the partial
 wave amplitude in the scalar-isoscalar channel, $t_{J=0}^{I=0}(s)$,
  has been carried out using precise inputs
 of the $\pi-\pi$ scattering lengths 
 obtained from chiral perturbation theory.  
 The mass and the width of $\sigma$ corresponding to the  second sheet pole
 are then deduced with high accuracy \cite{Caprini:2005zr,Leutwyler:2008xd}:
\beq       
m_{\sigma}=441_{-8}^{+16} \ {\rm MeV}, \ \ \ 
\Gamma_{\sigma} =544_{-25}^{+18} \ {\rm MeV}.
\label{eq:sigma_MG}
\eeq  
Although the existence of  $\sigma$ is established,
its quark-gluon structure is still unknown and is actively 
studied   theoretically, experimentally  \cite{Pennington:2007yt}
and also numerically in lattice QCD simulations \cite{Kunihiro:2008dc,Prelovsek:2008qu}.

\begin{figure}
\includegraphics[width=0.8\columnwidth]{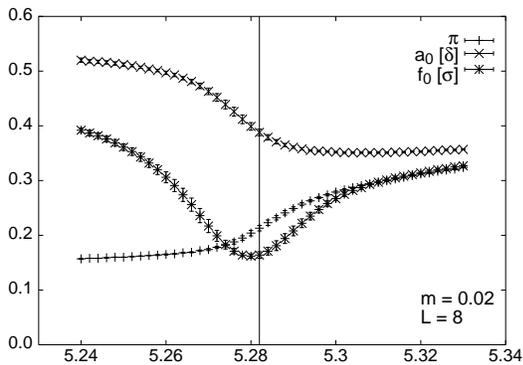}
\caption{
Thermal susceptibilities in three different channels ($\pi$,$\sigma$, and $a_0$)
for two-flavor QCD with staggered fermion on
the $8^3\times 4$ lattice with $m_{u,d} a$ = 0.02.
The vertical (horizontal) axis denotes $\sqrt{1/\chi_{_{\cal O}}}$
(the lattice coupling : $6/g^2$)  \cite{Karsch:2001cy}. Low (high) $T$ corresponds to
 the left (right) of the figure.}
\label{fig:sigma-spect} 
\end{figure}

The medium modification of $\sigma$ has not been  
established yet even at low temperature and density unlike the case of 
the pion. Nevertheless, we may expect from general grounds that there 
would be a  
 partial degeneracy between $\sigma$ and $\pi$ 
 if the system approaches to  the point of chiral symmetry
 restoration 
\cite{Hatsuda:1985eb,Hatsuda:1986gu,Hatsuda:1987kg}
\cite{Bernard:1987im}
\cite{Chiku:1997va,Hatsuda:1999kd,Hatsuda:2001da}.
At finite $T$, such a chiral degeneracy  can be  
detected, e.g., by the  
thermal hadronic susceptibilities associated with the operator ${\cal O}(\tau,\bx)$
 defined in the Euclidean time $\tau$;
\beq
\label{eq:sus-1}
\chi_{_{\cal O}} &=& \int_0^{1/T}d\tau \int d^3 x \langle
{\cal O}(\tau, {\bf x}) O^{\dagger}(0,{\bf 0}) \rangle_T ,\\
& = & \int_0^{\infty} d\omega^2 \frac{\rho_{_{\cal O}}(\omega)}{\omega^2}.
\label{eq:sus-2}
\eeq
Shown in Fig.\ref{fig:sigma-spect}
is the lattice QCD simulation of $\sqrt{1/\chi_{_{\cal O}}}$
with dynamical quarks in two-flavor \cite{Karsch:2001cy}. 
One can see the degeneracy between the susceptibilities in the $\sigma$ channel 
and the $\pi$ channel  as $T$ increases to the left. 
Also, $\sqrt{1/\chi_{_{\cal O}}}$ for $\sigma$ is smaller than that for
$a_0$ (scalar-isovector meson, traditionally called $\delta$) at low $T$, 
which indicates that the spectral strength in the $\sigma$ channel has more
weight in the low frequency region ($\sigma$ is lighter than $a_0$) as can
be seen from Eq.(\ref{eq:sus-2}).
 Splitting between  $a_0$ and $\pi$
  even when $\sigma$-$\pi$ degeneracy is realized at high $T$ 
   reflects the  explicit breaking of $U(1)_{\rm A}$ symmetry.
We note that the close relevance of the scalar-isoscalar susceptibility 
at finite baryon density to the nuclear matter properties 
is also pointed out  \cite{Ericson:2007mx}. 

\subsubsection{In-medium vector meson}   
\label{sec:SP-vector}

Unlike the case of  Eq.(\ref{eq:chiral-SR}) which relates  the chiral condensate
$\la \bar{q}q \ra$ and the spectral function in the pion channel,
no such relation is known in the vector channel. Still, one can derive  
useful relations
 by using the in-medium generalization of the QCD sum rules
\cite{Hatsuda:1991ez,Hatsuda:1992bv}.  In particular, the in-medium
  Weinberg relations \cite{Kapusta:1993hq} for 
 the vector and axial-vector spectral functions at
zero spatial momentum read
\beq
\label{eq:Weinberg-I}
\! \! \! \! \! \! \! \! \! \! \! \! 
& &\int_0^{\infty} \frac{d\omega^2}{\omega^2} (\rho_{_V}(\omega)-\rho_{_A}(\omega)) =0, \\
\label{eq:Weinberg-II}
\! \! \! \! \! \! \! \! \! \! \! \! 
& &\int_0^{\infty} {d\omega^2} (\rho_{_V}(\omega)-\rho_{_A}(\omega)) =0, \\
\label{eq:Weinberg-III}
\! \! \! \! \! \! \! \! \! \! \! \! 
& &\int_0^{\infty} {d\omega^2}{\omega^2}
(\rho_{_V}(\omega)-\rho_{_A}(\omega)) = -\frac{4\pi}{3}\als
\la {\cal O}_{4q} \ra  , 
\eeq
where ${\cal O}_{4q}= {\cal O}_{\mu}^{\mu} + 2 {\cal O}^{00}$ with
\beq
\label{eq:4-quark}
{\cal O}_{\mu \nu} 
=\frac{4}{3} (\bqL \gamma_{\mu} t_{_{\rm C}}^{\alpha} t_{_{\rm F}}^a \qL)
(\bqR \gamma_{\nu} t_{_{\rm C}}^{\alpha} t_{_{\rm F}}^a \qR) .
\eeq
Note that there are longitudinal and transverse spectral functions 
 in the medium, but they coincide at zero spatial momentum, so that 
such a distinction is not made in the above formula.  As is obvious from Eq.(\ref{eq:4-quark}),
 the DBCS in the vector and axial-vector channels 
 is manifested as the higher dimensional four-quark operator and not by the
 simple bilinear operator $\bar{q}q= \bqL \qR + \bqR \qL$: 
 The in-medium changes of $\la {\cal O}_{4q} \ra $ and 
$\la \bar{q}q \ra^2$ are different in general \cite{Hatsuda:1992bv,Eletsky:1992xd}.
At finite $T$ with zero baryon density, it has been proven that there
 is  no exotic phase in which
 $\la \bar{q}q \ra=0$ and  $\la {\cal O}_{4q} \ra \neq 0 $ take place simultaneously
 \cite{Kogan:1998zc}. However, such a phase is not ruled out at finite baryon density
 and is indeed realized in the color superconducting phase 
 \cite{Hatsuda:2008is}.

\begin{figure}
\includegraphics[width=0.8\columnwidth]{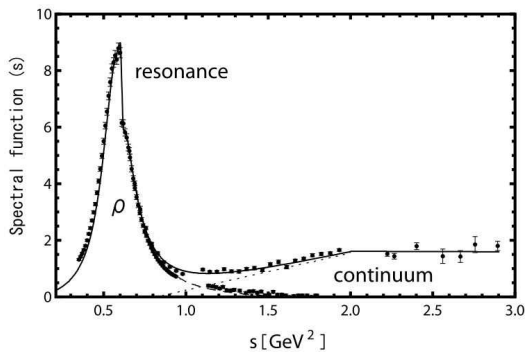}
\caption{Spectral function in the isovector channel in the vacuum
($\rho_{_V}^{\rm vac}(s)$ with $s=\omega^2$) 
obtained from the $e^+e^-$ annihilation into even numbers of pions
 \cite{Kwon:2008vq}.}
\label{fig:rho-spect} 
\end{figure}

The spectral modifications of vector and axial-vector channels at low $T$ with zero $\mu$
are realized as a mixing of the two channels
 due to thermal pions.  In the leading order of the virial expansion in 
 the two-flavor system, one finds \cite{Dey:1990ba}
\beq
\label{eq:vector-T}
\rho_{_V} &=& (1-\Theta(T)) \rho_{_V}^{\rm vac} + \Theta(T)  \rho_{_A}^{\rm vac} ,\\
\label{eq:axial-T}
\rho_{_A} &=& (1-\Theta(T)) \rho_{_A}^{\rm vac} + \Theta(T)  \rho_{_V}^{\rm vac} ,
\eeq
with $\Theta(T)$ given by Eq.(\ref{eq:qqbar-T}).
The spectral functions in the vacuum are measured experimentally (see, e.g., 
Fig.\ref{fig:rho-spect} in the vector channel).
The above mixing formulas show that
the pole positions of the correlation functions do not change at low $T$,
while the pole residues are modified as $(f_{\rho}^t(T)/f_{\rho})^2=1-\Theta(T)$.
Note also that these formulas satisfy the  Weinberg
sum rules. In fact,  
 the $T$-dependence of $\la {\cal O}_{4q} \ra_T$  in the 
 r.h.s. of Eq.(\ref{eq:Weinberg-III}) calculated by using
  the soft pion theorem coincides with that obtained from
  Eqs.(\ref{eq:vector-T},\ref{eq:axial-T}) 
 \cite{Hatsuda:1992bv}.

The density dependence of the four-quark condensate $\left<{\cal O}_{4q}\right>_\rho$ is not
known precisely.  In the leading order of the virial expansion in terms of the 
baryon density, we have 
$\la {\cal O} \ra_{\rho} \simeq  \la {\cal O} \ra_0+ \la {\cal O} \ra_{N} \rho  $.
The  
nucleon matrix element of ${\cal O} = {\cal O}_{4q}$ corresponds to  
 higher twist terms 
 in the deep inelastic lepton-nucleon scattering, but the value is  
 still uncertain. A crude approximation originally made
was a factorization ansatz \cite{Hatsuda:1991ez}: 
$\la \bar{q}_i q_j \bar{q}_k q_l \ra_N
\propto  \la \bar{q}q \ra_0 \la \bar{q}q \ra_N$ with  
appropriate Fierz coefficients.
It is not obvious, however, whether this estimate 
is accurate enough and  further  studies are necessary \cite{Thomas:2007gx}.

\subsection{Dynamical approaches to in-medium hadrons}
\label{sec:DA}

Although there are numerous attempts to relate hadronic spectral functions
 to $\la \bar{q}q \ra$ in the medium \cite{Cassing:1999es,Rapp:1999ej,Alam:1999sc,Mosel:2008tz},
 no rigorous relations 
 have been established yet except for the pion.
 In the following, we briefly outline various theoretical
 approaches that have been attempted so far.

\subsubsection{Naive quark model}
\label{sec:DA-QM}
 Assuming that
  the constituent quark mass $M$ originates mainly from DBCS
  according to the idea of  Nambu and Jona-Lasinio, and assuming further
 that  the vector meson mass follows the additive rule $m_{_V} \simeq 2 M$,
 one may expect a reduction of  $m_{_V}$ associated with the partial restoration
 of chiral symmetry. Such a shift could be detected through the 
 decay of the neutral vector meson into dileptons 
  \cite{Pisarski:1981mq}. 

\subsubsection{Nambu$-$Jona-Lasinio model}  
\label{sec:DA-NJL} 
  As a field-theoretical model
 to treat  the meson properties in the medium
 beyond the simple additive rule, the Nambu$-$Jona-Lasinio (NJL) model
  at finite temperature and density has been studied. In particular,  the 
  spectral degeneracy between $\pi$ and  $\sigma$
   in hot and/or dense matter was  explicitly demonstrated
     \cite{Hatsuda:1985eb,Hatsuda:1986gu} \cite{Bernard:1987im}.
 Further progress along these lines and a  similar dynamical model based on the 
 Dyson-Schwinger equation  can be seen in \cite{Vogl:1991qt,Klevansky:1992qe,Hatsuda:1994pi,Buballa:2003qv}
 and in  \cite{Roberts:2000aa}, respectively. 

\subsubsection{QCD sum rules}  
\label{sec:DA-QSR}
 The QCD sum rule (QSR) is a method which can 
 relate the hadronic spectral functions to the
  QCD condensates through the operator product expansion and the 
   dispersion relation \cite{Shifman:1978bx,Shifman:1978by}.
 This approach has been  generalized to attack the problems of
 hadron properties in the hot/dense medium \cite{Bochkarev:1985ex,Hatsuda:1991ez,Hatsuda:1992bv},
 and further theoretical elaborations were made 
  \cite{Asakawa:1993pq,Jin:1995qg,Koike:1996ga,Klingl:1997kf,Leupold:1997dg,
  Ruppert:2005id}.
 A major difference of the in-medium QSR from the in-vacuum QSR is that 
 there arise Lorentz-tensor condensates. 
 The  weighted average of the spectral function 
  $\langle \rho \rangle_{_W} = \int d\omega \rho(\omega) W(\omega)$ 
   obtained from the in-medium QSR gives useful
  QCD constraints on various models  \cite{Thomas:2005dc,Eichstaedt:2007zp,Kwon:2008vq}.

\subsubsection{Hadron mass scaling}  
\label{sec:DA-scaling}
It was conjectured that the masses of light vector mesons ($\rho, \omega$)
  scale universally as a function of density and/or temperature \cite{Brown:1991kk}.
Near the chiral restoration point,  
$\langle \bar{q}q \rangle /\langle \bar{q}q \rangle_0 \ll 1$,
 the scaling law reads
\beq  
\frac{m_{\rho}^*}{m_{\rho}} \simeq \frac{m_{\omega}^*}{m_{\omega}}
 \simeq 
\frac{\langle \bar{q}q \rangle }{\langle \bar{q}q \rangle_0},
\eeq
where $m^*$ denote the pole mass of in-medium vector mesons
\cite{Brown:2008xh}. The theoretical foundation of 
such a scaling law may be obtained by an approach in which 
 the vector mesons are considered to be the gauge bosons associated with the 
 hidden local symmetry of the chiral effective Lagrangian
   \cite{Harada:2001it,Harada:2001qz,Harada:2003jx,Harada:2005br,Hidaka:2005ac}.

\subsubsection{Bag model}  
\label{sec:DA-bag}
 The bag model is a phenomenological approach in which quarks and gluons are confined in 
 a ``bag'' inside non-perturbative QCD vacuum.  
 In the nuclear medium, the non-vanishing expectation values
  of the scalar-isoscalar meson $\langle \sigma \rangle $  and the time-component of the 
  vector-isoscalar meson $\langle \omega_0 \rangle $ develop, so that they act as 
   effective scalar and vector potentials on  the quarks inside
   the bag. In such a quark-meson coupling model,
  in-medium baryons composed of three valence quarks 
  feel both scalar and vector potentials with opposite sign, 
   while the in-medium mesons composed of quark and anti-quark feel only the scalar  
  potential and obey a universal scaling law dictated by the
  density dependence of $\langle \sigma \rangle $
    \cite{Saito:1994kg,Saito:1996yb,Saito:2005rv}.

\subsubsection{Hadronic models}  
\label{sec:DA-resonance}
 There are purely hadronic descriptions of the in-medium vector mesons.
In the  Walecka type models, the one-loop self-energy of vector mesons in nuclear matter
 receives contributions from the low-energy particle-hole ($p$-$h$) excitations and the 
 high energy nucleon$-$anti-nucleon ($N$-$\bar{N}$)
 excitations. The former gives the standard plasmon effect which increases the 
 pole mass of the vector meson, while the latter tends
  to decrease the pole mass: These effects are simply 
   understood from the  quantum mechanical level repulsion.  
 The net effect with a standard  parameter set of the Walecka model shows that the
 $N$-$\bar{N}$ effect wins and the pole mass decreases
 \cite{Saito:1989jv,Kurasawa:1990ks,Jean:1993bq,Caillon:1993vr,Shiomi:1994gu}.
 For more details of such a model, see the review \cite{Hatsuda:1996xt}.
  The actual vector meson  in nuclear matter receives 
 not only the nucleon effect  but also effects from resonances such as 
 $\Delta$ and $N^*$.
 They  modify the spectral structure non-trivially 
 and cause spectral shift, spectral broadening, and even new  peaks
  \cite{Asakawa:1992ht,Herrmann:1993za,Chanfray:1993ue,
  Friman:1997tc,Rapp:1997fs,Rapp:1999ej,Post:2003hu}.

\subsubsection{Chiral effective theories}  
\label{sec:DA-scalar}
 Spectral change of the broad resonance $\sigma$
  in the hot and/or dense medium is one of the interesting signals of 
 chiral symmetry restoration  \cite{Hatsuda:1985eb,Hatsuda:1986gu}.
 An example is the  enhancement of the spectral function in the scalar-isoscalar
  channel near the 2$\pi$ threshold at finite $T$ \cite{Chiku:1997va,Chiku:1998kd,Volkov:1997dx}
 and at finite baryon density \cite{Schuck:1988jn,Aouissat:1994sx,Hatsuda:1999kd}.
 Since  $\sigma$ and $\rho$ are both resonances in the $\pi$-$\pi$ system, it is
 necessary to take into account multiple scattering of the pions inside the medium
 to study their spectral structures. For this purpose,
  unitarization of the in-medium $\pi$-$\pi$ 
 scattering amplitude of the chiral Lagrangian 
 on the basis of the $N/D$ method and the inverse amplitude method has
 been extensively explored
 \cite{Jido:2000bw,Oller:2000ma,Yokokawa:2002pw,Cabrera:2005wz,FernandezFraile:2007fv,Cabrera:2008tj}.
 The large $N$ expansion of the $O(N)$ symmetric scalar model has also been studied
as a toy model  \cite{Patkos:2002vr,Hidaka:2004ht}.   

\subsubsection{Lattice QCD}  
\label{sec:DA-lattice}
 The best quantitative method to study the medium modification of hadrons
 should be the lattice QCD simulations.  The maximum entropy method (MEM), which
  allows us  to  extract
 the in-medium spectral functions from the Euclidean correlation functions measured
 on the 
 lattice has been proposed  and tested 
  in the quenched lattice QCD simulations 
 at finite temperature \cite{Asakawa:2000tr,Asakawa:2002xj,Karsch:2003jg}.
 Its application to the lattice data  with dynamical quarks has been started 
  at finite $T$ although  it is still limited to heavy quarkoniums at the moment \cite{Aarts:2007pk}.
 The application of this method to QCD at
  finite density is  still not attainable due to the 
 sign problem which invalidates the importance sampling in 
 Monte Carlo simulations  \cite{Muroya:2003qs}.  In the mean time,
  it would be useful to study in-medium hadronic properties by 
   analytic lattice approaches such as the strong 
  coupling expansion at finite $T$ and $\mu$ \cite{Ohnishi:2008yk}.

\section{PSEUDOSCALAR MESON: $\pi$ IN NUCLEI}

\subsection{Theoretical background}

As we have shown in the introduction, a sizable reduction of the
chiral condensate in nuclear matter of about 30\% is theoretically expected.
A possible way to detect this reduction is to study the 
in-medium pion properties  
 through  precision spectroscopy of deeply bound pionic atoms 
 and through precision measurements of the low-energy
  pion-nucleus scattering.

The basic tool to relate the in-medium chiral condensate and the
 experimental data is the pion-nucleus optical potential $U_{\rm opt}({\bf r})$.
The starting point is the pion propagator in asymmetric nuclear matter,
\beq
 [D_{\pi}(\omega, {\bf q})]^{-1}
=  \omega^2 - {\bf q}^2 - m_{\pi}^2 - \Pi(\omega, {\bf q})  .
\label{eq:pion-prop}
\eeq
Here $\Pi$ denotes the in-medium pion self-energy  which 
is written in the linear density approximation as 
\beq
\Pi (\omega,{\bf q})
 = - {\cal T}^{+} \rho -  \epsilon {\cal T}^{-} \delta \rho ,
\label{eq:pion-self}
\eeq
where $\rho=\rho_p+\rho_n$ (total baryon density),
$\delta \rho=\rho_p-\rho_n$ (total isospin density), and 
$\epsilon = +1, 0, -1$ for $\pi^{-}, \pi^0, \pi^{+}$.
The off-shell isoscalar (isovector) forward scattering amplitude
${\cal T}^{+(-)}$ for small $\omega$ and  at ${\bf q}=0$ reads \cite{Ericson:1988gk}
\beq 
{\cal T}^{+}(\omega)= \frac{\sigma_{\pi N}-\beta \omega^2}{f_{\pi}^2},
\ \ \ {\cal T}^{-}(\omega) = \frac{\omega}{2f_{\pi}^2}.
\eeq
From the constraints ${\cal T}^{+}(\omega=m_{\pi},{\bf 0})
= 4\pi (1+m_{\pi}/M) a_{\pi N}$ with the accidentally small isoscalar scattering
length
$a_{\pi N} = (0.0016 \pm 0.0013)m_{\pi}^{-1}$ \cite{Schroder:1999uq}, we have
$\beta \simeq \sigma_{\pi N}/m_{\pi}^2$. This approximation is valid
in 5\% accuracy to $\beta$. The $\pi^0$ does not have
a mass shift in the nuclear medium in the same approximation.

One may introduce the energy independent optical potential as
$\omega^2 = m_{\pi}^2 + 2m_{\pi} U_{\rm opt}$ which is a solution
 of the dispersion relation $\omega^2 - m_{\pi}^2 - \Pi(\omega)=0$.
 As long as $U_{\rm opt} \ll m_{\pi}$ and the linear density approximation
  is valid, one finds
\beq   
U_{\rm opt} \simeq \epsilon\  \frac{\delta \rho}{4[f_{\pi}^{t}(\rho)]^2} .
\label{eq:V_opt}
\eeq
This implies that there is an extra repulsion (attraction) for $\pi^-$ ($\pi^+$) 
from the medium  associated 
with the reduction of the pion decay constant \cite{Weise:2001sg,Weise:2000xp}.

In the local density approximation, $\rho_{p,n} \rightarrow \rho_{p,n}({\bf r})$,
the Klein-Gordon equation corresponding to  Eq.(\ref{eq:pion-prop}) reads
$\left[\bar{\omega}^2 + \nabla^2 - m_{\pi}^2 - 
\Pi(\bar{\omega},\nabla ;\rho_{p,n} ({\bf r})) \right] \Phi ({\bf r}) =0,$
or equivalently near the mass shell, 
\beq
\label{eq:KG-V_opt}
\! \! \! \! \! 
\left[ \bar{\omega}^2 + \nabla^2 - m_{\pi}^2 - 
2 m_{\pi}U_{\rm opt}(\rho_{p,n} ({\bf r})) \right] \Phi ({\bf r}) =0,
\eeq 
with $\bar{\omega}=\omega - V_{\rm Coul}({\bf r})$ where 
 $V_{\rm Coul}$ is the 
Coulomb potential between $\pi^{\pm}$ and the nucleus.
There are several important contributions to $U_{\rm opt}$ other than
Eq.(\ref{eq:V_opt}) in studying the experimental data of the 
deeply bound pionic atoms and the low energy $\pi$-nucleus scattering:
(i) higher order terms in $\omega$ and $m_{\pi}$ to the s-wave
part of   ${\cal T}^+$ and ${\cal T}^-$ 
\cite{Kolomeitsev:2002gc,Kolomeitsev:2002mm,Doring:2007qi},
(ii) the two-nucleon absorption of the pion, and (iii)
the $p$-wave contribution with spatial derivatives  \cite{Ericson:1988gk,Friedman:2007zz}.
Their explicit forms are given in the next section. 

\subsection{Pion-nucleus optical potential}
\label{sec:pion-nucleus-potential}
\subsubsection{$s$-wave and $p$-wave parts}
The pion-nucleus potential \cite{Ericson:1966fm} is composed of the $s$-wave and $p$-wave parts:
\begin{eqnarray}
\label{eqn:OpticalPotential}
&&U_{\rm opt} (r) = U_{\rm s} (r) + U_{\rm p} (r),\hspace*{4cm}\\
&&U_{\rm s} (r) = - \frac{2\pi}{m_\pi} [b(r) + \varepsilon_2 B_0 \rho^2(r)],\\
&&U_{\rm p} (r) = \frac{2\pi}{m_\pi} \vec{\nabla} \cdot
[c(r) +\varepsilon_2^{-1} C_0 \rho^2(r)]L(r) \vec{\nabla},
\end{eqnarray}
with
\begin{eqnarray}
b(r) &=& \varepsilon_1 \{b_0 \rho (r) + b_1 [\rho_n (r) - \rho_p
(r)]\},\hspace*{1.5cm}\label{eq:swave}\\
c(r) &=& \varepsilon_1^{-1} \{c_0 \rho (r) + c_1 [\rho_n (r) - \rho_p(r)]\},\\
L(r) &=& \frac{1}{1 + \frac{4}{3} \pi \lambda [c(r) + \varepsilon_2^{-1} C_0
\rho^2
(r)]},
\end{eqnarray}
where $\lambda$ is the
Lorentz-Lorenz-Ericson-Ericson correction parameter.
The kinematical factors $\varepsilon_1$ and $\varepsilon_2$ are defined as
$\varepsilon_1 = 1
+ m_\pi / m_N$ and $\varepsilon_2 = 1
+ m_\pi / 2m_N$ with the nucleon mass $m_N$.
The complex parameters $B_0$ and $C_0$ are the $s$-wave and $p$-wave absorption parameters, respectively.

\begin{figure}
\includegraphics[width=0.7\columnwidth]{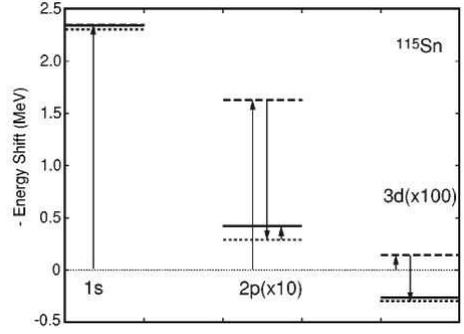}
\caption{\label{fig:umemoto-fig1}
The binding energies with finite-size Coulomb potential 
only, $B_{\rm Coul}$, and Coulomb plus optical potential, $B_{\rm full}$, are calculated. 
The energy shifts $B_{\rm Coul}-B_{\rm full}$ are shown as the solid bars for pionic 
$1s, 2p$, and $3d$ states for $^{115}$Sn. The shifts due to the real 
local terms in the potential are shown by dashed bars. Dotted bars 
are the results with all real terms (local plus nonlocal) in the optical 
potential \cite{Umemoto:2000aw}.
}
\end{figure}

From the pionic-atom x-ray data, the $p$-wave parameters have been fairly precisely determined, while the x-ray data are less sensitive to the $s$-wave parameters. This situation can be understood from Fig.~\ref{fig:umemoto-fig1}, which shows that the $1s$ strong-interaction shift is dominated by the $s$-wave potential while the $3d$ (and higher) level shifts are dominated by the $p$-wave potential. It is therefore important to obtain experimental information on the $1s$ level in order to precisely determine the $s$-wave parameters.

\subsubsection{Pionic hydrogen - the $\pi N$ scattering lengths at threshold}
\label{sec:pionic-hydrogen}

In the low-density limit, the potential is described by just two parameters, $b_0$ and $b_1$ in Eq.(\ref{eq:swave}), which are nothing but isoscalar and isovector $\pi N$ scattering lengths, respectively\footnote{\citet{Ericson:1966fm} defines $b_0\equiv (a_{1/2}+2 a_{3_2})/3, b_1\equiv (a_{3/2}-a_{1/2})/3$, while isoscalar $a^+$ and isovector $a^-$ scattering lengths are $a^+\equiv (a_{1/2}+2 a_{3_2})/3, a^-\equiv (a_{1/2}-a_{3/2})/3$ in pionic hydrogen literatures (e.g., \cite{Gotta:2008gd}). Here, $a_{1/2}$ and $a_{3/2}$ respectively are the isospin 1/2 and 3/2 scattering lengths. Note therefore $b_0=a^+$ and $b_1=-a^-$.}. These have been very precisely determined by the pionic hydrogen x-ray spectroscopy at  the Paul Scherrer Institut (PSI) \cite{Schroder:1999uq,Gotta:2008gd}.

The $\pi N$ scattering lengths can be determined from the observed strong-interaction shift $\epsilon_{1s}$ and width $\Gamma_{1s}$ of pionic hydrogen:
\begin{eqnarray}
\frac{\epsilon_{1s}}{B_{1s}} &=& -\frac{4}{r_B} a_{\pi^- p \rightarrow \pi^- p} (1+\delta_\epsilon)\nonumber\\
& \propto& b_0-b_1,\\
\frac{\Gamma_{1s}}{B_{1s}} &=& 8 \frac{q_0}{r_B}\left(1+\frac{1}{P}\right) \left[ a_{\pi^- p\rightarrow \pi^0 n} (1+\delta_\Gamma)\right]^2\nonumber\\
& \propto& (b_1)^2,
\end{eqnarray}
where $B_{1s}=3.24$ keV and $r_B=216$ fm respectively are the $1s$ binding energy and the `Bohr radius' of pionic hydrogen,  $q_0= 0.1421$ fm$^{-1}$ 
 the center-of-mass momentum of the $\pi^0$ in the charge-exchange reaction 
$\pi^- p\rightarrow \pi^0 n$
and $P= 1.546 \pm 0.009$ the branching ratio of charge exchange and radiative capture (Panofsky ratio). The quantities $\delta_{\epsilon, \Gamma}$ represent the corrections to be applied to the 
experimentally determined scattering length in order to obtain pure strong-interaction quantities.


\begin{figure}
\includegraphics[width=.85\columnwidth]{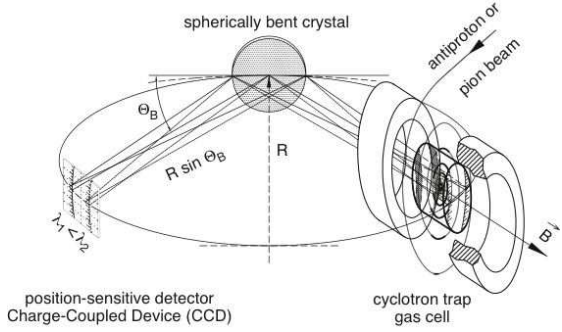}
\caption{\label{fig:pionic-hydrogen-setup}  
A schematic drawing of the PSI pionic hydrogen setup
\cite{Gotta:2008gd}.
}
\end{figure}
The PSI experiment  R-98.01 (Fig.~\ref{fig:pionic-hydrogen-setup}) used a superconducting cyclotron trap to 
produce a high stop density for pions in a hydrogen gas target, a
Johann-type bent-crystal spectrometer, and a two-dimensional CCD focal-plane detector to determine the 
pionic-hydrogen x-rays. The $1s$ strong-interaction shift was determined from the $3p-1s$ energy to be 
$\epsilon_{1s}^{\pi H} = 7120 \pm 11$ meV. The $1s$ hadronic broadening was deduced from the widths of $4p-1s$, $3p-1s$ and $2p-1s$ transitions, so as to correct for the kinematical broadening caused by the preceding transitions width, to be $\Gamma_{1s}^{\pi H} = 823\pm 18$ meV. With the corrections $\delta_\epsilon$ and $\delta_\Gamma$ obtained within chiral perturbation theory \cite{Gasser:2002ly,zemp:2003}, the scattering lengths have been obtained to be \cite{Gotta:2008gd}
\begin{eqnarray}
b_0&=&  0.0069 \pm 0.0031 \mbox{ $m_\pi^{-1}$ },\\
b_1&=& -0.0864\pm 0.0012\mbox{ $m_\pi^{-1}$ }.
\end{eqnarray}
These values are near the  leading-order result derived from current algebra (called Tomozawa-Weinberg (TW) values) \cite{Tomozawa:1966jm,Weinberg:1966kf}
of 
\begin{eqnarray}
b_0^{TW} &=&0 ,\\
b_1^{TW} &=&-\frac{1}{4 \pi \epsilon_1}\frac{m_\pi}{2 f_\pi^2}=-0.079 \mbox{ $m_\pi^{-1}$ },
\end{eqnarray} 
with $\epsilon_1=1+m_\pi/m_N=1.149$, $m_\pi=139.57{\rm~MeV}$ and $f_\pi=92.4\rm~MeV$,
revealing an important feature of the underlying chiral 
symmetry. The $\sim 10\%$ gap between the experimental value of $b_1$ and $b_1^{TW}$ gets closed by pion-loop corrections of order $m_{\pi}^3$ \cite{PhysRevC.52.2185,Bernard1993421}.

\subsubsection{The missing repulsion problem}
It has long been known that  the available pionic-atom x-ray data cannot be fitted with $b_0$ and $b_1$ values, but some enhancement of the $s$-wave repulsive strength is required. This is known as the `missing repulsion' problem.

\begin{figure}
\includegraphics[width=0.8\columnwidth]{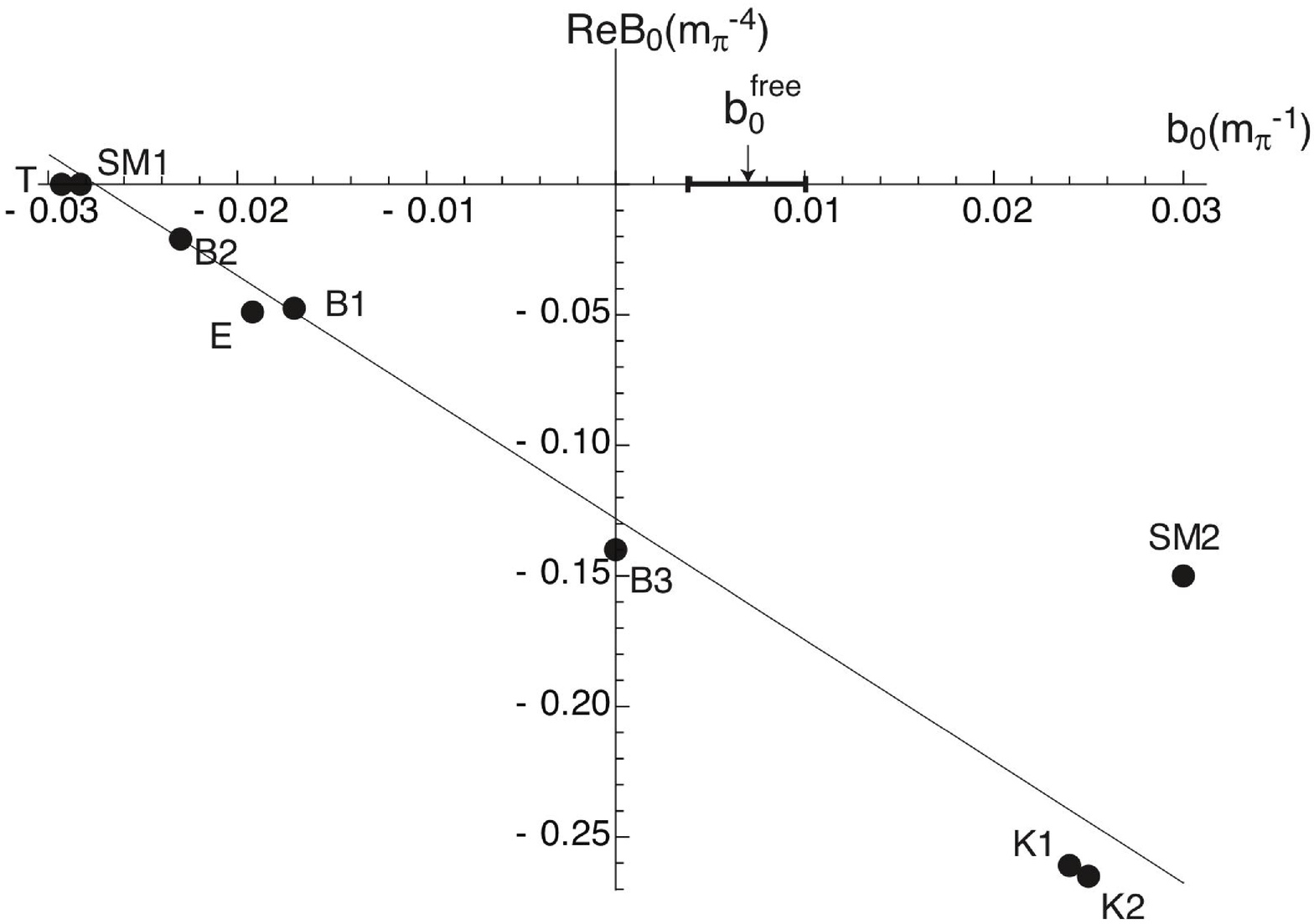}
\caption{\label{fig:b0-ReB0}  
The $s$-wave parameters $b_0$ vs ${\rm Re} B_0$ 
obtained by fitting the x-ray data; T: \cite{Tauscher:1971}, B1-3:  \cite{Batty:1997zp,Batty:1979cr,Batty:1983uv}, SM1-2:  \cite{Seki:1983sh}, K1-2: \cite{Konijn:1990uy}, E:  \cite{Ericson:1981hs}.  The line is $b_0+ 0.215 \, {\rm Re} B_0 =-0.028$. 
}
\end{figure}

The situation is illustrated in 
Fig.~\ref{fig:b0-ReB0}, in which the $s$-wave parameters $b_0$ and ${\rm Re} B_0$ obtained by various authors are plotted. We note, firstly, that there is a strong correlation between $b_0$ and ${\rm Re}B_0$, approximated by $b_0+ 0.215 \, {\rm Re} B_0 =-0.028$ (known as the Seki-Masutani relation \cite{Seki:1983sh}) \footnote{This linear relation arises since the mean density probed by the $\pi^-$ happens to be about $0.5-0.6\rho_0$ regardless of element, so that $b_0 \rho + {\rm Re}B_0 \rho^2$ can be linearized as $b_0 \rho +({\rm Re}B_0 \rho_0 /2 )\rho$ \cite{YH03}.}.
Secondly, the Seki-Masutani relation implies that the ${\rm Re}B_0$ value must be around $-0.15$ if we are to take the `free' $b_0$ value,  which is much  larger in magnitude than expected from the pion deuteron scattering length \cite{Chatellard:1997kc}. If on the other hand ${\rm Re}B_0$ is forced to be small ($\simeq 0$), the fitted $b_0$ is about $-0.03$, quite far away from the `free' value. 

Part of the repulsion is known to be provided by the `double scattering' correction to $b_0$ \cite{Loiseau:2001xr,Ericson:1966fm},
\begin{eqnarray}
\label{eq:doublescattering}
b_0 &\rightarrow& \bar{b}_0 = b_0 -[b_0^2+2 b_1^2]\left< \frac{1}{r}\right>,
\end{eqnarray}
where the inverse correlation length is expressed by the Fermi momentum $k_{\rm F}$ as
\begin{eqnarray}
\left<\frac{1}{r}\right> &=& \frac{3}{2\pi} k_{\rm F}(\rho) = \frac{3}{2\pi} \left[ \frac{3 \pi^2}{2} \rho(r)\right]^{1/3},
\end{eqnarray}
but this correction is still insufficient to account for the missing repulsion.
The experiment S236 at GSI, discussed in the next section, established that the in-medium value of $b_1$ is enhanced relative to $b_1$, which, through Eq.(\ref{eq:doublescattering}), provides the extra $s$-wave repulsion.


\subsection{Deeply-bound pionic atom spectroscopy}
\begin{figure}
\includegraphics[width=.65\columnwidth]{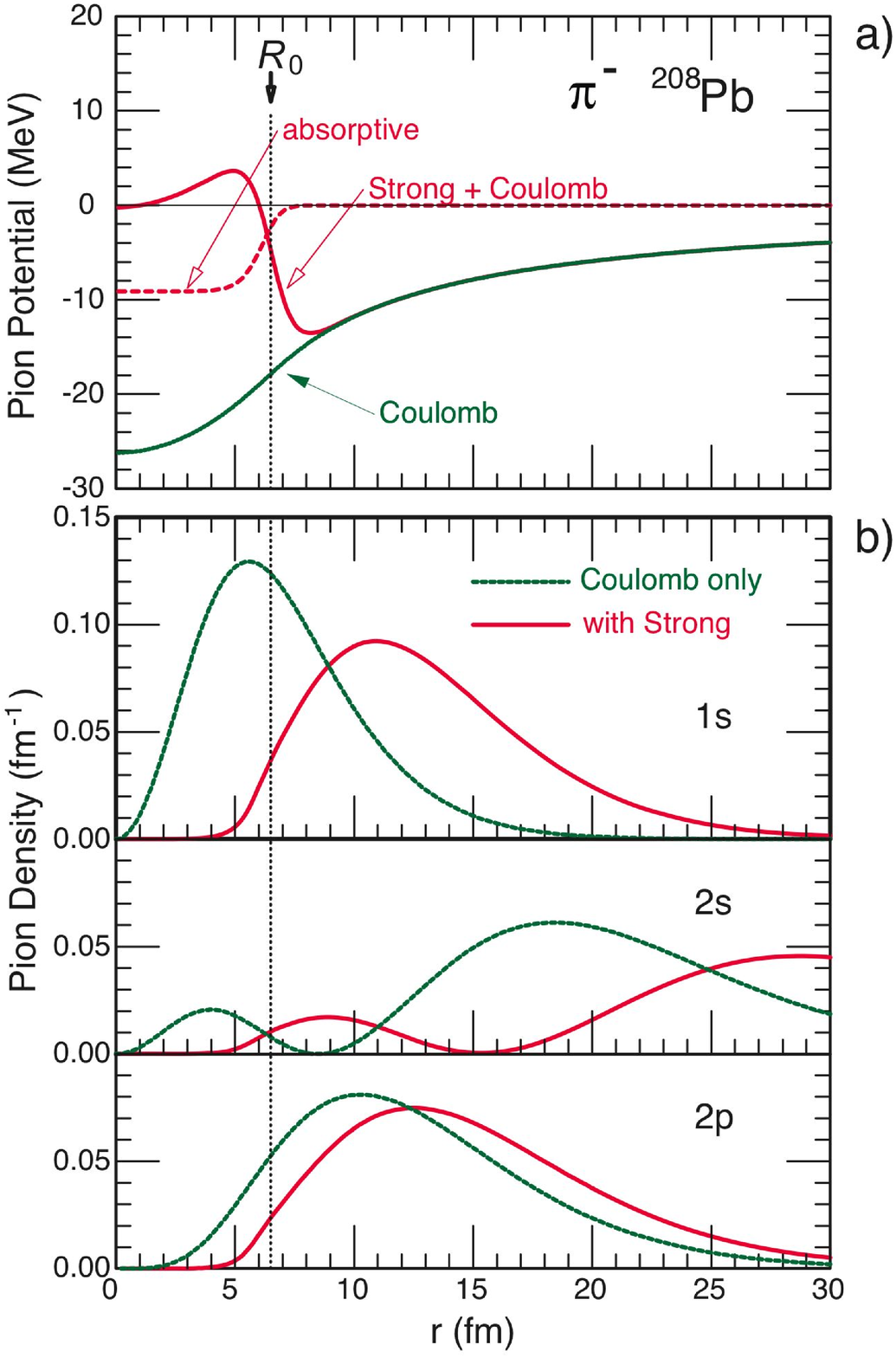}
\caption{\label{fig:pionic-pb208}  
a) The pion optical potential for $^{208}$Pb. The 
finite-size Coulomb potential is expressed by the dotted curve and the one with the optical potential 
by the solid curve. The imaginary part is depicted by the dashed curve. 
b) The pionic wavefunctions of the $1s$, $2s$ and $2p$ states in coordinate space. The dashed curves 
and the solid curves are obtained with the finite-size Coulomb potential and with the optical 
potential. The half-density radius $R_0$ of $^{208}$Pb is indicated by the broken line \cite{toki1991saf}.
}
\end{figure}

In 1996,  the experiment S160 at GSI reported the first observation of the  {\em deeply bound pionic states} in $^{207}$Pb \cite{yamazaki1996ddb}, whose existence and formation had 
been predicted previously \cite{Friedman:1984yg,Toki:1988nn,Toki:1989wq,Hirenzaki:1991us,Toki:1991kl}. 
This discovery
opened an entirely new way to study the hadron properties in the nuclear medium. 
As illustrated in Fig.~\ref{fig:pionic-pb208}, the $1s$ wavefunction of $\pi^-$ bound to a heavy nucleus overlaps appreciably with the nuclear density distribution, and hence the in-medium modification of the pion properties may have detectable effects on the binding energy and/or width.  

Although this was not clearly recognized when the deeply-bound pionic-atom spectroscopy was initially conceived \cite{Toki:1988nn}, it was later established
 that the in-medium value of the isovector $\pi N$ scattering length $b_1$, derived from the $1s$ binding energies, is connected to the (temporal part of the) in-medium pion decay constant $f^{t}_\pi$ via the in-medium Tomozawa-Weinberg relation \cite{Tomozawa:1966jm,Weinberg:1966kf},\cite{Kolomeitsev:2002gc}, 
\begin{equation}
b_1 (\rho) = -\frac{4\pi}{ 1+m_\pi/m_N}\frac{m_\pi}{2 \left[{f^{t}_\pi} (\rho)\right]^2}
\end{equation}
which is in turn connected to the  in-medium  chiral condensate via the  in-medium Gell-Mann$-$Oakes$-$Renner (GOR) relation Eqs. (\ref{eq:GOR-rho}) and (\ref{eq:GOR-ratio}), or via the Glashow-Weinberg (GW) relation Eq.(\ref{eq:TW-ratio}).

In the case of deeply-bound pionic atoms, the $\pi^-$ quantum number is well defined. Moreover, as will be shown, the use of recoilless $(d,^3\rm He)$ kinematics ensures that the nucleus is in the ground state. The pion wavefunction and its overlap with the nuclear density are therefore precisely calculable by solving the Klein-Gordon equation. This makes it possible to compare the experimental results with theoretical predictions, and to quantitatively deduce the effects of partial restoration of chiral symmetry.

\begin{figure}
\includegraphics[width=.75\columnwidth]{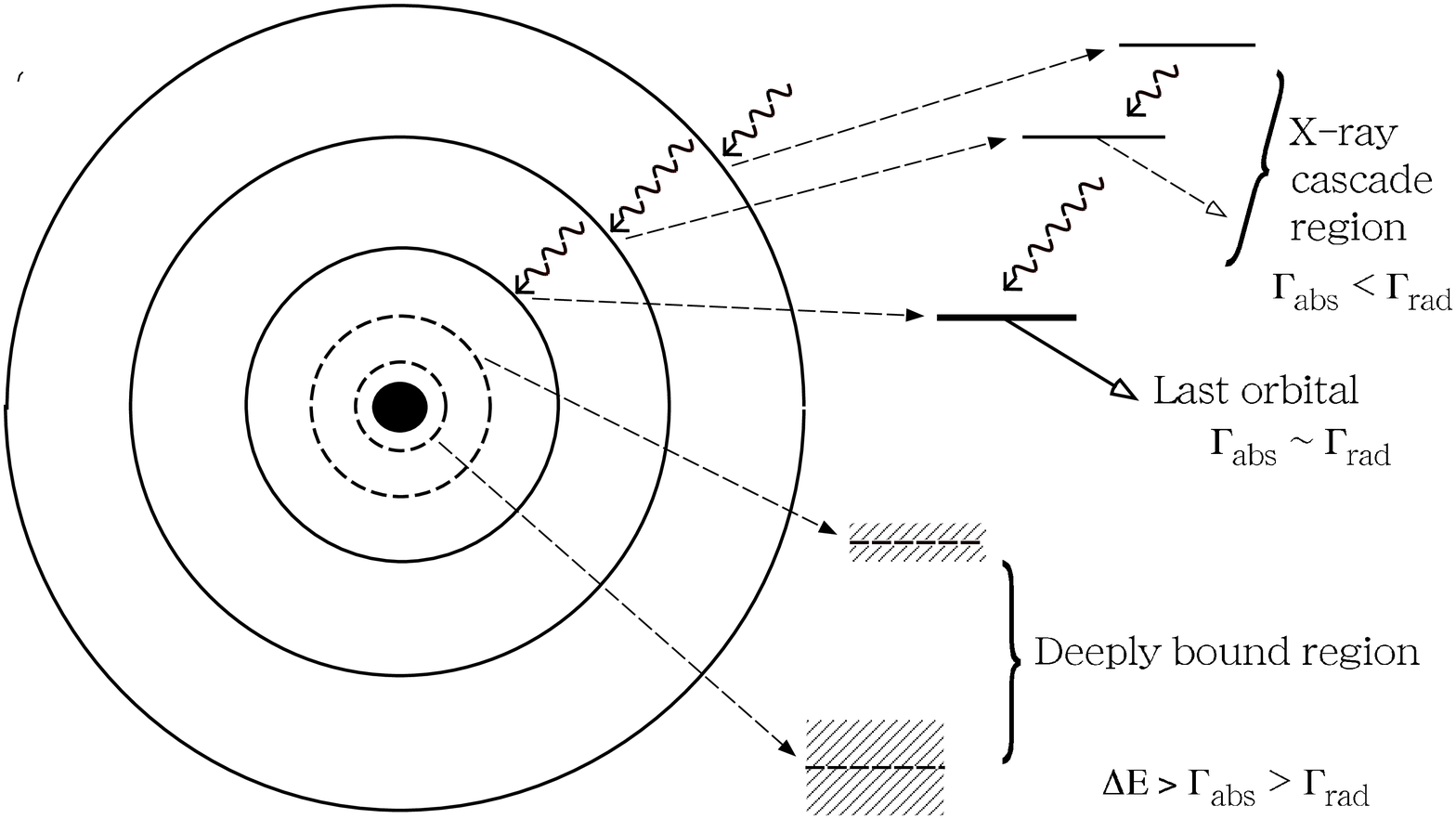}
\caption{\label{fig:PiatomX-ray11}  
A schematic figure of pionic atom states 
with x-ray transitions down to the last orbital and deeply bound inner orbits with large widths 
which cannot be populated following the x-ray cascade. They have large widths due to nuclear 
absorption but are still discrete states with $\Gamma_n < E_n-E_{n-1}$.  \cite{PhysRepPion}.
}
\end{figure}

A wealth of data on pionic atoms has been collected by means of pionic x-ray spectroscopy \cite{Batty:1997zp}.
In the x-ray spectroscopy experiments  (e.g., \citet{delaat1991ssi}), pions, injected into a target, slow down and form pionic atoms, and x-rays emitted in the cascade are measured, as schematically depicted in Fig.~\ref{fig:PiatomX-ray11}. While high-lying states do not show detectable strong-interaction effects, the energy levels of low-lying states get shifted from those calculated by using the electromagnetic interaction. The level widths also become larger due to pion absorption on the nucleus. From the measured strong-interaction shifts ($\epsilon$) and widths ($\Gamma$), the pion-nucleus strong-interaction potential parameters can be deduced.

\begin{figure}
\includegraphics[width=.8\columnwidth]{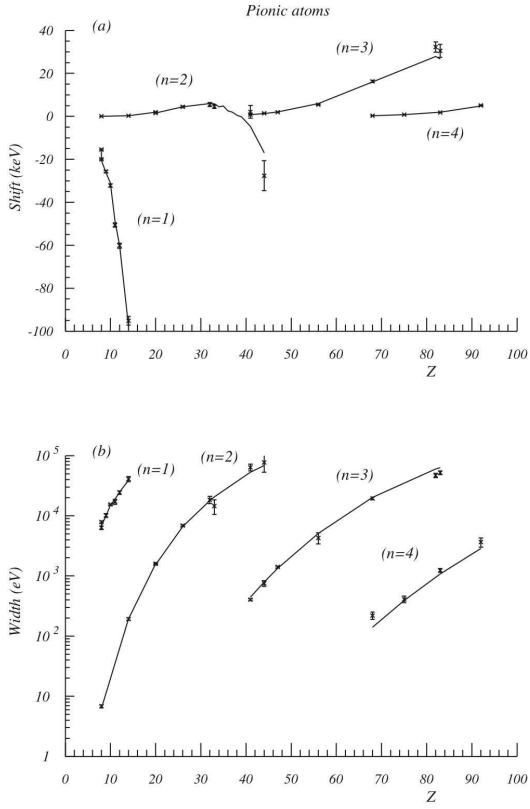}
\caption{\label{fig:bfg-pion-shift}  
Strong-interaction shift (top) and width (bottom) of pionic atoms. The continuous lines join points calculated with the best-fit optical potential
\cite{Batty:1997zp}.
}
\end{figure}

The measured pion-nucleus strong-interaction shifts are shown in Fig.~\ref{fig:bfg-pion-shift} as function of atomic number $Z$. This plot shows that the information on the $1s$ level shift is available only up to the atomic number of $Z=14$. This is because beyond the {\em last orbital} (Fig.~\ref{fig:PiatomX-ray11}) x-rays cannot be observed due to the increase of the pion-nucleus absorption width. As shown in Fig.~\ref{fig:bfg-pion-shift},  $3d$ is the last orbital for medium-to-heavy nuclei, meaning that it is not possible to obtain the information on $1s$ and $2p$ levels by means of x-ray spectroscopy. The $1s$ and $2p$ levels which cannot be populated by x-ray cascade are often referred to as the {\em deeply-bound pionic states}. 

\begin{figure}
\includegraphics[width=\columnwidth]{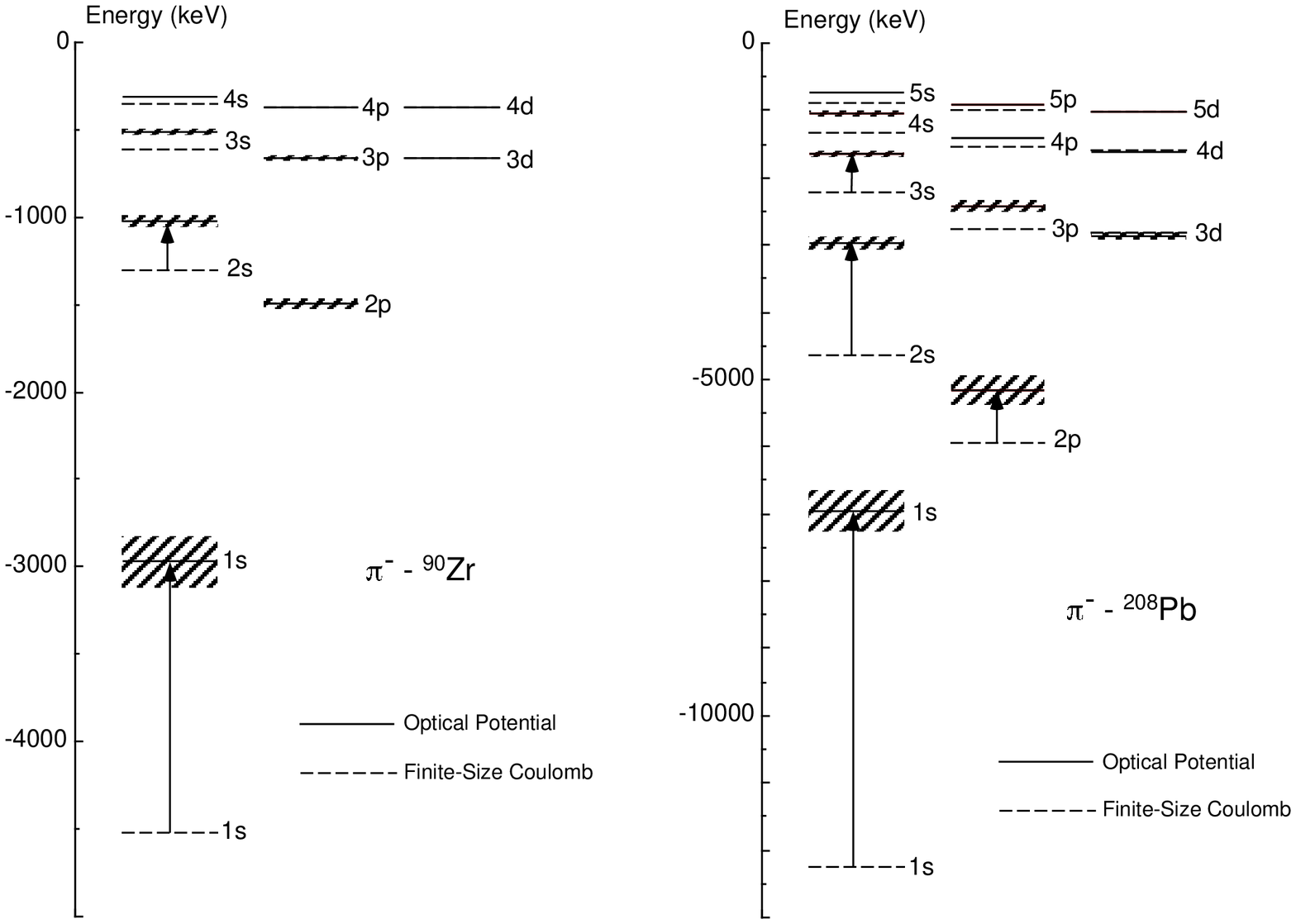}
\caption{\label{fig:ZrPbspectra}  
Energy levels of pionic atoms of $^{90}$Zr and $^{208}$Pb. The results 
with the finite-size Coulomb potential (i.e., taking account of the nuclear charge distribution) are shown by dashed bars, while those with the optical 
potential by solid bars with hatched area indicating the level widths
 \cite{Toki:1988nn}.
}
\end{figure}

\subsubsection{Structure of deeply-bound pionic atoms}

This, however, does not mean that those states do not exist, as was recognized by \citet{Friedman:1984yg} and by \citet{Toki:1988nn}.  In Fig.~\ref{fig:ZrPbspectra}, a typical pion-nucleus optical potential set (see section \ref{sec:pion-nucleus-potential}) was used to calculate the binding energies and widths of pionic Zr (left) and Pb (right) atoms. The $1s$ level width $\Gamma_{1s}$ is found to be much smaller than the $2p-1s$ level interval ($\Delta E_{2p-1s}<\Gamma _{1s}$) even for heavy pionic atoms such as pionic $^{208}$Pb. Namely, the deeply-bound pionic atoms are metastable, despite the strong pion-nucleus absorption.

This somewhat counter-intuitive result arises due to the repulsive nature of the pion-nucleus $s$-wave interaction, which causes the $1s$ binding energies to decrease (Fig.~\ref{fig:ZrPbspectra}) and also pushes the wavefunction outwards (see the difference between the solid and dashed curves in Fig.~\ref{fig:pionic-pb208}(b)). This then reduces the pion-nucleus overlap, making the level width narrower.

\subsubsection{Formation of deeply-bound pionic atoms}

\begin{figure}
\includegraphics[width=0.6\columnwidth]{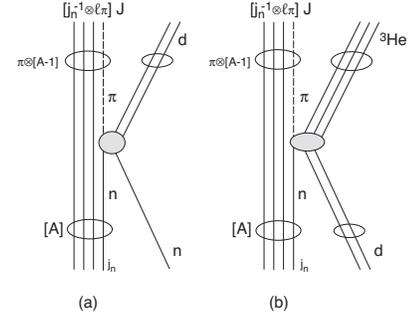}
\caption{\label{fig:nd-d3he}  
Diagrams for proton-pick-up pion-transfer (a) $(n,d)$ and (b) ($d$,$^3$He) 
reactions to form pionic bound states on a neutron-hole state 
\cite{Hirenzaki:1991us}.
}
\end{figure}

\begin{figure}
\includegraphics[width=0.75\columnwidth]{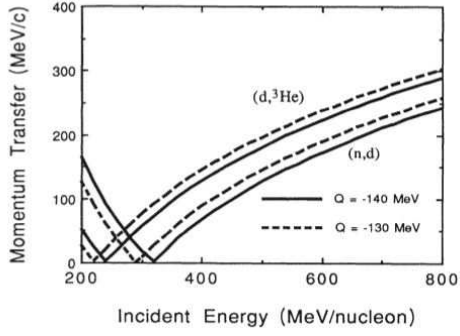}
\caption{\label{fig:q-trans}  
Momentum transfers in the (n,d) and (d,$^3$He) reactions on Pb for 
$Q=-130$ and $-140$~MeV as a functions of the incident energy per nucleon
\cite{Hirenzaki:1991us}.
}
\end{figure}

The original deeply-bound pionic atom formation scheme proposed by \citet{Toki:1988nn} was to use the $^{208}{\rm Pb}(n,p)$ reaction. This was tested at TRIUMF \cite{iwasaki1991sdb}, but no bound-state peak was observed.
It was soon recognized that  $(n,d)$ or $(d,^3\rm He)$ reactions (see Fig.~\ref{fig:nd-d3he}) are more suitable \cite{Hirenzaki:1991us}. These are {\em recoilless} as well as {\em substitutional} reactions, in which a neutron in the $s$ shell is picked up and the produced $\pi^-$ is left in the $1s$ orbit with a small momentum transfer (Fig.~\ref{fig:q-trans}). Due to the substitutional nature of the reaction, the angular momentum transfer is $\Delta L = 0$. The small momentum transfer makes it possible to satisfy the angular momentum matching in surface reactions, thereby minimizing the nuclear distortion.

\begin{figure}
\includegraphics[width=0.6\columnwidth]{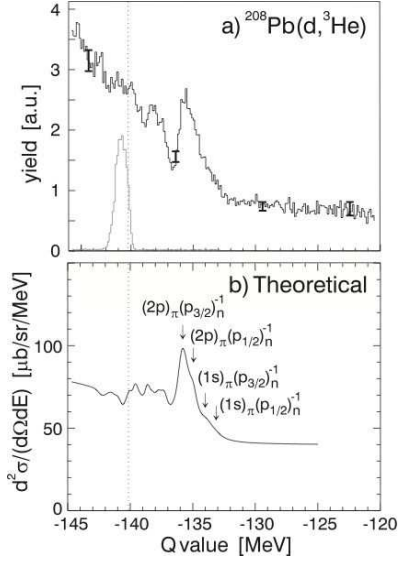}
\caption{\label{fig:pionic-pb207}  
(a) The $^{208}{\rm Pb}(d,^3{\rm  He})$ spectrum. The abscissa is the reaction $Q$ value. 
and the $\pi^-$ emission threshold $(Q = -140.14 MeV)$ is shown by a dotted line. 
The $p(d, ^3{\rm He})\pi^0$ reaction peak used as a calibration is also shown in 
the $Q$ value scale of the $^{208}{\rm Pb}(d,^3{\rm He})$ reaction kinematics. 
(b) Theoretical prediction by \citet{Hirenzaki:1991us}. A FWHM resolution of 0.5 MeV is assumed \cite{yamazaki1996ddb}.
}
\end{figure}

Using the $(d,^3{\rm He})$ reaction, the experiment S160 at GSI succeeded to observe the $2p$ and $1s$ states of pionic $^{207}$Pb \cite{yamazaki1996ddb}.  As shown in Fig.~\ref{fig:pionic-pb207}, the experimental result agrees remarkably well with the theoretical prediction. This firmly established the methodology of deeply-bound pionic atom spectroscopy.

\subsubsection{GSI S236 -  Sn(d,$^3$He)}

In principle,  the $s$-wave parameters may be obtained by analyzing
the $1s$-level energies and widths of light pionic atom x-rays ($Z\leq 14$, Fig.~\ref{fig:bfg-pion-shift}).
However, since the available data are on $N=Z$ nuclei, the sensitivity of the dataset to the isovector parameter $b_1$ is quite limited.

\begin{figure}
\includegraphics[width=.8\columnwidth]{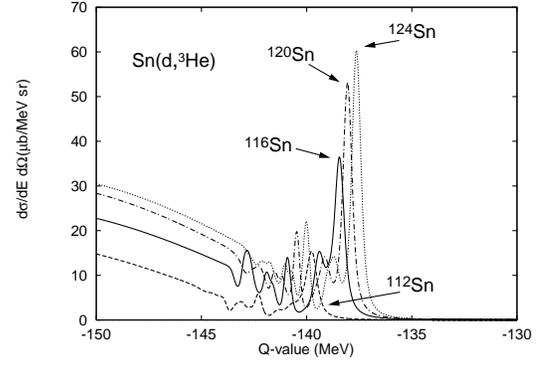}
\caption{\label{fig:Sn-spectra}  
The Sn isotope dependence of the total  $(d,^3{\rm He})$ spectrum for the pionic atom formation 
at $T_d=500$~MeV with 300~keV experimental resolution. The target nucleus is indicated in the figure. 
\cite{Umemoto:2000aw}.
}
\end{figure}


The S236 experiment at GSI, a successor to S160, measured the $1s$ binding energies and widths of pionic  $^{115,119,123}\rm Sn$ using the $^{116,120,124}{\rm Sn}(d, ^3{\rm He})$ reactions. 
Sn isotopes were chosen since the
$1s$  states are expected to be produced 
as the most dominant quasi-substitutional states, $(1s)_{\pi^-}(3s)^{-1}_n$, 
because of the presence of the $3s$
orbital near the Fermi surface (Fig.~\ref{fig:Sn-spectra}).  
Another merit is to make use of isotopes over a wide
range of $(N- Z)/A$ to sensitively deduce the isovector parameter $b_1$.

\begin{figure}
\includegraphics[width=\columnwidth]{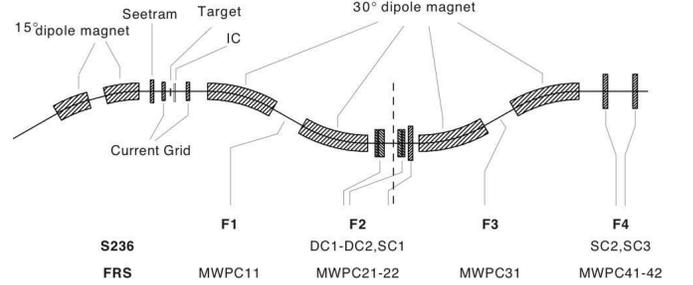}
\caption{\label{fig:frs-s236-configuration} 
 The Fragment Separator of GSI used for the (d,$^3$He) spectroscopy. 
}
\end{figure}

The experiment S236 used a deuteron beam from the heavy-ion synchrotron
SIS at GSI, Darmstadt, combined with the fragment
separator (FRS) as a high-resolution forward spectrometer (Fig.~\ref{fig:frs-s236-configuration}).
They chose the exact recoilless condition to
suppress minor states other than the enhanced $1s$ 
states with quasi-substitutional $1s$  states, with a deuteron
beam of a small momentum spread and an accurately
measured energy of $503.388 \pm 0.100$~MeV. 
%
The $Q$-value resolution was $394\pm 33$~keV (FWHM), and the absolute $Q$-value scale was calibrated to an accuracy of $\pm 7$ keV using the $p(d,^3$He)$\pi^0$ reaction, where a thin Mylar layer put on the surface of each Sn target was used as the proton source.

\begin{figure}
\includegraphics[width=0.7\columnwidth]{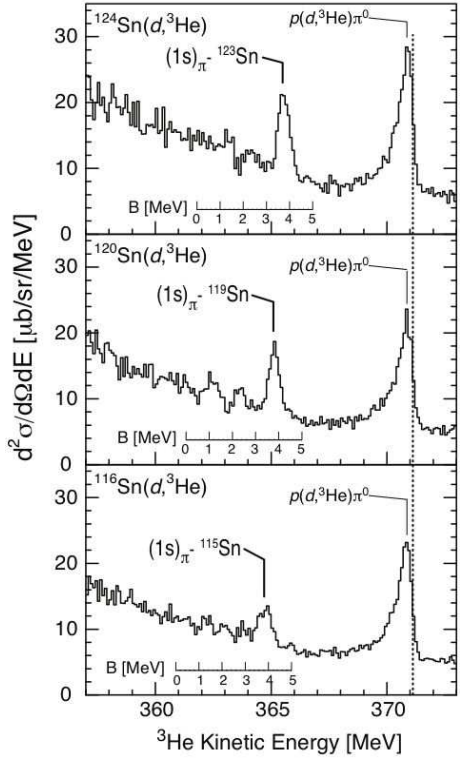}
\caption{\label{fig:suzuki-pionic-atom-fig1} 
Double differential cross sections versus the $^3$He 
kinetic energy of the $^{124,120,116}$Sn$(d,^3$He) reactions measured 
at the incident deuteron energy of 503.388~MeV. The scales of 
the $\pi^-$ binding energies in $^{123,119,115}$Sn are also indicated. 
From \citet{suzuki2004psp}.
}
\end{figure}

The observed spectra,
$d^2\sigma/d E/d \Omega$, on Mylar-covered $^{116}$Sn, $^{120}$Sn, $^{124}$Sn
targets as a function of the $^3$He kinetic energy, are shown
in Fig. \ref{fig:suzuki-pionic-atom-fig1}.
In each spectrum of Fig.~\ref{fig:suzuki-pionic-atom-fig1} a distinct peak at around
365~MeV was observed, which was assigned to a dominant configuration of $(1s)_\pi (3s)^{-1}_n$. 
The skewed peaks at around 371 MeV arise from
$p(d,^3$He)$\pi^0$. 
The overall spectrum
shapes for the three Sn targets were found to be in good
agreement with the predicted ones (Fig.~\ref{fig:Sn-spectra}). The spectra were
decomposed according to the theoretical prescription of
\citet{Umemoto:2000aw}, from which the $1s$
binding energies $(B_{1s})$ and widths $(\Gamma_{1s})$ were determined as:
\begin{eqnarray*}
^{115}{\rm Sn:} B_{1s} &=& 3.906 \pm 0.021\mbox{ (stat)}\pm 0.012\mbox{ (syst) MeV},\\
\Gamma_{1s} &=& 0.441 \pm 0.068\mbox{ (stat)}\pm 0.054\mbox{ (syst) MeV},\\
^{119}{\rm Sn:} B_{1s} &=& 3.820 \pm 0.013\mbox{ (stat)}\pm 0.012\mbox{ (syst) MeV},\\
\Gamma_{1s} &=& 0.326 \pm 0.047\mbox{ (stat)}\pm 0.065\mbox{ (syst) MeV}\\
^{123}{\rm Sn:} B_{1s} &=& 3.744 \pm 0.013\mbox{ (stat)}\pm 0.012\mbox{ (syst) MeV},\\
\Gamma_{1s} &=& 0.341 \pm 0.036\mbox{ (stat)}\pm 0.063\mbox{ (syst) MeV}.\\
\end{eqnarray*}

\subsubsection*{In-medium isovector scattering length $b_1$}

The $s$-wave potential parameters $\{b_0$, $b_1$, ${\rm Re}B_0$, ${\rm Im}B_0\}$ were deduced by simultaneously fitting $B_{1s}$ and $\Gamma_{1s}$ of the three Sn isotopes together with those of symmetric light nuclei ($^{16}$O, $^{20}$Ne and $^{28}$Si), with the $p$-wave parameters fixed to the known values from pionic x-ray data listed in \citet{Batty:1997zp}.

\begin{figure}
\includegraphics[width=0.7\columnwidth]{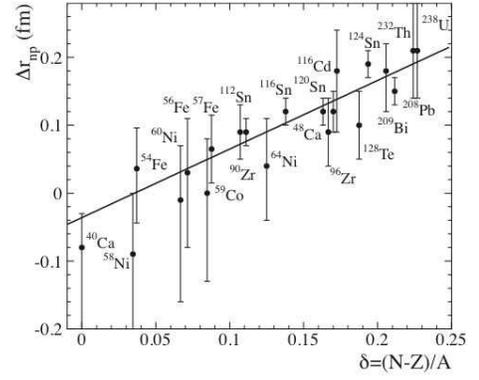}
\caption{\label{fig:rnp-from-pbar} 
Difference $\Delta r_{np}$ between the rms radii of the neutron 
and proton distributions as deduced from the antiprotonic atom 
x-ray data, as a function of $\delta=(N-Z)/A$. The proton distributions were obtained from electron scattering data \cite{vries87} (Sn 
nuclei) or from muonic atom data \cite{fricke1995ngs} (other nuclei). The 
full line represents the linear relationship between $\delta$ and $\Delta r_{np}$ 
as obtained from a fit to the experimental data.  
From \citet{Trzcinska:2001sy}.
}
\end{figure}

\begin{figure}
\includegraphics[width=0.7\columnwidth]{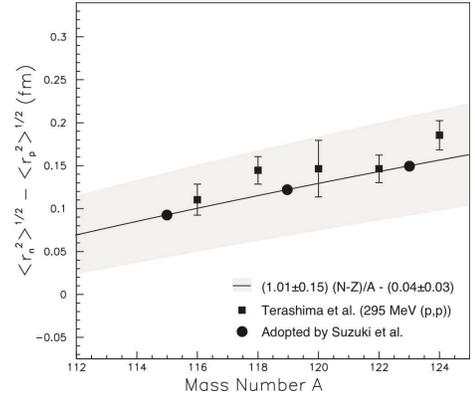}
\caption{\label{fig:terashima} 
Difference $\Delta r_{np}$ between the rms radii of the neutron 
and proton distributions for various tin isotopes deduced from proton elastic scattering at 295 MeV  \cite{terashima:024317} (squares) compared with the values adopted in the analysis of  \citet{suzuki2004psp} (filled circles), which used the relation $\Delta r_{np} = (1.01 \pm 0.15)(N-Z)/A+(-0.04\pm 0.03)$ obtained by  \citet{Trzcinska:2001sy}.
}
\end{figure}

\begin{figure}
\includegraphics[width=0.8\columnwidth]{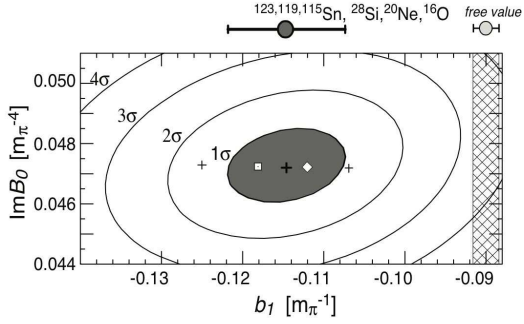}
\caption{\label{fig:suzuki-pionic-atom-fig3} 
Likelihood contours in the $\{ b_1, {\rm Im} B_0 \}$ plane from the 
simultaneous fitting of $\{ B_{1s}, \Gamma_{1s} \}$ of t he $1s$ pionic states in the 
three Sn isotopes and three light symmetric nuclei \cite{suzuki2004psp}.
}
\end{figure}

The obtained values are:
\begin{eqnarray*}
b_0 &=& -0.0233\pm 0.0038 \mbox{ $m_\pi^{-1}$ },\\
b_1 &=& -0.1149\pm 0.0074 \mbox{ $m_\pi^{-1}$ },\\
{\rm Re}B_0 &=& -0.019\pm 0.017 \mbox{ $m_\pi^{-4}$ },\\
{\rm Im}B_0 &=& 0.0472\pm 0.0013 \mbox{ $m_\pi^{-4}$ }.
\end{eqnarray*}
The effective density probed by the $\pi^-$ was $\rho\simeq 0.6\rho_0$ (obtained from the overlap of the nuclear density and the pion wavefunction, such as shown in Fig.~\ref{fig:pionic-pb208}~\cite{YH03}).
Figure \ref{fig:suzuki-pionic-atom-fig3} shows the likelihood contours in the plane of $\{ b_1, {\rm Im}B_0 \}$. As shown, the best-fit $b_1$ value deviates significantly from the `free' value obtained from pionic hydrogen.

Also shown therein is how the best-fit values move if the shape of the neutron distribution was changed.
In fact, the most serious problem in the analysis was the relatively
poor knowledge concerning the neutron distribution 
$\rho_n (r)$ in Sn isotopes, whereas the proton distribution 
$\rho_p (r)$ is well known \cite{fricke1995ngs}. 
In the fit, the neutron diffuseness $(a_n)$ and half-density radius $(c_n)$ parameters in the two-parameter Fermi distribution were chosen so as to satisfy
the difference between the neutron and proton rms
radii, $\Delta r_{np} = (1.01 \pm 0.15)(N-Z)/A+(-0.04\pm 0.03)$ fm  (Fig.~\ref{fig:rnp-from-pbar}),
based on experimental data of antiprotonic atoms of Sn isotopes \cite{Trzcinska:2001sy} as well as of
many other nuclei. The two extreme assumptions, the ``skin'' type $(c_p<c_n, a_p=a_n)$ and the ``halo'' type $(c_p=c_a, a_p<a_n)$, where $c_p$ ($c_n$) is the proton (neutron) half-density radius and $a_p$ ($a_n$) is the proton (neutron) diffuseness of the two-parameter Fermi distribution, are shown in Fig.~\ref{fig:suzuki-pionic-atom-fig3} in open diamonds and open squares, respectively. The best-fit values were obtained by using the parameters halfway between these two assumptions. The two crosses in the figure indicate the dependence of $b_1$ on the uncertainty of $\pm 0.04$~fm in $\Delta_{np}$. The quoted errors on the best-fit values do not include these uncertainties in the neutron distribution\footnote{Recently, \citet{terashima:024317} have determined the neutron rms radii of Sn isotopes using proton elastic scattering at 295 MeV. Their results are in good agreement with the radii assumed in \cite{suzuki2004psp}. See Fig.~\ref{fig:terashima}}.

\subsubsection*{In-medium quark condensate}
\begin{figure}
\includegraphics[width=0.8\columnwidth]{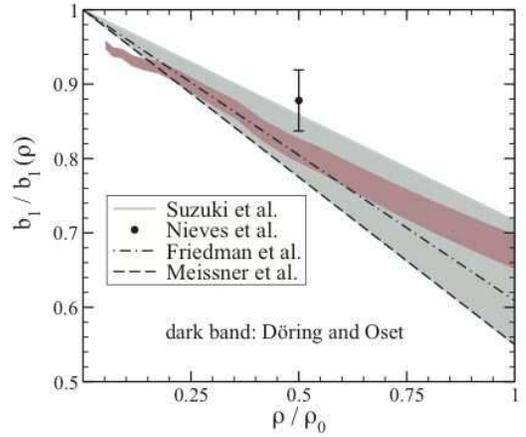}
\caption{\label{fig:doring-oset} 
In-medium isovector $b_1 (\rho)$ compared to the vacuum isovector term $b_{1}$. The gray band, the experimental result from  \citet{suzuki2004psp}, is compared with chiral calculations by \citet{Doring:2007qi} (dark band), \citet{Meissner:2001gz} (dashed line) and \citet{Friedman:2004jh} (dash-dot line). The point is a phenomenological fit by \citet{Nieves1993509}. From  \citet{Doring:2007qi}}.
\end{figure}

From the observed enhancement of the $b_1$ parameter relative to the free value, $b_1/b_1 (\rho) = 0.78\pm 0.05$, \citet{suzuki2004psp} deduced that the chiral order parameter $f_\pi$ is subject to the in-medium reduction of  $\left(f_\pi^t (\rho) /f_{\pi} \right)^2 \simeq 0.64$ at the normal nuclear density $\rho=\rho_0$, based on the suggestion \cite{Weise:2001oa,Weise:2000xp,Kienle:2001jh} that the missing repulsion may be explained in terms of a possible in-medium change of the pion decay constant. A global fit \cite{Friedman:2002um,Friedman:2002ix} to the pionic x-ray as well as the deeply-bound $^{205}$Pb data \cite{geissel2002dba} supported this view. Recent direct calculation of $b_1/b_1 (\rho)$ in the unitarized chiral approach (Fig.~\ref{fig:doring-oset}) is also consistent with the above result \cite{Doring:2007qi}.

Here, we adopt a new model independent relation obtained by exploiting operator relations in QCD, Eq.(\ref{eq:TW-ratio}).   \citet{Jido:2008bk}  showed that $Z_\pi^{1/2} (\rho) \simeq 
\left( 1 - \gamma \frac{\rho}{\rho_0}\right)$, 
with $\gamma=0.184$. Using $b_1/b_1 (\rho) = 0.78$ at $\rho=0.6\rho_0$, the ratio of the quark condensates is found to be 
\begin{equation}
\frac{\left<\bar q q\right>_\rho}{\left< \bar q q\right>} \simeq 1-0.37 \frac{\rho}{\rho_0}.
\end{equation}

\noindent
A similar conclusion, $\left<\bar q q\right>_\rho / \left< \bar q q\right> \simeq 1-0.39 \rho / \rho_0$, was obtained from the differential cross sections for $\pi^\pm$-nucleus elastic scattering data at 21.5 MeV \cite{Friedman:2004jh,Friedman:2005pt}. 





\section{SCALAR MESON: $\sigma$ IN NUCLEI}
\label{sec:sigma}

\subsection{Theoretical background}

The $\sigma$-meson in the vacuum is a very broad resonance
in the scalar-isoscalar channel as  discussed in Sec.\ref{sec:SP-scalar}.
It is an excitation having the same quantum number as the vacuum, and
may be interpreted as the Higgs boson in QCD.
The fate of such scalar-isoscalar excitation in hot and/or dense medium is
strongly correlated with 
chiral symmetry restoration  \cite{Hatsuda:1985eb,Hatsuda:1986gu}.
The basic idea is rather simple and general:
If  a sizable reduction of the chiral condensate
takes place in the medium, the ground state becomes ``soft" against the
amplitude fluctuation of the order parameter
due to the reduction of the stiffness.
 In fact, the frequency of the $\sigma$ excitation
(the amplitude fluctuation) is red-shifted toward the frequency of the  $\pi$ excitation
(the phase fluctuation). Similar softening phenomena are well known
  in solid state  physics, e.g., the soft phonon modes
  in anti ferro-elastic crystals such as SrTiO$_3$ \cite{Gebhardt:1980} and in
  ferro-electric crystals such as SbSI     \cite{Kittel:2004}.

In the case of QCD, an interesting signal associated with the
chiral softening is the spectral enhancement in the scalar-isoscalar channel
near the 2$\pi$ threshold at finite temperature and density
as mentioned in Sec.\ref{sec:DA-scalar}. 
Consider the retarded propagator of $\sigma$ at rest in the medium,
$D_{\sigma}(\omega)$. The spectral function is defined as
$\rho_{\sigma} = - \frac{1}{\pi} {\rm Im} D_{\sigma}
=  \frac{1}{\pi}
{\rm Im} D^{-1}_{\sigma}/[({\rm Re} D^{-1}_{\sigma})^2+({\rm Im} D^{-1}_{\sigma})^2]$.
Because of the softening and strong $\sigma \pi \pi$ coupling,
${\rm Re} D^{-1}_{\sigma}(\omega\simeq 2 m_{\pi})$ becomes small or even vanishes
at a certain temperature or baryon density.  In that situation, the
spectral function is dominated by the imaginary part of the inverse propagator with the
 phase space factor \cite{Chiku:1997va,Chiku:1998kd,Hatsuda:1999kd,Volkov:1997dx},
 \beq
 \label{eq:sigma-threshold}
\rho_{\sigma}(\omega) \simeq   \frac{1}{\pi} \frac{1}{{\rm Im} D^{-1}_{\sigma}(\omega)}
\propto \frac{\theta(\omega - 2 m_{\pi})}{\sqrt{ 1-\frac{{4m_{\pi}}^2}{\omega^2} }}.
\eeq
This implies
a large enhancement of the spectral function near 2$\pi$ threshold.
Such an enhancement may be seen in, e.g., the dipion production and diphoton production
from the hot/dense medium. Also, a $\sigma$-mesic nuclei (or
 the bound dipion in nuclei)
could be formed by $(d,t)$, $(d,{\rm ^{3}He})$ and $(\gamma,p)$ reactions, if
there is large enough softening \cite{Hirenzaki:2002zm,Nagahiro:2005gf}.

\begin{figure}
\includegraphics[width=0.8\columnwidth]{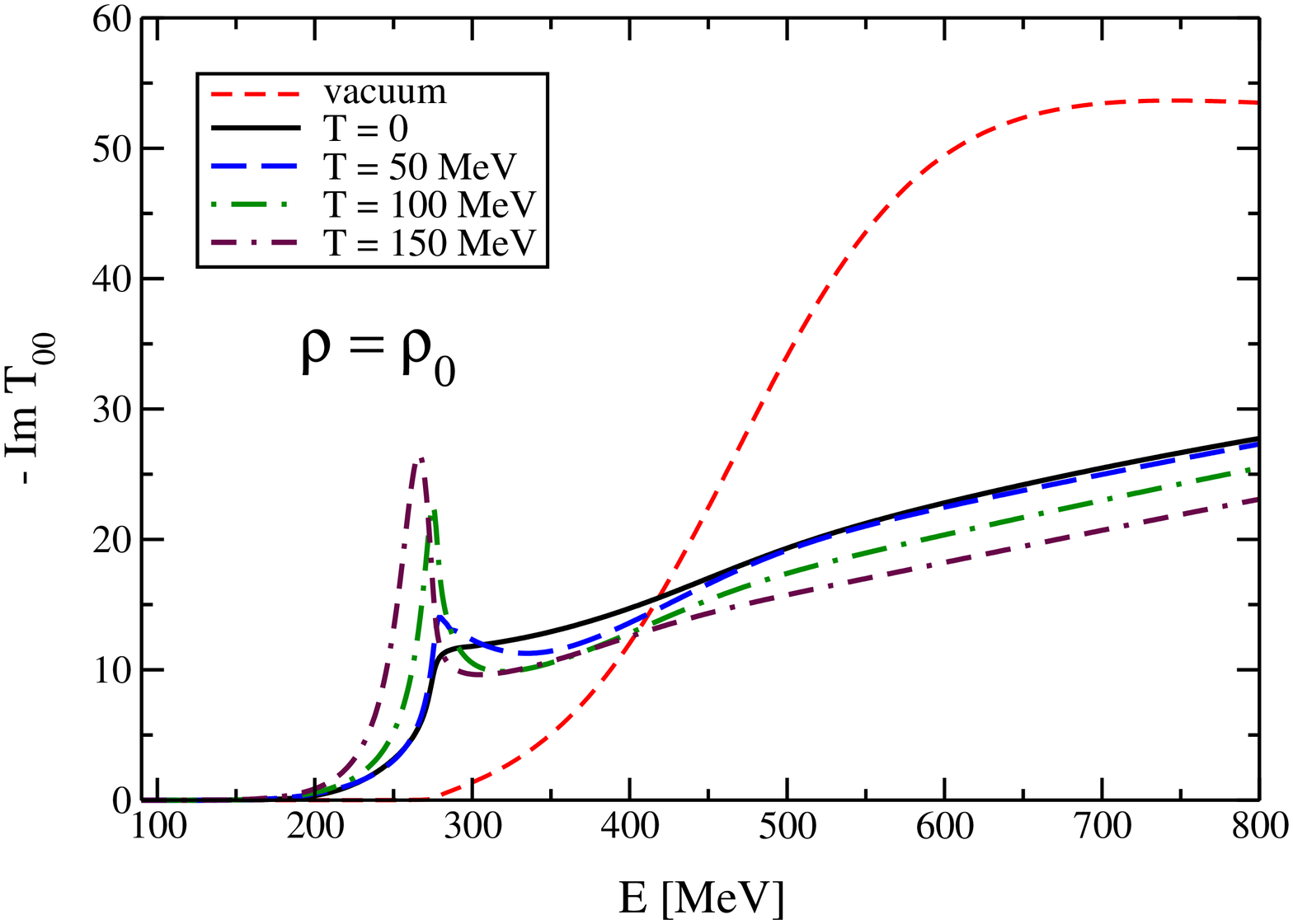}
\includegraphics[width=0.8\columnwidth]{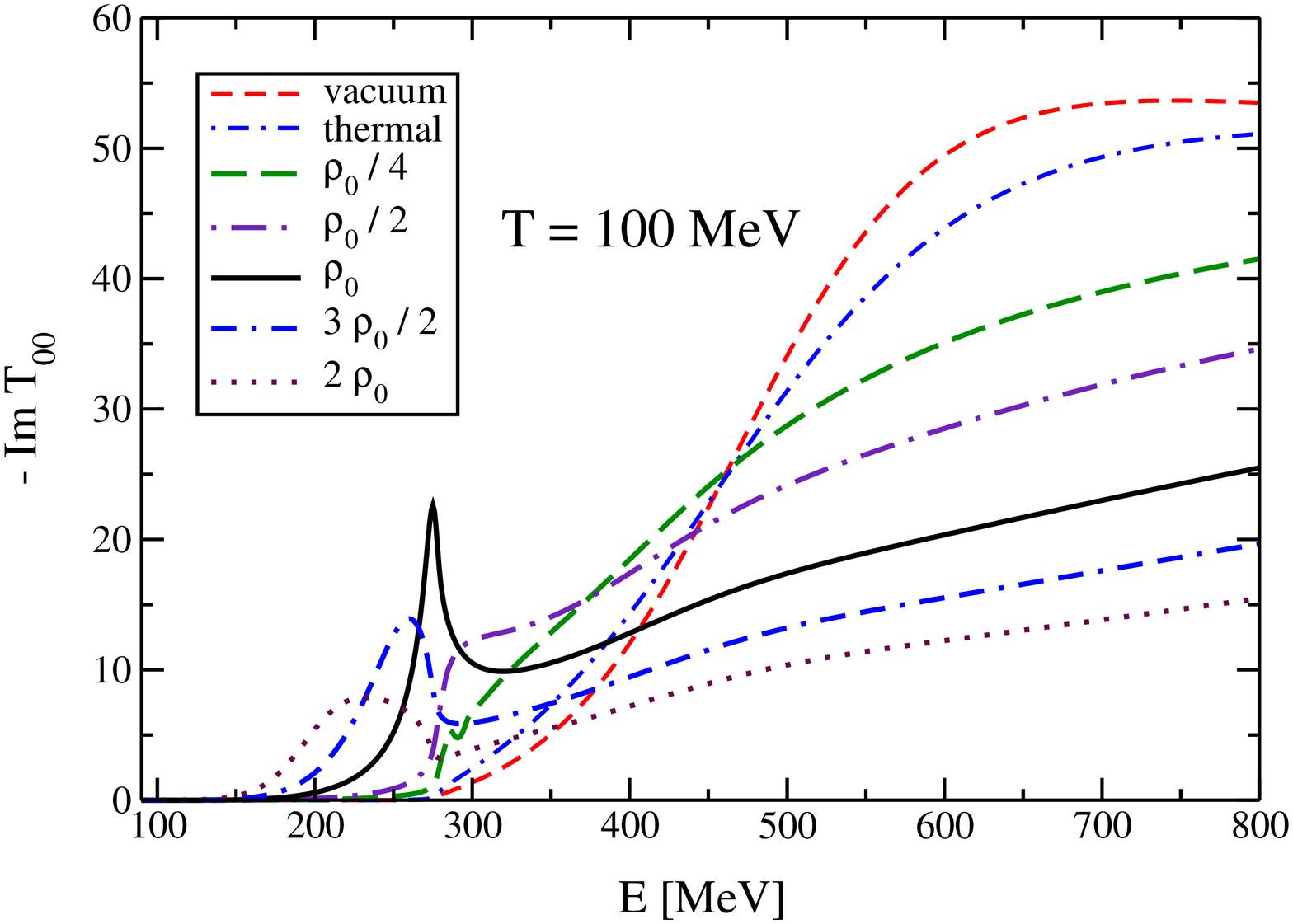}
\caption{Imaginary part of the $\pi-\pi$ amplitude $T_{00}(E)$
in the scalar-isoscalar channel calculated in the chiral unitary model
\cite{Cabrera:2008tj}. }
\label{fig:sigma-T-rho}
\end{figure}

In reality,  the pion has a width inside the medium, so that the
spectral function in the $\sigma$-channel 
does not have a simple form such as Eq.(\ref{eq:sigma-threshold}).
Nevertheless, more sophisticated approaches 
 indicate similar enhancement of the in-medium $\pi\pi$ scattering
amplitude in the scalar-isoscalar channel
near the 2$\pi$ threshold (see the references cited 
 in Sec.\ref{sec:DA-scalar}). Shown in Fig. \ref{fig:sigma-T-rho}
 is one of such examples obtained by using  
 the chiral unitary model. Here we note 
 that the expansion parameter of the non-linear chiral models 
 is a ratio, [typical momentum of the pion or nucleon fields]/
 [chiral symmetry breaking scale $4\pi f_{\pi}$].
 Therefore, non-linear chiral approaches loose their predictive 
 power at high temperature and/or high baryon density.

\subsection{Experiments}

\subsubsection{CHAOS}
The CHAOS (Canadian high acceptance orbit spectrometer) collaboration at TRIUMF was the first to report such a low-mass enhancement in the $(\pi\pi)_{I=J=0}$ channel.
They used $\pi^+ A \rightarrow \pi^+\pi^- X$ reactions 
on hydrogen \cite{Kermani:1998pipi} 
and on nuclear targets \cite{bonutti1996dpi,Bonutti:1998pia,bonutti:018201,Bonutti:2000tt,camerini2001ppr,Camerini:2004dg,Grion:2005bq} at  pion kinetic energy $T_\pi = 243-305$~MeV, and observed the spectral 
``softening'' in the 
$\pi^+\pi^-$ channel but not in the $\pi^+\pi^+$ (i.e., $I=2$) channel. 
The nuclear data were taken at $T_\pi=283$~MeV on $^2$H, $^{12}$C, $^{40}$Ca and $^{208}$Pb targets, as well as at $T_\pi=243, 264, 284, 305$~MeV on $^{45}$Sc.

\begin{figure}
\includegraphics[width=0.65\columnwidth]{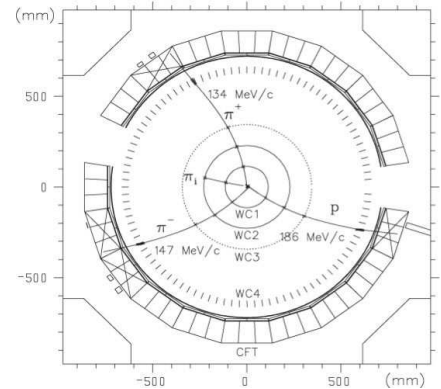}
\caption{\label{fig:chaos-event} 
A typical reconstructed event  in CHAOS for the $\pi_{i}^{+} \rightarrow \pi^+\pi^- p$ 
reaction on $^{12}$C. WCs are the four wire chambers, CFTs are the CHAOS First-level Trigger counters, which are also used for particle identification. 
Of the 20 CFTs, two are removed to free the pion beam path.
 \cite{Bonutti:2000tt}.
}
\end{figure}

\begin{figure}
\includegraphics[width=0.85\columnwidth]{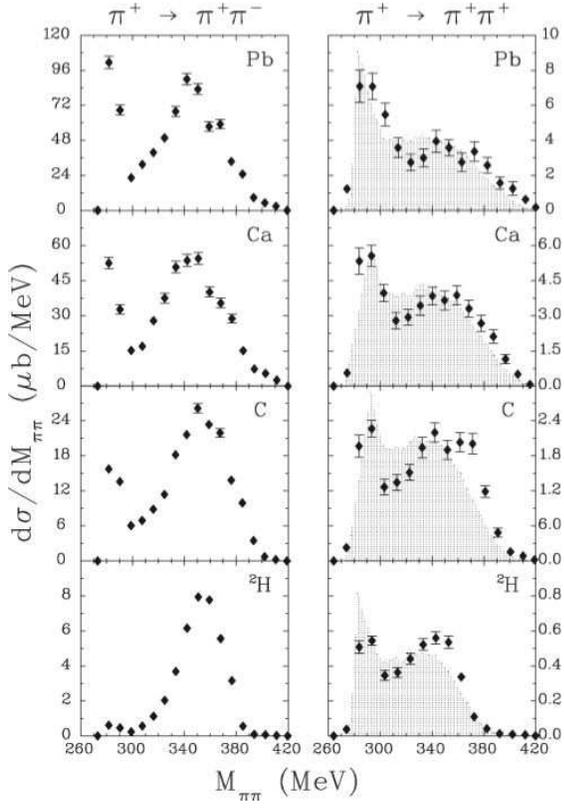}
\caption{\label{fig:chaos-pipi-invmass} 
 Invariant mass distributions (diamonds) for the $\pi^+ \rightarrow \pi^+\pi^-$ and  
$\pi^+ \rightarrow \pi^+\pi^+$ reactions
on $^2$H, $^{12}$C, $^{40}$Ca and $^{208}$Pb. The shaded regions represent the results of phase-space simulations for the pion-production reaction $\pi A \rightarrow \pi \pi N[A-1]$
 \cite{Bonutti:2000tt}.
}
\end{figure}

Figure~\ref{fig:chaos-event} shows a typical reconstructed $\pi^+\pi^-$ event. 
The spectrometer is based on a cylindrical dipole magnet producing vertical magnetic fields up to 1.6 T (0.5~T for the $2\pi$ experiments)\cite{Smith:1995nx}. The target is located in the center of the magnet. Charged particle tracks produced by pion interactions are identified using four concentric cylindrical wire chambers (WC1, 2, 3, 4) surrounding the target. 
Particles are identified by cylindrical layers of scintillation counters and lead-glass \v Cerenkov counters, which also provide a first level trigger (CFTs).  The detector subtends approximately 10\% of 4$\pi$. The momentum resolution delivered by the detector system is 1\%.
The pion-detection threshold energy is 11 MeV.

Fig.~\ref{fig:chaos-pipi-invmass} shows the $\pi^+\pi^-$ (left) and $\pi^+\pi^+$ (right) invariant-mass spectra taken on nuclear targets. 
The distributions span the range from $2 m_\pi$ up to the 420 MeV, the maximum allowed by the reaction.
While the $\pi^+\pi^+$ spectra can be fairly well represented by the phase-space simulations (shaded region), the $\pi^+\pi^-$ spectra show a peak in the low-mass region, 
increasing with mass number $A$.

\begin{figure}
\includegraphics[width=0.85\columnwidth]{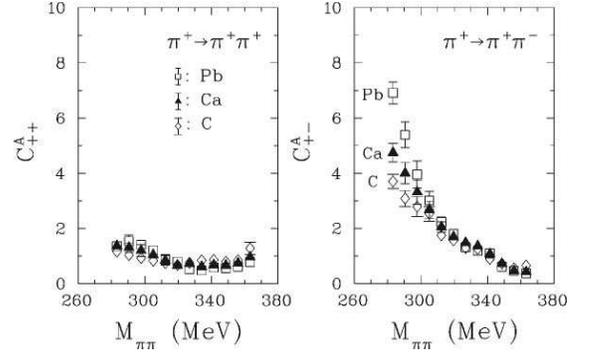}
\caption{\label{fig:bonutti-prc-fig1-fig2} 
(left): $C^{A}_{\pi\pi}$, the bin-by-bin ratio (see text) of $\pi\pi$ invariant mass distributions for the two reactions $\pi^+ A\rightarrow \pi^+ \pi^+ A'$ and 
$\pi^+ p \rightarrow \pi^+ \pi^+ n$, as a function of the $M_{\pi\pi}$ energy. 
The nuclei $(A)$ 
examined are $^2$H (which plays the role of a proton, $p$), $^{12}$C (open 
diamonds), $^{40}$Ca (full triangles), and $^{208}$Pb (open squares). 
(right): Same as the left panel but for the $\pi^+ \rightarrow \pi^+\pi^-$ reaction channel
 \cite{bonutti:018201}.
}
\end{figure}

Instead of comparing the raw spectra with theoretical predictions or with results of other experiments,  a composite observable was used:
$$C^A_{\pi\pi} = \frac{\sigma(M^A_{\pi\pi})/\sigma^A_T}{\sigma(M^N_{\pi\pi})/\sigma^N_T},$$
where $\sigma(M^A_{\pi\pi})$ ($\sigma(M^N_{\pi\pi})$) is the triple differential cross section 
$d^3\sigma/d M_{\pi\pi} d\Omega_\pi d\Omega_\pi$ for nuclei (nucleon), $M_{\pi\pi}$ represents the $\pi\pi$ invariant mass, $\Omega_\pi$ denotes the pion-detection solid angle, and $\sigma^A_T$ ($\sigma^N_T$) is the total cross section in nuclei (nucleon). The ratios are presented in Fig.~\ref{fig:bonutti-prc-fig1-fig2}. The ratios $C^A_{\pi^+\pi^-}$ (right panel) show that 
the low-mass $\pi^+\pi^-$ pairs are more abundant in heavier targets, 
while no such trend can be seen in $C^A_{\pi^+\pi^+}$ (left panel).

\subsubsection{Crystal Ball}
\begin{figure}
\includegraphics[width=0.45\columnwidth]{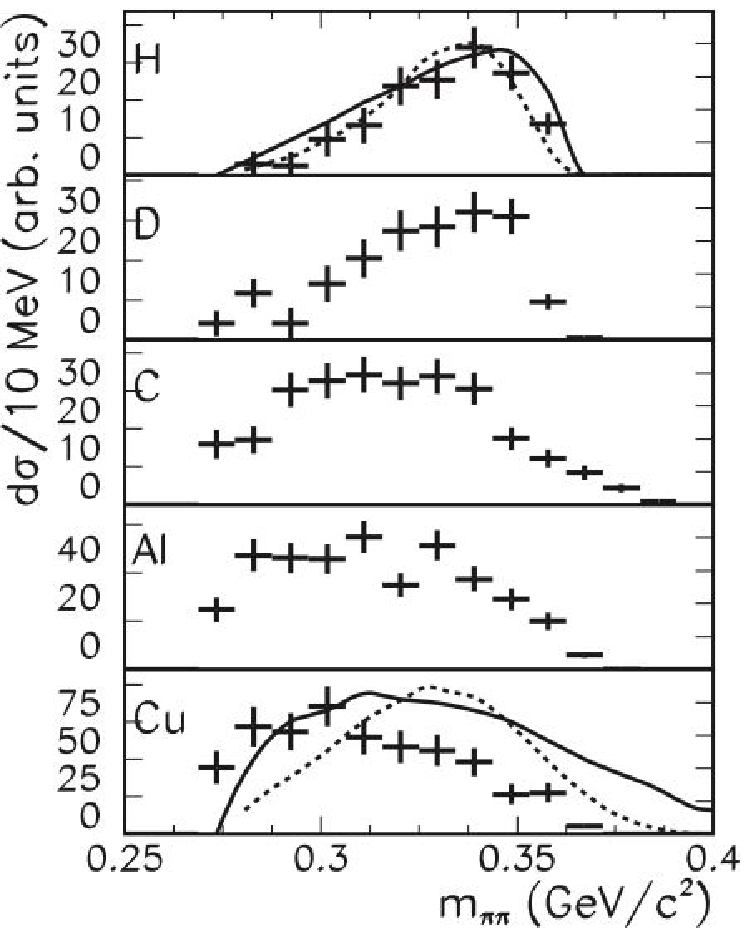}
\hfill
\includegraphics[width=0.45\columnwidth]{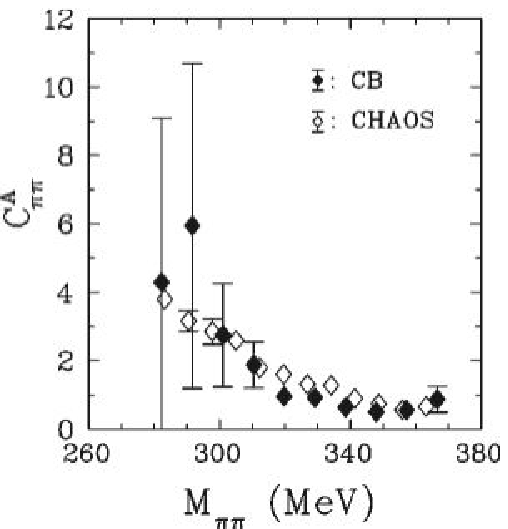}
\caption{\label{fig:cb-pi0pi0} 
(Left) Experimental results for the $2\pi^0$ invariant mass distributions obtained for the H, D, C, Al, and Cu targets corrected 
for Crystal Ball acceptance. The vertical scale is in arbitrary 
units. The solid lines show the results of calculations made for the TAPS group by Rapp, and the dashed line is the prediction by Vicente Vacas. 
\cite{starostin2000mpipi}. 
(Right) The composite ratios $C^A_{\pi\pi}$
as a function of the $\pi\pi$ invariant mass for the $^{12}$C target. 
Full diamonds, the $C^C_{\pi^0\pi^0}$
distribution deduced from the CB data of \citet{starostin2000mpipi}; open 
diamonds, the CHAOS $C^C_{\pi^+\pi^-}$ 
distribution taken from \citet{bonutti:018201} \cite{camerini2001ppr}.}
\end{figure}

The Crystal Ball (CB) collaboration at the AGS investigated
the $\pi^- A \rightarrow \pi^0\pi^0 A'$ reaction on CH$_2$, CD$_2$, C, Al and Cu targets
at $p_{\pi^-} = 408$ MeV/$c$ \cite{starostin2000mpipi}.
The Crystal Ball comprises 672 optically isolated NaI(Tl) crystals that cover 93\% of $4\pi$, and
has a $\pi^0\pi^0$ acceptance of $11-17\%$. The invariant mass resolution is about 1.2\% at $m_{\pi^0\pi^0} = 2 m_{\pi^0}$ and reaches a plateau of 2.2\% at $m_{\pi^0\pi^0} \simeq 0.3$~GeV/$c^2$.
The measured $\pi^0\pi^0$ invariant-mass spectra (left panel of Fig.~\ref{fig:cb-pi0pi0}) show a gradual 
shift of intensity towards lower $m_{\pi\pi}$ for heavier targets, but a sharp strong peak near $2m_\pi$ as reported by the CHAOS collaboration cannot be seen.

However, \citet{camerini2001ppr} pointed out that if the composite ratios $C^A_{\pi\pi}$ are used to
compare the CHAOS and CB results, so as to (mostly) remove uncertainties arising from acceptance corrections\footnote{Note that the CHAOS acceptance is about 10\% of 4$\pi$ while that of Crystal Ball is 93\% of 4$\pi$. The small acceptance of CHAOS may be the origin of the sharp peaks close to the threshold.}, the two results are not statistically inconsistent, at least in the case of $^{12}$C 
(the only nucleus common to the two experiments). See Fig.~\ref{fig:cb-pi0pi0} right panel.

\subsubsection{TAPS}
The TAPS (Two Arms Photon Spectrometer) collaboration used the tagged photon facility at
the MAMI accelerator \cite{Anthony:1991lo} to measure $A(\gamma, \pi^0\pi^0)$ as well as $A(\gamma, \pi^0\pi^{+/-})$ cross sections \cite{messchendorp2002mmpi,bloch2007dpp}.

\begin{figure}
\includegraphics[width=0.65\columnwidth]{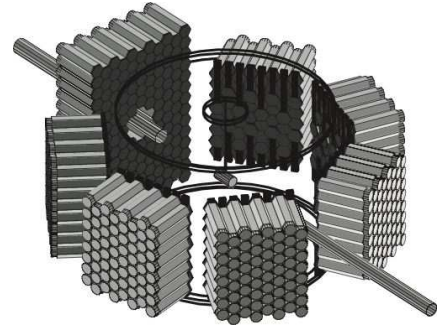}
\caption{\label{fig:taps-pipi-configuration} 
Setup of the TAPS detector at the Mainz MAMI accelerator. The beam entered the target chamber from the lower right edge. 
 \cite{Kermani:1998pipi,bloch2007dpp}}
\end{figure}

The experiment covered the photon energy range 
from 200–800 MeV with an energy resolution of 
2 MeV per tagger channel. 
The targets used were $^{1}$H, $^{12}$C, $^{\rm nat}$Pb \cite{messchendorp2002mmpi}
as well as $^{40}$Ca \cite{bloch2007dpp}.
The reaction products from the target were 
detected with the electromagnetic calorimeter TAPS, comprising
510 hexagonally shaped BaF$_2$ 
crystals of 25 cm length with an inner diameter of 5.9 cm. 
They were arranged in six blocks of 64 
modules and a larger forward wall of 138 modules (see Fig.~\ref{fig:taps-pipi-configuration}). 
The 
blocks were arranged in one plane around the target at 
a distance of 55 cm from the target center and at polar 
angles of $\pm 54^\circ, \pm 103^\circ$, and $\pm 153^\circ$,
while the forward wall
was placed 60 cm away from the target center at $0^\circ$ 
and 
the photon beam passed through a hole in the center of 
the forward wall. Each detector module was equipped with 
an individual plastic veto detector, read out by a separate 
photomultiplier. The setup covered 37\% of 
the full solid angle. The 
two-$\pi^0$ invariant-mass resolution was between $2.0\%$ and $2.5\%$  in the 
incident-photon energy range of interest. 

\begin{figure}
\includegraphics[width=0.85\columnwidth]{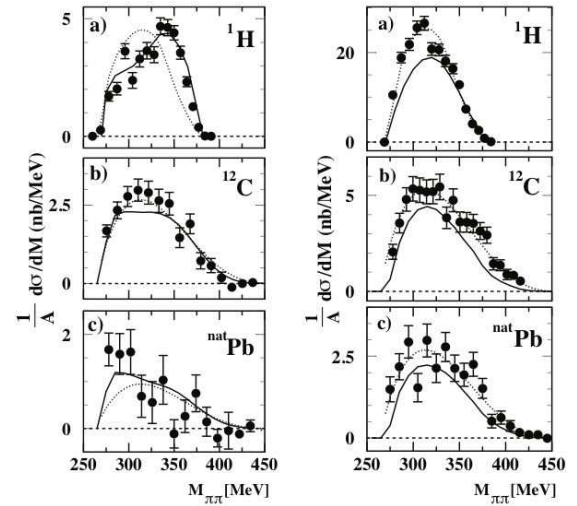}
\caption{\label{fig:messchendorp-pipi-invmass} 
(Left) Mass-number-normalized differential cross sections of  the reaction $A(\gamma, \pi^0\pi^0)$ with 
$A={\rm ^{1}H}, {\rm ^{12}C}, {\rm ^{nat}Pb}$.
(Right)  The same for the reaction
$A(\gamma, \pi^0\pi^+)$.
Both panels are for incident photons in the energy 
range of $400 - 460$~MeV (solid circles). Error bars denote 
statistical uncertainties. The dotted curves indicated phase-space distributions determined by the Monte Carlo model, and the solid curves are predictions by \cite{Roca:2002hl}. From 
\citet{messchendorp2002mmpi}.
}
\end{figure}

Figure \ref{fig:messchendorp-pipi-invmass} shows the $\pi^0\pi^0$ (left) and $\pi^0\pi^{+/-}$ (right) spectra
measured at the incident photon energy range of $400-460$ MeV.
This energy range was chosen so that its centroid corresponds to the same 
center-of-mass energy as was used in the pion-induced experiments, 
enabling a direct comparison. Since this range is below the $\eta$-production threshold of 550 MeV,
the event identification is clean. 
Figure \ref{fig:messchendorp-pipi-invmass} indicates that the strength in the distribution of $M_{\pi^0\pi^0}$ 
(but not $M_{\pi^0\pi^{+/-}}$)
is shifted towards smaller invariant masses with increasing $A$.

\begin{figure}
\includegraphics[width=0.85\columnwidth]{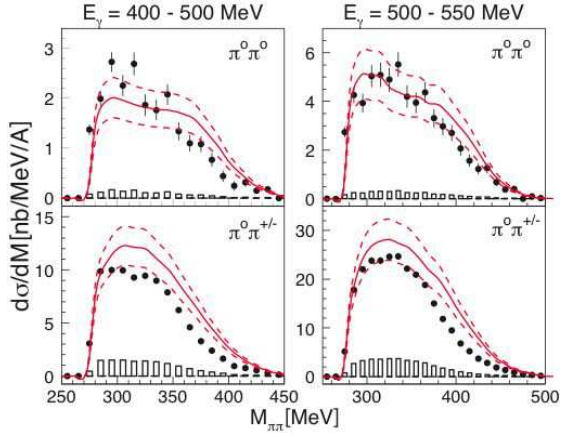}
\caption{\label{fig:taps-ca40-pi0pi0} 
Pion-pion invariant-mass distributions compared to 
results of the BUU model \cite{Buss:2006vh}. The bars at the bottom represent the systematic uncertainty of the data, the dashed lines represent the error band for the BUU calculation  \cite{bloch2007dpp}.
}
\end{figure}

A more recent analysis of $^{40}$Ca$(\gamma,\pi\pi)$, with higher statistics by \citet{bloch2007dpp},  shown in Fig.~\ref{fig:taps-ca40-pi0pi0}, revealed that the invariant-mass spectra show a similar 
softening effect 
as already found in \citet{messchendorp2002mmpi} for carbon and lead nuclei \footnote{Note that 
the cross-section ratios $\sigma(\pi^0\pi^{\pm})/\sigma(\pi^0\pi^0)$ are about 2 in Fig.~\ref{fig:messchendorp-pipi-invmass}, while they are about 5 in Fig.~\ref{fig:taps-ca40-pi0pi0} 
This is likely due to the larger systematic errors in the $\pi^0\pi^{\pm}$ cross sections in \citet{messchendorp2002mmpi} \cite{Krusche:2008priv}.}, 
and that the strength of the effect is comparable to carbon, but they also found that 
a sizable part of the 
in-medium effects can be explained by  final-state 
interaction effects, namely, pion rescattering, as discussed below.
 
\subsection{Final-state interaction (FSI) effects}
\label{sec:pipirescattering}
The solid curves in Fig.~\ref{fig:messchendorp-pipi-invmass} are predictions by \citet{Roca:2002hl}.
Here, the meson-meson interaction in the $I=J=0$ channel is studied in the framework of a 
chiral-unitary approach at finite baryon density. The model dynamically 
generates the $\sigma$ resonance, reproducing the meson-meson phase shifts in 
vacuum and accounts for the absorption of the pions in the nucleus. In the model, the $\pi-\pi$
FSI modified by the nuclear medium 
produces a shift of strength of the $\pi\pi$ invariant mass distribution induced by the moving of the 
$\sigma$ pole to lower masses and widths as the nuclear density increases.
The data are described well by the model considering a 
theoretical uncertainty of 20\%.

On the other hand, the curves in Fig.~\ref{fig:taps-ca40-pi0pi0} are the results of the
semi-classical Boltzmann-Uehling-Uhlenbeck (BUU) calculation \cite{Buss:2006vh}, 
which reproduce both $\pi^0\pi^0$ and $\pi^0\pi^{+/-}$ data reasonably well.  
This model does not 
contain the $\pi-\pi$ final-state interactions, but the ``softening'' of the $\pi^0\pi^0$ spectra
is due to charge-exchange pion-nucleon scattering (i.e., $\pi-N$ FSI) which mixes the contributions 
from the different charge channels. Since the total cross-section for $\pi^0 \pi^{\pm}$ 
production is much larger than the $\pi^0\pi^0$ 
cross-section, the latter receives significant side feeding 
from the mixed charge channel via $\pi^\pm N \rightarrow  \pi^0 N$
scattering, which increases the fraction of re-scattered low-energy 
pions in this channel. In the same way, re-scattering of $\pi^+\pi^-$ 
contributes to the $\pi^0 \pi^\pm$ 
channel.

\begin{figure}
\includegraphics[width=0.5\columnwidth]{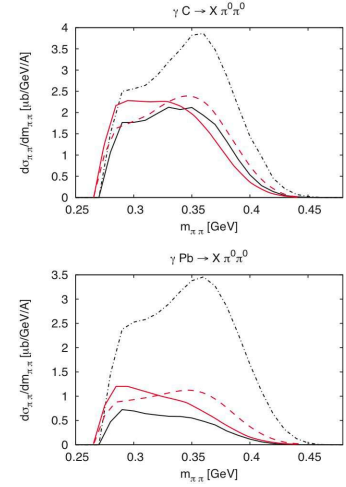}
\caption{\label{fig:buss-roca} 
(color) Two-pion invariant mass distributions for $\pi^0\pi^0$ photoproduction off $^{12}$C and $^{208}$Pb
for $E_\gamma = 0.4-0.46$~GeV, calculated by \citet{Buss:2006vh}. Results without final-state interactions (black dash-dotted lines), and with FSI (black solid lines). These are compared with the results obtained by \cite{Roca:2002hl}, with in-medium final $\pi\pi$ interaction (red solid lines) and
with the free-space $\pi\pi$ interaction (red dashed lines).}
\end{figure}

\section{VECTOR MESONS: $\rho, \omega, \phi$ IN NULCEI}

\subsection{Theoretical background}

As is discussed in Sec.\ref{sec:SP-vector},
the direct signature of  chiral restoration
 is the
degeneracy between the  vector spectral function $\rho_{_V}$
and axial-vector spectral function $\rho_{_A}$.
Since the vector current couples  to virtual photons which eventually decay
into dileptons ($l^+l^-$), $\rho_{_V}$ is directly related to the
physical observable. For example, the emission rate of dileptons
 (number of dileptons emitted per space-time volume $d^4 x$
  and per energy-momentum volume $d^4p$) from the
hot/dense matter reads
\beq
\frac{d^8 N_{l^+l^-}}{d^4 x d^4 p}
= \frac{\alpha^2}{3\pi^2 p^2}
\frac{(2\rho_{_V}^{\rm T}+ \rho_{_V}^{\rm L})(\omega,{\bf p})}
{e^{\omega/T}-1} I(m_l^2/p^2) .
\eeq
Here $p^{\mu}=(\omega,{\bf p})$ is the total four momentum of $l^+$ and $l^-$,
the superscript T (L) implies transverse (longitudinal) and
$I(z)=(1+2z)(1-4z)^{1/4}\theta(1-4z)$ with
 $z=m_l^2/p^2$ denotes the phase space correction
from the finite lepton mass, $m_l$.
Unlike the case of $\rho_{_V}$, it is difficult to measure $\rho_{_A}$
by dilepton, since the decay occurs through  $Z^0$ and is
highly suppressed at low energies.

The spectral constraints on  $\rho_{_V}$ itself are  obtained from
the operator product expansion
 \cite{Hatsuda:1992bv,Hatsuda:1991ez} similar to the derivation of the 
Weinberg-type sum rules in Sec.\ref{sec:SP-vector}:
\beq
\label{eq:vector-I}
\! \! \! \! \! \! \! \! \! \! \! \! \! \! \! \! \!
& &\int_0^{\infty} \frac{d\omega^2}{\omega^2} (\rho_{_V}(\omega)-\rho_{_{\rm pQCD}}(\omega)) =0, \\
\label{eq:vector-II}
\! \! \! \! \! \! \! \! \! \! \! \! \! \! \! \! \!
& &\int_0^{\infty} {d\omega^2} (\rho_{_V}(\omega)-\rho_{_{\rm pQCD}}(\omega))
= \sum_i C_4^i \la {\cal O}_{4}^i \ra , \\
\label{eq:vector-III}
\! \! \! \! \! \! \! \! \! \! \! \! \! \! \! \! \!
& &\int_0^{\infty} {d\omega^2}{\omega^2}
(\rho_{_V}(\omega)-\rho_{_{\rm pQCD}}(\omega)) = \sum_i C_6^i
\la {\cal O}_{6}^i \ra  .
\eeq
Here ${\cal O}_n^i$ are the local composite operators with dimension $n$
with Lorentz indices in general,
and $C_n^i$ are the corresponding Wilson coefficients.
In the SU(2) chiral limit $m_{u,d}=0$, ${\cal O}_4^i$ are all chirally
 symmetric, while
${\cal O}_4^i$ contain both chirally symmetric and non-symmetric operators.
Also, $\rho_{_{\rm pQCD}}(\omega)$ is the spectral function
which reproduces the perturbative calculation of the 
 correlation function (the l.h.s. of Eq.(\ref{eq:R-correlation-def}))
 in the deep Euclidean region ($\omega^2 \rightarrow - \infty$). 
 Then, $\rho_{_{\rm pQCD}}(\omega)$  is chirally symmetric by definition.
  These are the reasons why the Weinberg-type
sum rules with only chirally asymmetric condensates in Sec.\ref{sec:SP-vector}
are obtained by taking the difference between vector and axial-vector
correlations.  Since chirally symmetric condensates do not have to vanish
at the critical point of the chiral transition,
 one cannot immediately relate the
spectral modification of vector mesons to the restoration of chiral symmetry. 

Even if in-medium changes of $\la {\cal O}_{4,6}^i \ra $ are obtained
exactly from lattice QCD simulations,  the above sum rules  only supply
  information on the weighted averages of the spectral function
  and not on the exact spectral shape.   
  Nevertheless, 
    these sum rules are useful to make a consistency check in
     various models of QCD  \cite{Klingl:1997kf,Kwon:2008vq}.
 Also, these sum rules may be used to extract the information
 on $\la {\cal O}_{4,6}^i \ra $ by adopting
 the experimental dilepton spectrum after background subtraction 
 in the l.h.s. of Eqs.(\ref{eq:vector-II},\ref{eq:vector-III})
   \cite{Hatsuda:1997lbln}.

 In general, the  spectral function receives  peak-shift,
 broadening, new peaks, etc due to the complex interaction of the
vector current with the medium \cite{Rapp:1999ej}.
Also, such spectral changes may well depend
 on the spatial momentum of the current 
 \cite{Jean:1993bq,Eletsky:1996jg,Friman:1999wu}.
 Indeed, the transverse and longitudinal spectral
 functions $\rho_{_{V,A}}^{\rm T,L}(\omega,{\bf p})$
 obey different Weinberg-type sum rules for ${\bf p}\neq 0$
 \cite{Kapusta:1993hq}.
 Because of these reasons, it is not appropriate to
 oversimplify the problem to ``mass shift vs. width broadening".
 Nevertheless, there is a theoretical suggestion that
the width broadening at low temperature and/or baryon density is eventually taken over
 by the mass shift near the critical point of chiral transition
 \cite{Yokokawa:2002pw}\cite{Brown:2008xh}.
Experimentally, it is important to measure the
full momentum dependence of the spectral function 
$\rho_{_V}^{\rm T,L}(\omega,{\bf p})$ instead of the  projected 
invariant-mass spectra. 

\subsection{Dileptons, why and how?}
Dileptons ($l^+ l^-$ pairs, where $l=e$ or $\mu$)
are an excellent tool to study
possible in-medium modifications of vector mesons (Table~\ref{tab:vectormesons}), $\rho, \omega$ and $\phi$ in nuclear media,
because of their negligible final-state interactions. 
Due to their short lifetime, $\rho^0$ mesons have larger probability of decaying in medium,
while $\omega$ and $\phi$ tend to decay outside. 
In order to study $\omega/\phi$ in-medium properties, it is important to choose a proper reaction and to select slow-moving mesons.

Obtaining dilepton distributions (usually presented in the form of $\l^+\l^-$ invariant-mass spectra $M_{\l^+\l^-}$) is technically very demanding because of the small dilepton-decay
branching ratios of these mesons (Tab. \ref{tab:vectormesons}), while there are many hadronic sources which can produce leptons.
The detector therefore must have an excellent lepton-identification capability, and must also provide means to suppress combinatorial background, the background caused by an $l^+$ being erroneously paired up with an $l^-$ from other origin (e.g., a $e^+$ from $\pi^0 \rightarrow \gamma e^+e^-$ paired up with an $e^-$ from $\gamma\rightarrow e^+e^-$ occurring in the same event).

Even with the start-of-the-art dilepton detectors, the combinatorial background is severe, especially in high-energy heavy-ion collisions. For example, at CERN SPS (158 $A$GeV central collisions), the NA60 experiment $(\mu^+\mu^-)$ reported a signal $S$ to background $B$ ratio of about 1/11 \cite{Arnaldi:2007oe}, and it was about 1/22 in the case of the CERES experiment $(e^+e^-)$ \cite{Adamova:ng}. At RHIC ($\sqrt{s_{_{NN}}}=200$ GeV Au+Au minimum bias), the PHENIX experiment $(e^+e^-)$ reported a signal-to-background ratio of about  $1/100$ \cite{Afnasiev:2007xw}.

In order to reliably extract meaningful results despite such small signal-to-background ratios, methods such as event mixing and like-sign pair subtraction have been developed to reliably subtract combinatorics, as discussed in section~\ref{sec:combinatorics}.

The combinatorics-subtracted $M_{l^+l^-}$ distribution still contains a broad continuous background due to Dalitz decays. In order to extract the vector-meson contributions, the measured distribution is compared with the ``hadronic cocktail'', which contains all known sources of $l^+l^-$ pairs produced in the detector acceptance.

\begin{table*}
\begin{center}
\caption{\label{tab:vectormesons} Properties of vector mesons. \cite{Amsler:2008pj}}
\label{tab:mesons}
\begin{small}
\begin{tabular}{c|r|r|r|c|r|r|r}
\hline
&\multicolumn{1}{c|}{Mass} & \multicolumn{1}{c|}{$\Gamma$} & \multicolumn{1}{c|}{$c \tau$} & Main & $\frac{\Gamma{e^+e^-}}{\Gamma_{\rm tot}}$ $^*$ & $\frac{\Gamma{\mu^+\mu^-}}{\Gamma_{\rm tot}}$ $^*$ &  $\frac{\Gamma{\pi^0 \gamma}}{\Gamma_{\rm tot}}$ $^*$\\
 & (MeV/$c^2$) & (MeV/$c^2$) & (fm)& decay & $(\times 10^{-5})$ & $(\times 10^{-5})$ & \\
\hline
$\rho^0$ & $775.49$ & $149.4$&1.3 & $\pi^+\pi^-$  & $4.7$ &4.6 & $6.0\times 10^{-4}$\\
              &$ \pm 0.34$ &  $ \pm 1.0$  &       &$(\sim 100\%$)&     &            &              \\
\hline
$\omega$ & $782.65$ & $8.49$&23.2 & $\pi^+\pi^-\pi^0$  & $7.2$ & 9.0 & 8.9\%\\
              &$ \pm 0.12$ &  $ \pm 0.08 $  &       &$(89\%$)&                  &             & \\
\hline
$\phi$ & $1019.455$ & $ 4.26$&46.2 & $K^+K^-$  & $29.7$&28.6 & $12.7\times 10^{-4}$\\
              &$ \pm 0.020$ &  $ \pm 0.04$  &       &$(49\%$)&         &              &         \\
\hline
\end{tabular}
\end{small}

\noindent
$^*$ These branching ratios are at the pole mass.
\end{center}
\end{table*}

\subsubsection{$\mu^+\mu^-$-pair detection}
Muons produced in pion and kaon decays (e.g., $\pi^+\rightarrow \mu^+\bar\nu$ and $K^+\rightarrow \mu^+\bar\nu$) are much more abundant than those from vector-meson decays, and they contribute to the combinatorial background. It is therefore essential to absorb hadrons as close as possible to the interaction point in a thick absorber.
Muon momenta are measured by magnetic spectrometer(s) placed behind the absorber, corrected for the energy loss and multiple scattering in the absorber, and the pair-mass distribution is reconstructed.
This is exactly how $\Upsilon (b \bar b)$ was first discovered by \citet{lederman:1977}.

This method works well for heavy-mass region ($\Upsilon$ as well as $J/\Psi$), 
where the decay-muons have high momenta, and the particle multiplicities are low,
but the measurement of low-mass vector mesons is difficult due to larger combinatorial background and larger multiple scattering in the absorber (hence lower mass resolution). 
HELIOS/3   and  NA60 at CERN SPS overcame these difficulties and successfully measured the dimuon spectra all the way down to the pair-mass threshold of $2 m_\mu$.\footnote{Since muons must be energetic enough to penetrate the hadron absorber, the rapidity of the reconstructed pairs are high, e.g., $3.3<y<4.3$ in the case of NA60 \cite{Arnaldi:2007oe}. For $e^+e^-$, pairs can be measured in the mid-rapidity region of $2.1<\eta<2.65$ \cite{ceresfullpaper:2005sh}.}

\subsubsection{$e^+e^-$-pair detection}
The spectrometer used in the  discovery of $J/\Psi (c \bar c)$ in the $p + {\rm Be} \rightarrow e^+ e^- X$ reaction \cite{ting:1974} contains the essence of $e^+e^-$ measurement, such as
(i) excellent electron identification (hadron rejection), 
(ii) good momentum (pair-mass) resolution,
and (iii) importance of rejecting $e^+e^-$ pairs from photon conversion and Dalitz decays.

In the $e^+e^-$ spectra, severe background sources are
photons from meson decays such as $\pi^0 \rightarrow 2 \gamma$ converting
in the target $(\gamma\rightarrow e^+ e^-$) and in detectors, and the Dalitz decays such as
$\pi^0\rightarrow \gamma e^+e^-$, $\eta \rightarrow \gamma e^+e^-$, $\eta' \rightarrow \gamma e^+e^-$ and  $\omega\rightarrow \pi^0 e^+e^-$.

Although pairs from these sources have small opening angles and low masses, the limited track reconstruction efficiency and acceptance lead to a combinatorial background for events in which two or more of these low-mass pairs are only partially reconstructed. 
This is {\em the} central problem of any low-mass $e^+e^-$-pair experiment.

\subsubsection{Combinatorial background}
\label{sec:combinatorics}
Uncorrelated sources can produce, in addition to unlike-sign $(l^+ l^-)$ pairs,  like-sign ($l^+ l^+$ and $l^- l^-$) pairs. Most experiments make use of this fact in evaluating the combinatorial background.

A typical method of subtracting the combinatorial background is as follows \cite{Toia:2006qd,Toia:2007yu}:
Under the assumption that electron and positron multiplicities are Poisson-distributed, and that the like sign pairs are uncorrelated, the combinatorial background $B$ can be accounted for by \begin{equation}
B=2 \sqrt{N^{++} N^{--}}, 
\label{eq:combinatorics}
\end{equation}
where $N^{++}$ and $N^{--}$ are the number of measured $l^+ l^+$ and $l^- l^-$ pairs, respectively. The number of signal pairs $S$ is then obtained as $S = N^{+-} -B$, where $N^{+-}$ is the number of measured unlike-sign pairs.

This would work if the detector acceptance is the same for like and unlike sign pairs, and if a sufficient number of like-sign pairs are collected.
This is in general not the case. 
A mixed event technique is then used to compute the combinatorial background.
In this method, unlike-sign tracks from different events (with similar event topology) are paired. 
Since the same track can be used many times, paired up with tracks from different events, the background spectra can be generated with high statistics. 
The accuracy of the technique can be tested
by comparing the shape of the measured like-sign pair spectrum with that of the mixed combinatorial background.
The generated background event distribution can be normalized to the number of expected unlike-sign pairs from Eq.(\ref{eq:combinatorics}).

%
%
%

\subsection{High Energy Heavy Ion Reactions}

Although the main subject of the present review is the behavior of mesons produced in nuclei with elementary reactions, we nevertheless 
touch upon the low-mass ($\lesssim 1 {\rm GeV}/c^2$) dileptons observed in heavy-ion collisions.

An enhanced yield of dilepton pairs in the low-mass region is ubiquitous from 1 $A$GeV (Bevalac/SIS), through SPS energies ($40-200 A$GeV), up to the RHIC energy of $\sqrt{s_{_{NN}}} = 200$ GeV. Here, the ``enhancement'' is defined as the excess of the observed $l^+l^-$  yield over the sum of the ``hadronic cocktail'' as discussed above.

\subsubsection{Bevalac/SIS energies ($1\sim 2 A$GeV)}

\subsubsection*{DLS}

\begin{figure}
\includegraphics[width=.8\columnwidth]{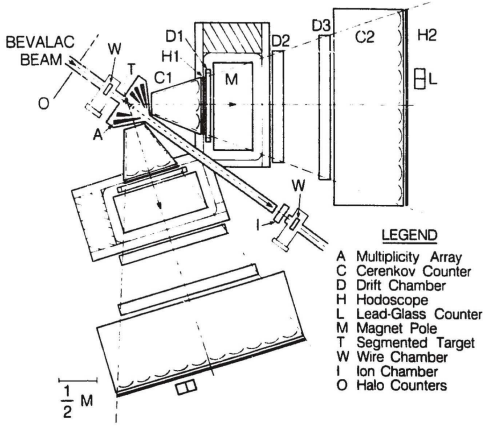}
\caption{\label{fig:dls-setup} Top view of the DLS (dilepton spectrometer) at Bevalac \cite{Yegneswaran:1990sf}.}
\end{figure}

\begin{figure}
\includegraphics[width=.8\columnwidth]{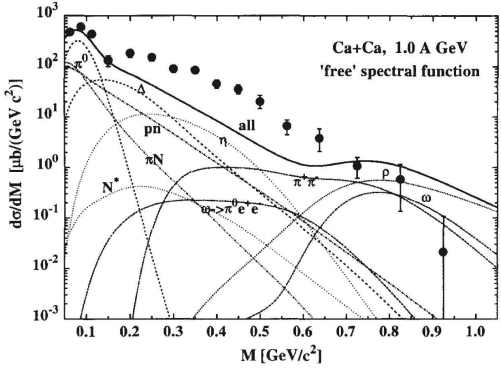}
\caption{\label{fig:dls-ca+ca} The dilepton spectrum for Ca+Ca at 1.0 $A$GeV measured by the DLS collaboration (circles) \cite{Porter:1997dls}, 
compared with the ``hadronic cocktail'' assuming the ``free'' $\rho$ spectral function
\cite{Bratkovskaya:1998zm}.}
\end{figure}

The first anomalous dilepton excess was reported by the DLS (dilepton spectrometer) experiment at Bevalac ($1 A$GeV). Using a two-arm spectrometer as shown in Fig.~\ref{fig:dls-setup}, with a pair-mass resolution of $\Delta M/M \sim 10\%$, they succeeded for the first time to measure the dielectron spectra in heavy-ion collisions \cite{Porter:1997dls}. When the measured spectra were compared with transport theory calculations \cite{Ernst:1998,Cassing:1999es,Shekhter:2003, Cozma2006170,Bratkovskaya:2008qd}, an excess of about a factor $6-7$ was found in the mass range of $0.15<M_{_{e^+e^-}}<0.4$ GeV (Fig.~\ref{fig:dls-ca+ca}). 
Including the $\rho$-meson modifications in the medium (mass dropping) did not eliminate the discrepancy (the observed yield was still higher by about a factor of 3 over the HSD curve). This has become known as the ``DLS puzzle''.

\subsubsection*{HADES}

\begin{figure}
\includegraphics[width=.65\columnwidth]{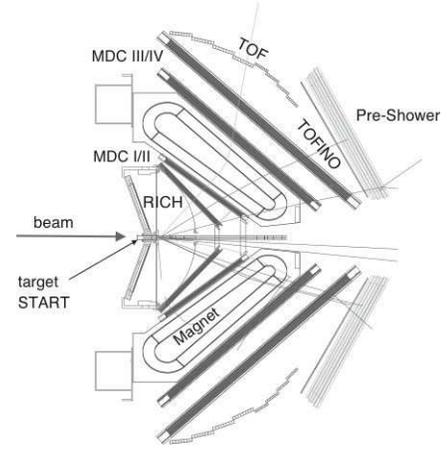}
\caption{\label{fig:hades-sideview} A side view of HADES. The RICH detector, consisting of a gaseous radiator, a carbon fiber mirror 
and a tilted photon detector, is used for electron identification. Two sets of multiwire drift chambers (MDCs) are 
placed in front and behind the magnetic field to measure particle momenta. A time of flight wall (TOF/TOFINO) 
accompanied by a Pre-shower detector at forward angles is used for an additional electron identification and 
trigger purposes. For a reaction time measurement, a start detector is located near the target
\cite{Salabura:2004wn}.}
\end{figure}

This DLS puzzle has been recently revisited by
the HADES (high acceptance dielectron spectrometer) collaboration at the heavy ion synchrotron SIS at GSI Darmstadt. HADES uses modern technologies such as a ring-imaging \v Cerenkov detector (RICH) to achieve good particle identification as well as a high mass resolution of $\Delta M/M \simeq 2.7 \%$ (Fig.~\ref{fig:hades-sideview}). They recently studied $1 A$GeV $\rm C+C$ collisions in a low-resolution mode ($\Delta M/M =8 \%$ at 0.8 GeV/$c^2$ to emulate that of DLS) and projected the measured spectra into the DLS acceptance. 

\begin{figure}
\includegraphics[width=.65\columnwidth]{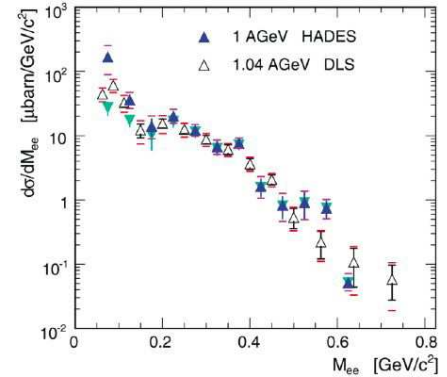}
\caption{\label{fig:dls-vs-hades} 
Direct comparison of the dilepton pair mass distributions measured in $\rm C + C$ at 1 $A$GeV by HADES (within the DLS acceptance) and at 1.04 $A$GeV by DLS \cite{Porter:1997dls}.  Statistical and systematic errors are shown. Overall normalization errors (not shown) are 20\% for the HADES and 30\% for 
the DLS data points
\cite{Agakishiev:2008qm}.}
\end{figure}

The resulting spectrum, shown in Fig.~\ref{fig:dls-vs-hades}, is consistent with
that measured by DLS. 
For masses of
$0.15 {\rm GeV}/c^2 < M_{_{e^+e^-}} < 0.50 {\rm GeV}/c^2$ it exceeds expectations
based on the known production and decay rates of hadrons (most important being the $\eta$
meson) by a factor of about 7, thereby reconfirming the DLS data.

However, recent HADES measurements of $p+p \rightarrow e^+e^-  X$ and $p+n \rightarrow e^+e^- X$
seem to show that the $\rm C+C$ spectrum at $1 A$GeV agree well with the $\frac{1}{2}(pp + np)$ spectrum at 1.25 GeV, when scaled by the $\pi^0$ yield \cite{Galatyuk:2009eb}. 
This may be indicating that the DLS/HADES effect does not have nuclear (in-medium) origin, but that the $NN$ bremsstrahlung cross sections are in fact larger than hitherto assumed \cite{Shyam:2003,Kaptari:2006la}. 
With the enhanced bremsstrahlung cross sections implemented in the HSD transport code, \citet{Bratkovskaya:2008qd} have recently shown that the calculated spectra agree well with DLS and HADES data.

\subsubsection{SPS energies ($40\sim 200 A$GeV)}

At the CERN Super Proton Synchrotron (SPS), low-mass dilepton spectra were studied in the $e^+e^-$ mode by the CERES collaboration, and in the $\mu^+\mu^-$ mode
by the HELIOS/3 collaboration 
and the NA60 collaboration. These experiments all reported a low-mass dilepton enhancement.

\subsubsection*{HELIOS/3 $(\mu^+\mu^-)$}

\begin{figure}
\includegraphics[width=\columnwidth]{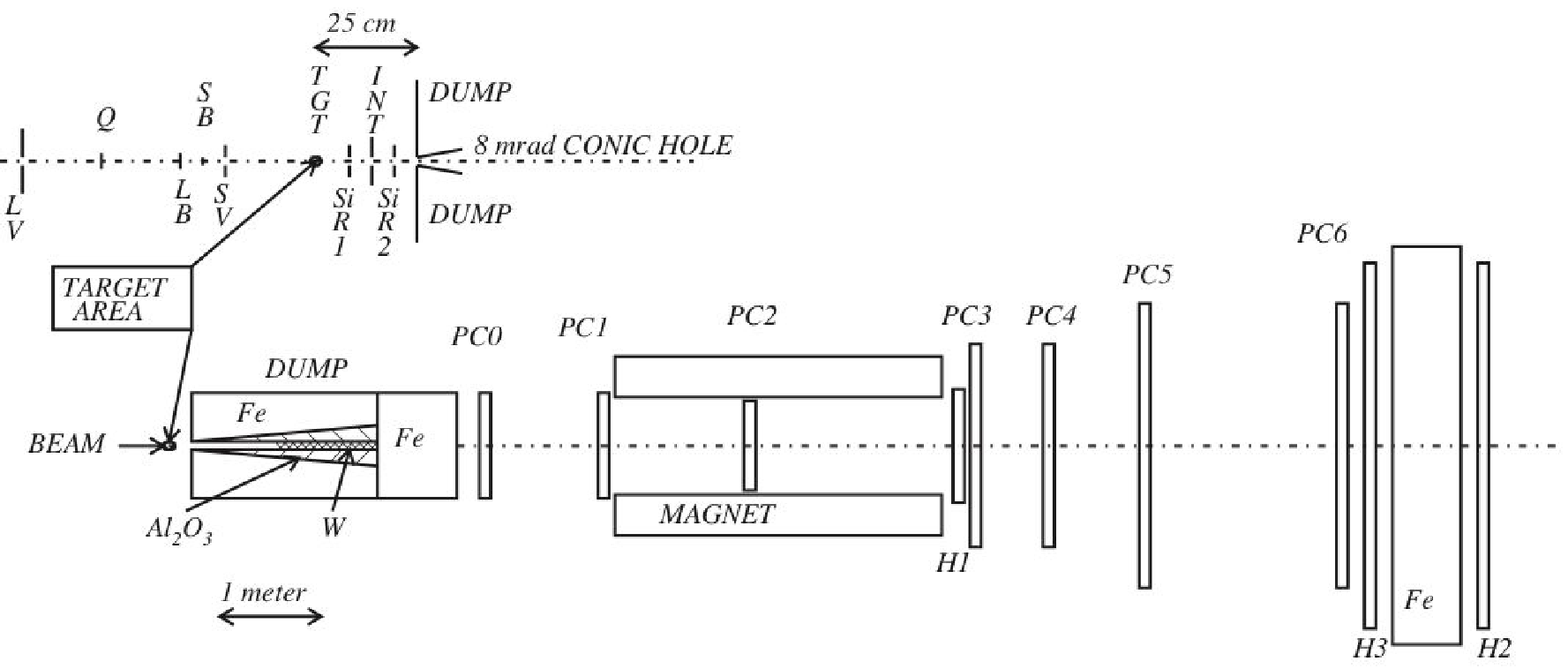}
\caption{\label{fig:helios3-setup} Helios-3 setup
\cite{Angelis:2000uf}.
}
\end{figure}

The HELIOS/3 experiment used a dimuon spectrometer shown in Fig.~\ref{fig:helios3-setup} to measure $\mu^+\mu^-$ distributions in proton on tungsten and sulphur on tungsten at 200 $A$GeV.
The
spectrometer consisted of a hadron absorber (placed 25 cm downstream from the target), six
interaction lengths of Al$_2$O$_3$ and 100 cm of Fe, followed
by a magnetic spectrometer and muon hodoscopes.

\begin{figure}
\includegraphics[width=.65\columnwidth]{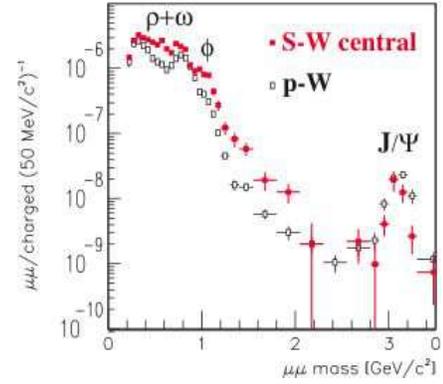}
\caption{\label{fig:helios-3-pw-sw} $M_{\mu\mu}$ distribution normalized to the charged-particle multiplicity obtained by the HELIOS/3 collaboration for S-W (filled squares) and $p$-W (open squares) collisions at 200 $A$GeV
\cite{Angelis:2000uf}.
}
\end{figure}

They found, by comparing the measured $p$-W and S-W dimuon distributions, each normalized to the charged-particle multiplicity (Fig.~\ref{fig:helios-3-pw-sw}), 
an excess in S-W interactions relative to minimum-bias $p$-W interactions. The 
observed excess is continuous over the explored mass range and has no apparent resonant structure. In the low mass ($<0.7$ GeV) region the dimuon yield increases by $76\pm 4 \%$ of the corresponding $p$-W dimuon spectrum (in the higher mass region, the excess was higher).

\subsubsection*{CERES (NA45) $(e^+e^-)$}

\begin{figure}
\includegraphics[width=.85\columnwidth]{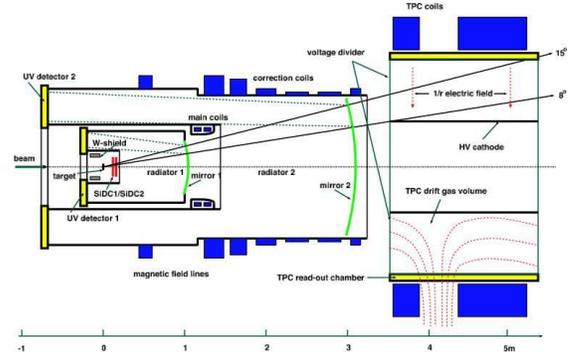}
\caption{\label{fig:ceres-with-tpc} Schematic view of the CERES spectrometer with a radial drift time projection chamber (TPC). The latter was added in NA45/2 for the Pb+Au runs \cite{Marin:2004ceres}.}
\end{figure}

CERES is an innovative `hadron-blind' axial-symmetric detector  (Fig.~\ref{fig:ceres-with-tpc}) dedicated to the measurement of electron pairs in the low-mass range (up to $\sim 1.5$ GeV/$c^2$). 
At the heart of CERES are the two coaxial ring-imaging \v Cerenkov detectors (shown in the left half of Fig.~\ref{fig:ceres-with-tpc}) having a high \v Cerenkov threshold of $\gamma_{\rm thr} \simeq 32$, placed in a superconducting solenoid. With this setup,  $e^+e^-$ pairs in a window of $\Delta \eta = 0.53$ around mid-rapidity were selectively detected and reconstructed. 

\begin{figure}
\includegraphics[width=\columnwidth]{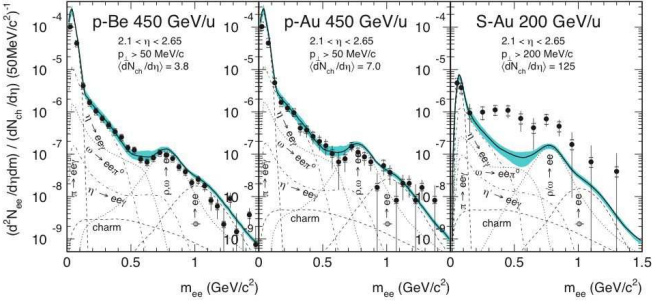}
\caption{\label{fig:ceres-fullpaper-fig1}
CERES inclusive $e^+e^-$ mass spectra of 450 GeV $p$-Be, $p$-Au, and 200 $A$GeV S-Au collisions \cite{agakichiev1995epl,Agakichiev:1998xr}. Plotted is the 
number of electron pairs per charged particle, both in the acceptance and per event. Contributions from various hadron decays 
as expected from $p$-$p$ collisions are shown together with their sum (thick line), and the systematic error on the latter is indicated 
by the shaded area \cite{Adamova:ng}. }
\end{figure}

The dielectron pairs per charged particles, shown in Fig.~\ref{fig:ceres-fullpaper-fig1} show that 
while the $p$-Be and $p$-Au data are reproduced within errors by Dalitz and direct decays of neutral mesons as known from $p$-$p$ collisions, dielectrons from S-Au collisions reveal a substantial enhancement in the mass region $0.2-1.5 {\rm GeV}/c^2$ of a factor $5$~\cite{agakichiev1995epl}. 

This observation generated lots of excitement in the community. It has been attributed to the pion annihilation in the fireball, $\pi^+\pi^- \rightarrow \rho \rightarrow e^+e^-$ with a strong in-medium modification of the intermediate $\rho$, such as mass dropping~\cite{Brown:2001nh} or broadening~\cite{Rapp:1999ej}.

\subsubsection*{NA60 $(\mu^+\mu^-)$}

\begin{figure}
\includegraphics[width=\columnwidth]{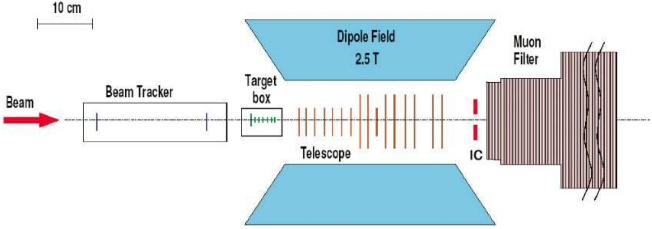}
\caption{\label{fig:na60-schematic}  Layout of the NA60 detectors in the vertex region  \cite{Usai:2005fy}.}
\end{figure}

The NA60 experiment added a telescope of radiation-tolerant silicon pixel detectors in between the target and the hadron absorber of the NA50 dimuon spectrometer (see Fig.~\ref{fig:na60-schematic}).
This enabled the collaboration to match muon tracks before and after the hadron absorber, both in angular and momentum space, thereby improving the dimuon mass resolution in the region of light vector mesons from $\sim 80$ to $\sim 20 $MeV$/c^2$ \cite{arnaldi:162302,Arnaldi:2007oe}.

\begin{figure}
\includegraphics[width=.65\columnwidth]{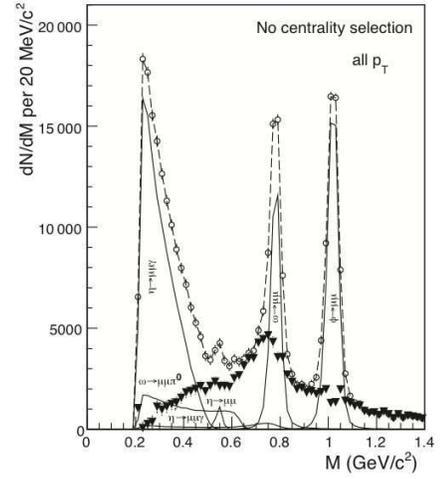}
\caption{\label{fig:na60-excess-isolation}
Isolation of an excess above the hadron decay cocktail 
(see text). Total data (open circles ), individual cocktail sources 
(solid ), difference data (thick triangles ), sum of cocktail sources 
and difference data (dashed ) 
\cite{arnaldi:162302,Arnaldi:2007oe}.}
\end{figure}

With this setup, NA60 succeeded to completely resolve $\omega$ and $\phi$ peaks in the In-In collisions at 158 $A$GeV,  for the first time in nuclear collisions. This is shown in Fig.~\ref{fig:na60-excess-isolation} (the mixed-event technique was employed here to subtract the combinatorial background).

By adjusting the cross
section ratios $\eta/\omega$, $\rho/\omega$ and $\phi/\omega$, as well as the level of $D$ meson
pair decays, the peripheral data could be fitted 
by the 
expected electromagnetic decays of the neutral mesons, i.e.,
the 2-body decays of the $\eta$, $\rho$, $\omega$ and $\phi$ resonances 
and the Dalitz decays of the $\eta$, $\eta'$ and $\omega$.
In the more central cases, a fit procedure is ruled out
due to the existence of a strong excess with a priori unknown characteristics.

\begin{figure}
\includegraphics[width=.6\columnwidth]{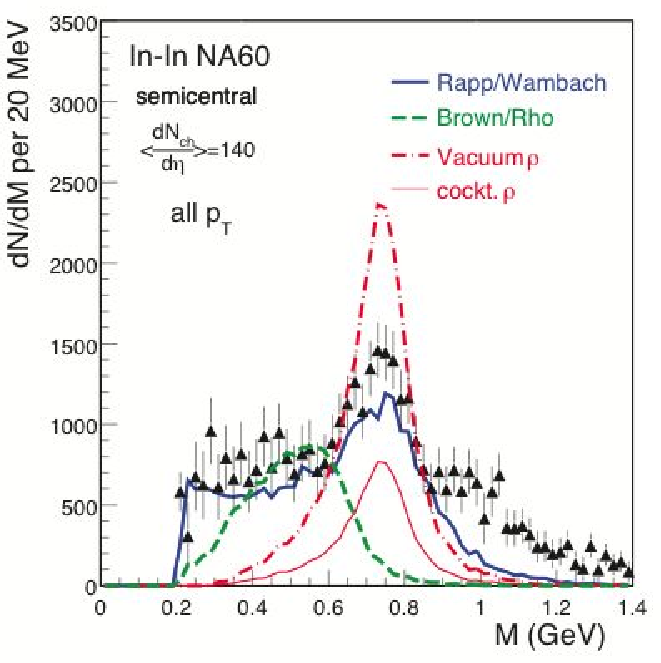}
\caption{\label{fig:na60-rho-shape} 
Comparison of the excess mass spectrum for the semi-central bin 
to model predictions, made for In-In at $dN_{ch} =d\eta |_{\eta=0} = 
140$. Cocktail $\rho$ (thin solid red line), unmodified $\rho$ (dashed-dotted red line), 
in-medium broadening $\rho$ \cite{chanfray:1996,Rapp:1997fs,Rapp:1999ej} (thick solid blue line), in-medium 
moving $\rho$ related to \cite{Brown:1991kk,LiKoBrown:1995,Brown:2001nh} (dashed green line). 
The errors are purely 
statistical. The systematic errors of the continuum are about 25\%. 
From 
 \citet{arnaldi:162302,Damjanovic:2008ta}.}
\end{figure}

The excess was therefore isolated by subtracting the measured decay cocktail,
without the $\rho$, from the data, as shown in Fig.~\ref{fig:na60-excess-isolation}. 
The resultant distribution shows some non-trivial centrality dependence, but is largely consistent with a dominant contribution from $ \rho\rightarrow \mu^+\mu^-$ annihilation.
Fig.~\ref{fig:na60-rho-shape} shows a distribution obtained for  semi-central collisions, compared with in-medium broadening~\cite{Rapp:1999ej} and mass-dropping~\cite{Brown:2001nh} scenarios. The observed distribution ($\rho$ spectral function) exhibits considerable broadening, but essentially no shift of the $\rho$-peak position.

\subsubsection*{CERES (NA45/2) $(e^+e^-)$}
In preparation for the lead beam acceleration in the SPS, CERES upgraded the detector by adding a cylindrical time projection chamber (TPC) with a radial electric field (right half of Fig.~\ref{fig:ceres-with-tpc}). Among other things, this improved the mass resolution $\Delta m/m$ in the region of the $\rho/\omega$ from $9\%$ to about $6\%$ \cite{Marin:2004ceres,ceresfullpaper:2005sh}.

\begin{figure}
\includegraphics[width=\columnwidth]{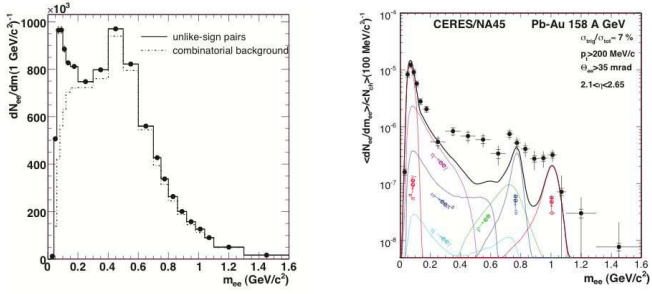}
\caption{\label{fig:ceres-pb-au} 
(left)  CERES
unlike-sign pair yield (histogram) and combinatorial background (dashed 
curve). 
(right)
Invariant $e^+ e^-$ mass spectrum compared to the expectation from hadronic decays  \cite{Adamova:ng}.}
\end{figure}

The dielectron distribution obtained in Pb-Au collisions at 158 $A$GeV before combinatorial subtraction, together with the normalized mixed-event background is shown in the left panel of Fig.~\ref{fig:ceres-pb-au}. The background-subtracted distribution
 is compared with the hadronic cocktail in the right panel of Fig.~\ref{fig:ceres-pb-au}. Here again, an enhancement over the cocktail is observed in the mass range $0.2 < m_{_{e^+e^-}} < 1.1 {\rm GeV}/c^2$, the enhancement factor being $2.45 \pm 0.21 \mbox{~(stat)} \pm 0.35 \mbox{~(syst)} \pm 0.45 \mbox{~(decays)}$, where the last error is from the systematic uncertainty in the cocktail calculation.

\begin{figure}
\includegraphics[width=\columnwidth]{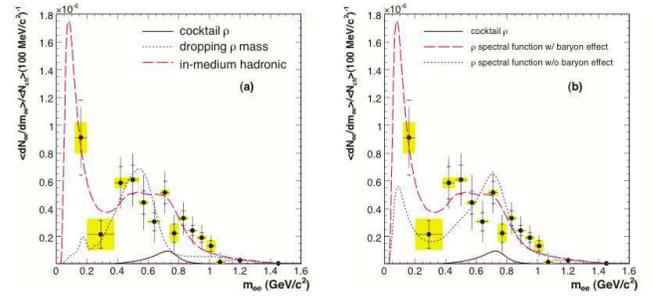}
\caption{\label{fig:ceres-plb-dropping-broadening} 
CERES $e^+ e^-$ pair yield after subtraction of the hadronic cocktail. In addition to the statistical error bars, systematic errors of the data (horizontal ticks) and the systematic
uncertainty of the subtracted cocktail (shaded boxes) are indicated. The broadening scenario (long-dashed line: \cite{Rapp:1999ej,hees:102301}) is compared to a calculation assuming a density dependent
dropping $\rho$ mass (dotted line in (a): \cite{Brown:1991kk,Brown:1995qt,Brown:2001nh}) and to a broadening scenario excluding baryon effects (dotted line in (b)). From 
 \citet{Adamova:ng}.}
\end{figure}

Fig.~\ref{fig:ceres-plb-dropping-broadening} shows the dielectron yield after the hadronic-cocktail subtraction, compared with the mass dropping (left) and width broadening (right) assumptions. 
Although the error bars are larger than those in the NA60 spectra, the authors concluded that a substantial in-medium broadening of the $\rho$ is favored over a density-dependent shift of the $\rho$ pole mass.

\subsubsection{RHIC ($\sqrt{s_{_{NN}}} = 200$ GeV)}

\begin{figure}
\includegraphics[width=.85\columnwidth]{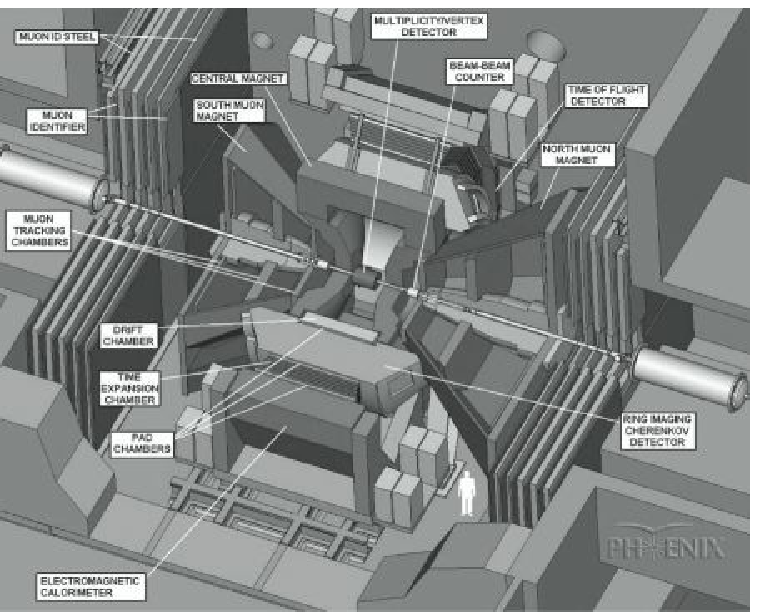}
\caption{\label{fig:phenix-overview} 
The PHENIX experiment at RHIC
 \cite{Adcox:2003ud}.}
\end{figure}

At RHIC, the PHENIX experiment (Fig.~\ref{fig:phenix-overview}) has been designed to measure dielectrons over a wide mass range. Electrons and positrons are reconstructed in the two central arm spectrometers using drift chambers, located outside an axial magnetic field. They are identified by hits in the ring imaging \v Cerenkov detector (RICH) and by matching the momentum with the energy measured in an electromagnetic calorimeter.

\begin{figure}
\includegraphics[width=.85\columnwidth]{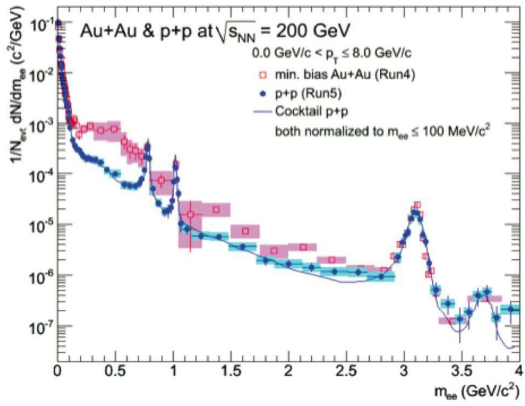}
\caption{\label{fig:phenix-dielectron-fig2} 
Invariant $e^+e^-$ pair yield observed by PHENIX in Au+Au minimum-bias and in $p+p$ collisions at $\sqrt{s_{_{NN}}}=200$ GeV. The curve is the hadronic cocktail for $p+p$.
The $p+p$ spectrum and the cocktail curve were normalized to the Au+Au 
yield in the $m_{_{e^+e^-}}<100$ MeV/$c^2$ region 
\cite{Afnasiev:2007xw,phenix:2008asa,Toia:2008dj}.}
\end{figure}

Fig.~\ref{fig:phenix-dielectron-fig2} shows a dielectron distribution observed by PHENIX in Au+Au minimum-bias collisions at $\sqrt{s_{_{NN}}}=200{\rm GeV}$, after combinatorial background subtraction using the mixed event shape normalized to the like-sign pair yields  \cite{Afnasiev:2007xw}.
The dielectron yield in the minimum bias collisions, in the mass range between 150 and 750 MeV$/c^2$, is enhanced over the cocktail by a factor of $3.4\pm 0.2 \mbox{(stat.)} \pm 1.3\mbox{(syst.)}\pm0.7\mbox{(model)}$. A clear increase with centrality is also observed. No detailed analysis of the excess is available yet.


\subsubsection{High energy heavy ion summary}
An enhanced yield of dilepton pairs over the hadronic sources in the low-mass region has been observed, regardless of the bombarding energy. 

At low beam energy of  $\sim 1 A$GeV, the long-standing DLS puzzle (excess) has been confirmed by the recent HADES experiment. However, recent indications of C+C dielectron distribution agreeing with the $(pp+pn)/2$ distribution, if confirmed, may rule out the possibility of in-medium modification effects at this energy.

The two SPS experiments, CERES measuring dileptons and NA60 measuring dimuons, both established that there is a dilepton enhancement in the low-mass region. The excess here is consistent with $\pi^+\pi^- \rightarrow \rho\rightarrow l^+l^-$ with the $\rho$ significantly broadened in the nuclear medium, while the data do not call for the simple $\rho$ mass change.

At RHIC, the PHENIX experiment showed that there is a dilepton enhancement in the low-mass region, the magnitude of which increases faster with the centrality of 
the collisions than the number of participating nucleons, but the statistical errors are still fairly large in order to draw firm conclusions based on the data.

\subsection{$\rho$, $\omega$ and $\phi$ mesons produced in nuclei with elementary reactions}
\subsubsection{TAGX at INS Electron Synchrotron}
\label{sec:tagx}
\begin{figure}
\includegraphics[width=.5\columnwidth]{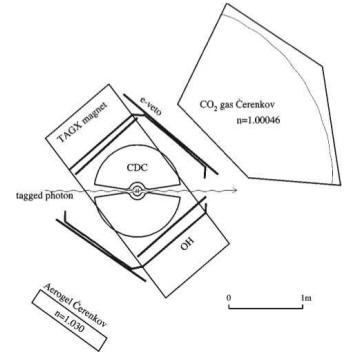}
\caption{\label{fig:tagx-setup} 
A plan view of the TAGX spectrometer  \cite{huber:065202}.
}
\end{figure}

The TAGX experiment (Fig.~\ref{fig:tagx-setup}) at the 1.3-GeV INS Electron Synchrotron (Institute for Nuclear Study, Tokyo University)  
used a tagged photon beam in the energy range of 600-1120 MeV to study the $\gamma A \rightarrow \pi^+\pi^- X $ reaction on $A=^2$H, $^3$He and $^{12}$C targets. This was a pioneering experiment which attempted to study in-medium modifications of $\rho^0$ with elementary reactions.

However, the claim of finding a $\rho^0$ mass shift of $-160\pm 35 {\rm MeV}/c^2$ in $^3$He \cite{lolos1998erhom}  was met with skepticism due to the inevitable pion rescattering effect even for light targets (the emitted pions being in the resonance region), the small target volume, and the much-larger-than-expected shift.


They later applied a helicity analysis to extract in-medium $\rho_L^0$ invariant mass distributions \cite{huber:065202}, compared the spectra with various simulations \cite{Rapp:1997fs,Post:2003hu,STT:1997}, 
and obtained a smaller but still sizable mass shift of $-65\sim -75{\rm MeV}/c^2$ in the photon energy bin of $E_\gamma = 800-900$ MeV, and $-45{\rm MeV}/c^2$ for $E_\gamma = 960-1120{\rm MeV}$. The $^{12}$C distributions on the other hand, were found to be consistent with quasi-free $\rho^0$ production.

Why TAGX observed such a large effect only in $^3$He is not yet understood, but in view of the fact that the $\gamma A \rightarrow e^+e^- X$ data of CLAS-g7 (section \ref{sec:g7}) do not show any sign of $\rho^0$ mass shift, this is most likely unrelated to the $\rho^0$ in-medium modification.\footnote{The incident photon energy range of TAGX was $0.6-1.12$ GeV while it was $0.61-3.82$ GeV in the CLAS-g7 experiment. Therefore, a more direct comparison would be to use low incident energy events of  the CLAS-g7 data sample. }


\subsubsection{E325 experiment at KEK}
\label{sec:e325}

\begin{figure}
\includegraphics[width=0.9\columnwidth]{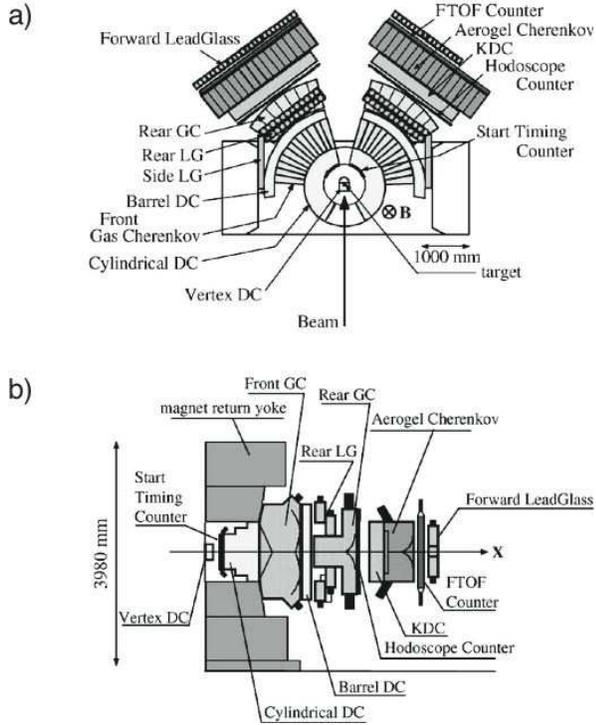}
\caption{\label{fig:ozawa-fig1}  Schematic view of the experimental setup of the E325 
spectrometer: (a) the top view and (b) the side view. The side 
view shows the cross section along the center of the kaon arm 
 \cite{ozawa2001orhoomega,Sekimoto:2004ek}.
}
\end{figure}

The experiment E325 at the KEK 12 GeV Proton Synchrotron was the first to measure
dileptons in search for  the modiﬁcation of the vector meson mass in a nucleus in elementary reactions.
They measured the invariant mass spectra of $e^+e^-$ pairs produced in 12 GeV proton-induced nuclear reactions. The setup is a two-arm spectrometer (Fig.~\ref{fig:ozawa-fig1}), and was designed to 
measure the decays of the vector mesons, $\phi\rightarrow e^+e^-$, $\rho/\omega \rightarrow e^+e^-$ as well as $\phi\rightarrow K^+K^-$.

For electron identification, two stages of electron-identification counters were used.
The first was the front gas-\v Cerenkov counters (FGC). 
The second stage consisted of the rear gas-\v Cerenkov 
counters (RGC), the rear lead-glass electromagnetic (EM) calorimeters 
(RLG),  the forward lead-glass EM calorimeters (FLG), and the side lead-glass EM calorimeters (SLG).
The overall electron efficiency was 78\% with a pion 
rejection power of $3\times 10^{-4}$ \cite{Sekimoto:2004ek}.

The mass resolution was estimated to be 8.0 MeV$/c^2$ and 10.7 MeV/$c^2$ for $\omega\rightarrow e^+e^-$ and $\phi \rightarrow e^+e^-$ decays, respectively.

\begin{figure}
\includegraphics[width=0.85\columnwidth]{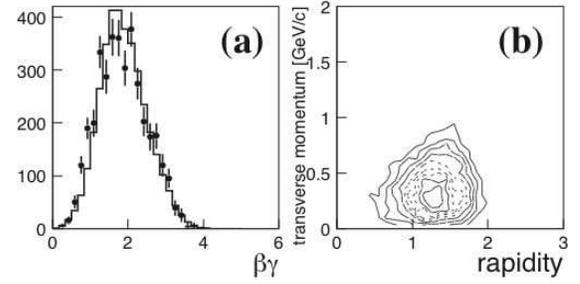}
\caption{\label{fig:muto-fig2}  Kinematical distributions of $e^+e^-$ pairs in the $\phi$ mass region
($0.95\rightarrow M_{_{e^+e^-}}<1.05{\rm GeV}/c^2$ detected in the E325 spectrometer (points with error bars),
together with the simulation result using the JAM nuclear cascade 
code (histogram,  \cite{JAM:1999}). 
(left) $\beta\gamma$ distribution. (right) Rapidity $y$ vs pair transverse momentum $p_T$
 \cite{muto:042501}.
}
\end{figure}

The kinematical 
region covered was $0.5 < y< 2$ and $1 <\beta\gamma<3$ for $e^+e^-$ pairs (Fig.~\ref{fig:muto-fig2}), where the decay probability inside the target 
nucleus was expected to be enhanced.
Assuming that the meson decay widths are unmodified in nuclei, the coverage would correspond to the in-nucleus decay fractions shown in Table ~\ref{tab:decayfraction}.

\begin{table}
\caption{\label{tab:decayfraction} Expected in-nucleus decay fractions of vector mesons in the E325 kinematics, assuming that the meson decay widths are unmodified in nuclei, obtained by using a Monte Carlo-type model calculation \cite{naruki:092301,muto:042501}.}
\begin{center}
\begin{tabular}{c|rr}
\hline
 & \hspace*{1cm}C  &\hspace*{1cm}Cu\\
\hline
$\rho$ & $46\%$ &$61\%$\\
$\omega$ & $5\%$ &$9\%$\\
$\phi$ & & $6\% ^*$\\
\hline
\end{tabular}

\noindent
* for slow $\phi$ mesons with $\beta\gamma<1.25$.
\end{center}
\end{table}

\begin{figure}
\includegraphics[width=0.6\columnwidth]{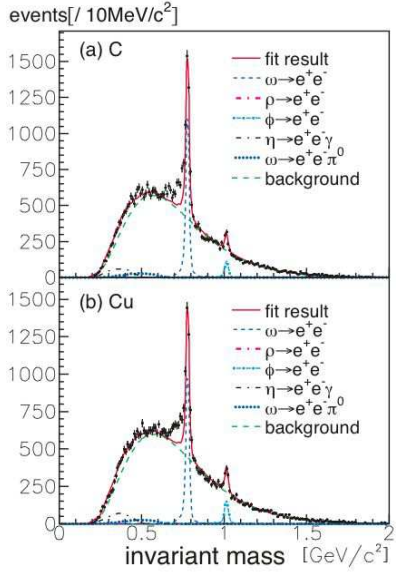}
\caption{\label{fig:naruki-fig1}  
Invariant mass spectra of $e^+ e^-$ for the (a) C 
and (b) Cu targets. The solid lines are the best-fit results, which 
is the sum of the known hadronic decays, $\omega \rightarrow e^+e^-$ 
(dashed line), $\phi \rightarrow e^+e^-$ (thick dash-dotted line), 
$\eta \rightarrow e^+e^- \gamma$ (dash-dotted line), and 
$\omega \rightarrow e^+ e^- \pi^0$ (dotted line) together with the 
combinatorial background (long-dashed line). 
$\rho \rightarrow  e^+ e^-$ is not visible
  \cite{naruki:092301}.
}
\end{figure}

The E325 invariant mass spectra for C and Cu targets are shown in Fig.~\ref{fig:naruki-fig1}. 
The data were taken with 
the ``unlike-sign-double-arm'' trigger condition, i.e., either a positron in the left arm and an
electron in the right arm (“LR event”) or vice versa (“RL event”), to suppress 
the background from Dalitz decays and conversions, thereby precluding the possibility of normalizing the background to the like-sign pair distribution.

The combinatorial background shape was obtained by the event-mixing method, and its normalization was obtained by fitting the data together with contributions from $\omega\rightarrow e^+e^-$, $\rho\rightarrow e^+e^-$, $\phi\rightarrow e^+e^-$, $\eta\rightarrow e^+e^-\gamma$ and $\omega \rightarrow e^+e^-\pi^0$. 
The relativistic Breit-Wigner distribution was used for the resonance shapes, and kinematical distributions of mesons were obtained by the nuclear cascade code JAM \cite{JAM:1999}, which is in  good agreement with 
the experimental data (see, e.g., Fig.~\ref{fig:muto-fig2} (a)).

\subsubsection*{E325 results on the $\rho/\omega$ mesons}
The striking features of the $\rho/\omega$ region of the E325 spectra are as follows. 
i) A significant excess can be seen on the low-mass side of the  $\omega$ peak, which could not be fitted with the cocktail. Therefore, in the fit shown in Fig.~\ref{fig:naruki-fig1}, the range $0.6 < m_{_{e^+e^-}}<0.76 {\rm GeV}/c^2$ was excluded from the fit. 
ii) The $\rho/\omega$ ratio, which is known to be close to unity in $pp$ collisions at this energy \cite{Blobel:1974fp}, is here $\rho/\omega < 0.15$ and $<0.31$ for C and Cu targets, respectively (95\% C.L.).

\begin{figure}
\includegraphics[width=\columnwidth]{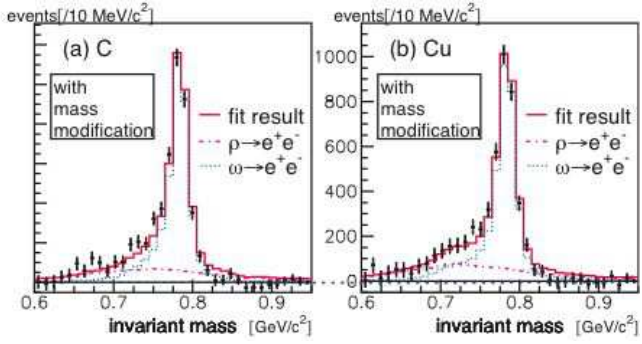}
\caption{\label{fig:naruki-fig2}  
Invariant mass spectra of $e^+e^-$. The combinatorial background and the shapes of
$\eta \rightarrow e^+e^- \gamma$
and $\omega \rightarrow e^+ e^- \pi^0$ 
were subtracted. The result of the model calculation considering the in-medium modification for the 
(a) C and (b) Cu targets.
The solid lines show the best-fit results. In (a) and (b), the shapes of 
$\omega \rightarrow e^+e^-$ (dotted line) and
$\rho \rightarrow e^+e^-$ (dash-dotted line) were modified according to the model using the formula $m_{V}(\rho)/m_{V}(0) = 1-k(\rho/\rho_0)$ with $k = 0.092$
  \cite{naruki:092301}.
}
\end{figure}

The disappearance of $\rho$ and the appearance of the excess may be due to the in-medium dropping of the $\rho$ mass\footnote{Alternatively, this may be due to the over-subtraction of the background component, as pointed out by the J-Lab CLAS g7 collaboration, whose results are in conflict with those of KEK E325. See Section \ref{sec:g7}.}, which would take away the strength from the normal $\rho$ and put them in the excess region. This assumption was tested by fitting the background-subtracted spectra using a Monte Carlo-type model including the mass-dropping model,
\begin{equation}
\label{eq:massdropping}
m_{V}(\rho)/m_{V}(0) = 1 - k(\rho/\rho_0).
\end{equation}
The vector mesons were generated on the surface of an incident hemisphere of the target (supported by the $A^{2/3}$ dependence of the $\omega$ production cross section \cite{tabaru2006nmn}), propagated through the nucleus which was modeled by a Woods-Saxon density distribution. The parameter $k$ was  common for $\omega$ and $\rho$ as well as for C and Cu targets. The $\rho/\omega$ ratio was also allowed to vary.

The best fit results are, $k = 0.092\pm 0.002$ and a $\rho/\omega$ ratio of $0.7\pm 0.1$ and $0.9\pm 0.2$ respectively, for C and Cu targets. 
The best-fit curves are superimposed on the background-subtracted spectra shown in Fig.~\ref{fig:naruki-fig2}. They also examined whether or not the $\rho-\omega$ interference can account for the observed shoulder, but found that the interference cannot explain the data even though the $\rho/\omega$ ratio and the mixing angle were scanned over a wide range.

So, for the $\rho/\omega$ region, E325 concluded that both $\rho$ and $\omega$ masses are shifted by 9\% at the normal nuclear density. 
Fits with density-proportional width broadening did not fit the data;
the fit results favored the 
zero-broadening case
 \cite{naruki:092301}.
This is in conflict with the J-Lab CLAS g7 result discussed in \ref{sec:g7}.

\subsubsection*{E325 results on the $\phi$ meson}

Due to the long lifetime of the $\phi$ meson, in-medium modification effects, if any, are expected only for the slow-moving mesons which have a chance to decay inside the target nucleus. E325 therefore divided the data in three parts based on the $\beta\gamma$ values of the observed $e^+e^-$ pairs, $\beta\gamma<1.25, 1.25<\beta\gamma<1.75$ and $1.75<\beta\gamma$ (see Fig.~\ref{fig:muto-fig2}).


\begin{figure}
\includegraphics[width=0.65\columnwidth]{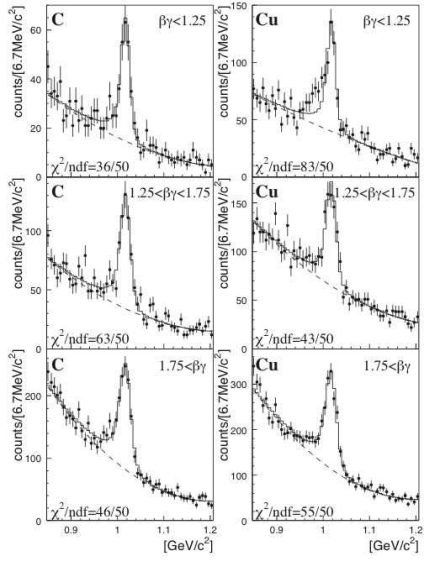}
\caption{\label{fig:muto-fig3}  
Obtained $e^+e^-$ distributions with the fit results.
The target and $\beta\gamma$ region are shown in each panel. The points with 
error bars represent the data. The solid lines represent the fit 
results with an expected $\phi\rightarrow e^+e^-$  shape and a quadratic 
background. The dashed lines represent the background
  \cite{muto:042501}.
}
\end{figure}

The $\beta\gamma$-selected spectra are shown in Fig.~\ref{fig:muto-fig3}, together with fit results. 
The $\phi$  was assumed to have in-vacuum mass and width, convoluted over the detector response in the simulation according to the JAM-generated kinematical distributions of the $\phi$ meson in each $\beta\gamma$ region. A quadratic background was added to the simulated peak, and the background parameters and the $\phi$ abundance were obtained from the fit.

\begin{figure}
\includegraphics[width=0.8\columnwidth]{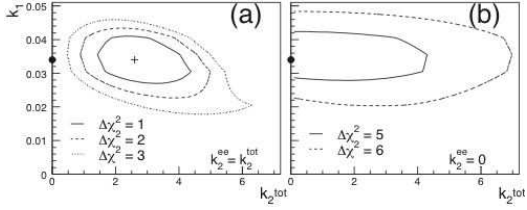}
\caption{\label{fig:muto-fig4}  
Confidence ellipsoids for the modification parameters 
$k_1$ and $k^{\rm tot}_2$ in cases (i) in (a) and (ii) in (b). 
The values of $\Delta\chi^2$s 
in both panels are the differences from the $\chi^2_{\rm min} (=316.4)$ at the 
best-fit point in case (i) which is shown by the cross in the panel 
(a). The best-fit point in case (ii) is shown by the closed circle in 
the panel (b), and also in (a) since the ordinates are common to 
both cases in parameter space 
  \cite{muto:042501}.
}
\end{figure}

The fits are satisfactory, except for the $\beta\gamma<1.25$ region of Cu data, in which a large excess of $N_{\rm excess}/(N_{\rm excess}+N_{\phi}) = 22\%$ was found. If this excess is to be ascribed to the in-medium $\phi$ modification, not only the mass but also the width need to be varied, since the JAM-based simulation indicates only 6\% of the $\phi$ meson produced in copper nuclei with $\beta\gamma<1.25$ would decay in the target nucleus if broadening is not introduced (see Tab. ~\ref{tab:decayfraction}).

It was thus attempted to fit the data by introducing both the density-linear mass shift 
$$m_\phi (\rho)/m_\phi (0) = 1-k_1 (\rho/\rho_0)$$
and the density-linear width broadenings 
\begin{eqnarray*}
\Gamma_\phi ^{\rm tot}(\rho)/\Gamma_\phi ^{\rm tot} (0) &=& 1 + k_2^{\rm tot} (\rho/\rho_0), \mbox{ and}\\ 
\Gamma_\phi ^{ee}(\rho)/\Gamma_\phi ^{ee} (0) &=& 1 + k_2^{ee} (\rho/\rho_0),\\ 
\end{eqnarray*}
where $\Gamma_\phi ^{\rm tot}$ is the total width and $\Gamma_\phi ^{ee}$ is the $e^+e^-$ partial-decay width.


Figure \ref{fig:muto-fig4} shows the fit result for the two cases examined, (a) $k_2^{ee} = k_2^{\rm tot}$ (i.e., the branching ratio $\Gamma^{ee}_\phi/\Gamma^{\rm tot}_\phi$ remains unchanged in the medium), and (b) $k_2^{ee}=0$ (i.e., $\Gamma^{\rm tot}_\phi$ increases but $\Gamma^{ee}_\phi$ does not increase in the medium). 

The fit favors the former case. The obtained values are $k_1=0.034^{+0.006}_{-0.007}$ and $k_2^{\rm tot} = k_2^{ee} = 2.6^{+1.8}_{-1.2}$, indicating the in-medium $\phi$-meson mass shift of 3.4\% and width increase of a factor of 3.6 ($\Gamma^{\rm tot}_\phi \simeq 15 {\rm MeV}/c^2$) at normal nuclear density.

\subsubsection{J-Lab E01-112 (g7) experiment}
\label{sec:g7}
\begin{figure}
\includegraphics[width=\columnwidth]{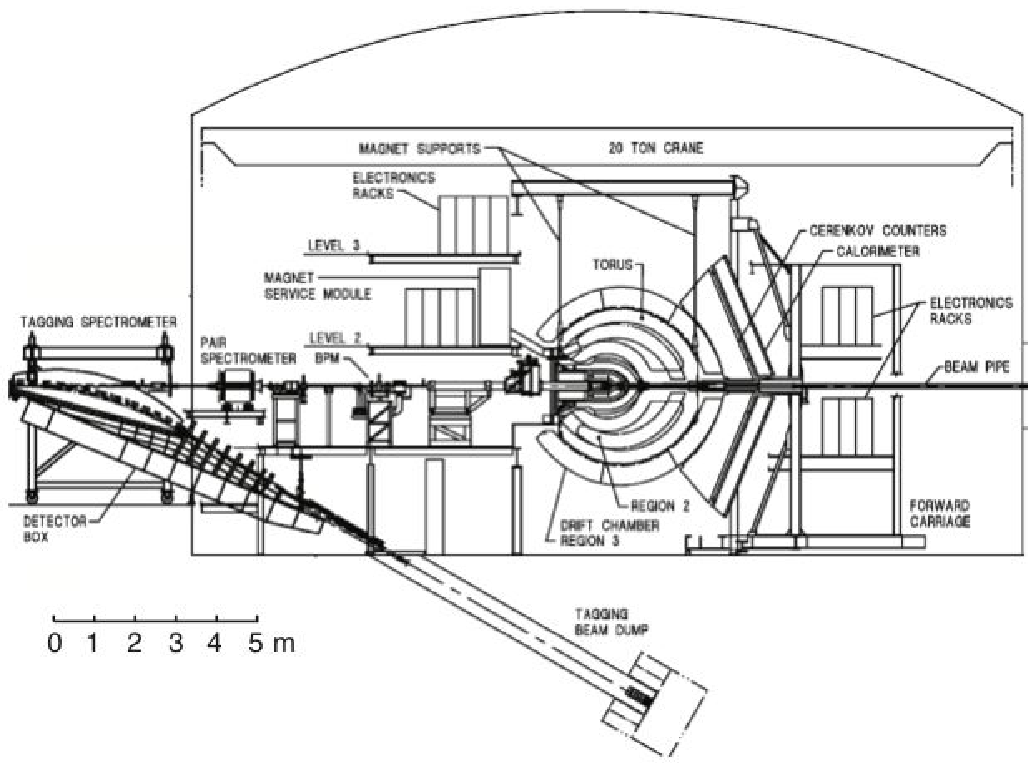}
\includegraphics[width=\columnwidth]{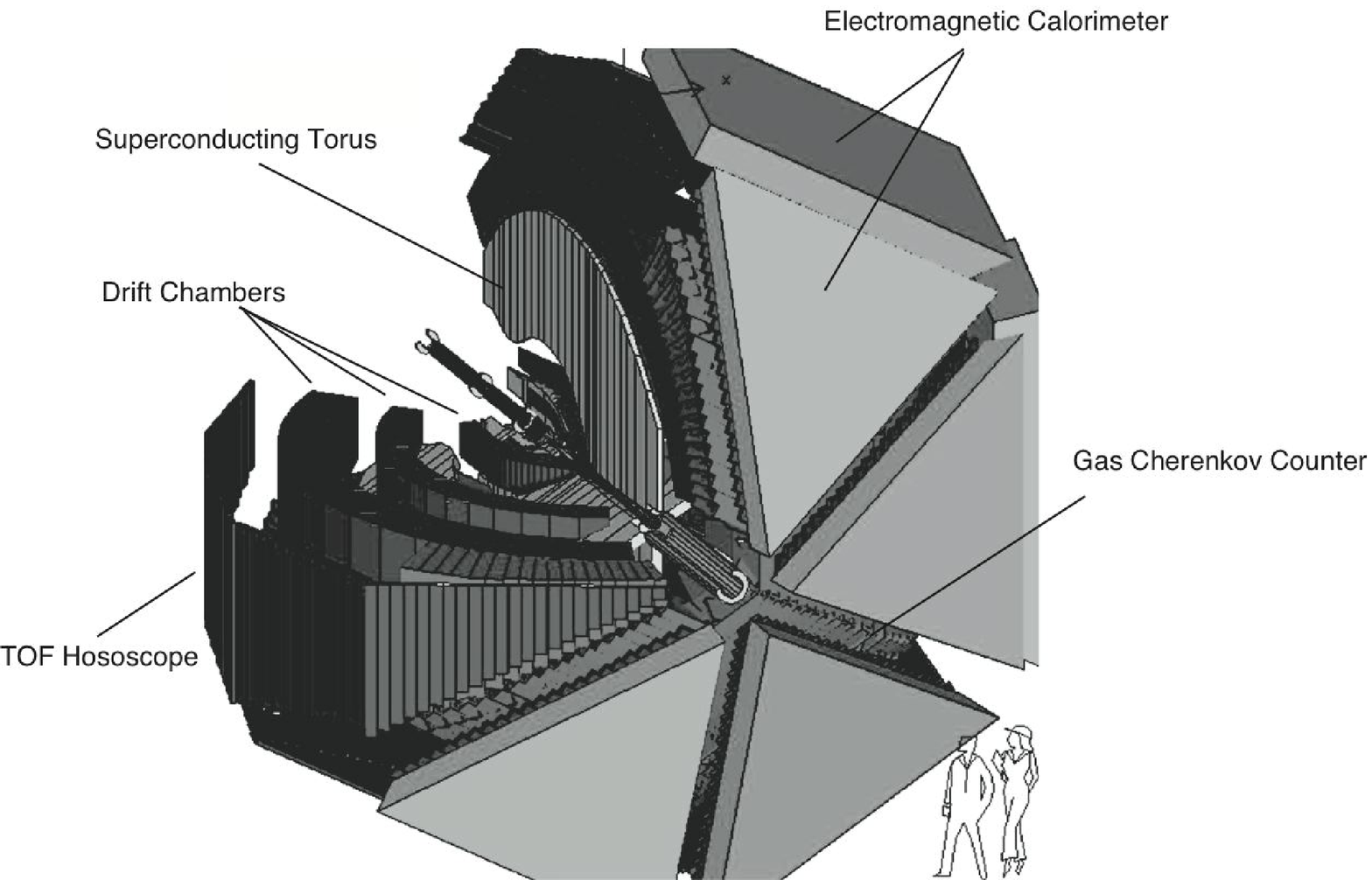}
\caption{\label{fig:clas-in-hall-b} (top) Side view of the CLAS detector in Hall B of
J-Lab, with the photon-tagging system. (bottom) CLAS cutaway view  \cite{Mecking:2003la}.
}
\end{figure}

J-Lab E01-112, better known as the CLAS experiment g7, was conducted in
 in Hall-B of Jefferson Laboratory (Fig.~\ref{fig:clas-in-hall-b}). An electron beam accelerated by the Continuous Electron Beam Accelerator Facility (CEBAF) was used to produce a tagged photon beam having an energy range of $0.61-3.82{\rm GeV}$.

The CEBAF Large Acceptance Spectrometer (CLAS) is a nearly $4\pi$-detector
based on a  six-coil superconducting toroidal magnet, and was designed to track charged particles with 
momenta greater than 200 MeV/$c$ over the polar angle range from $8^\circ$ to $142^\circ$, while covering up to 80\% of the azimuth. 
The CLAS detector is divided into six identical spectrometers (sectors), each made of three regions of drift chambers (DC), time-of-flight scintillators, \v Cerenkov counters  (CC) and electromagnetic calorimeters (EC) (Fig.~\ref{fig:clas-in-hall-b} bottom). 
The target materials were liquid deuterium (LD$_2$), carbon, titanium, iron, and lead (simultaneously in the beam).
To reduce the low-energy $e^-$ and $e^+$ background from pair production in the targets, a ``mini-torus'' magnet was situated just beyond the target region and inside the DC.

The $e^+e^-$ 
event selection and the rejection of the very large $\pi^+\pi^-$ background were done through cuts 
on the EC and the CC. The pion rejection factor was $5.4\times 10^{-4}$ per track, or $2.8\times 10^{-7}$ for the pair. The pair-mass resolution was 10 MeV$/c^2$ for the $\phi$ peak.
The pair momenta of $\sim 0.8 \sim 1.8 {\rm GeV}/c$ were accepted, similar to the KEK E325 acceptance \cite{Djalali2008priv}.

In reconstructing the $e^+e^-$ pairs, the two leptons were required to be detected in different sectors of the CLAS detector. This requirement removed the large background due to pair-production, Bethe-Heitler processes, and $\pi^0$ and $\eta$ Dalitz decays that have a small opening angle.

\begin{figure}
\includegraphics[width=\columnwidth]{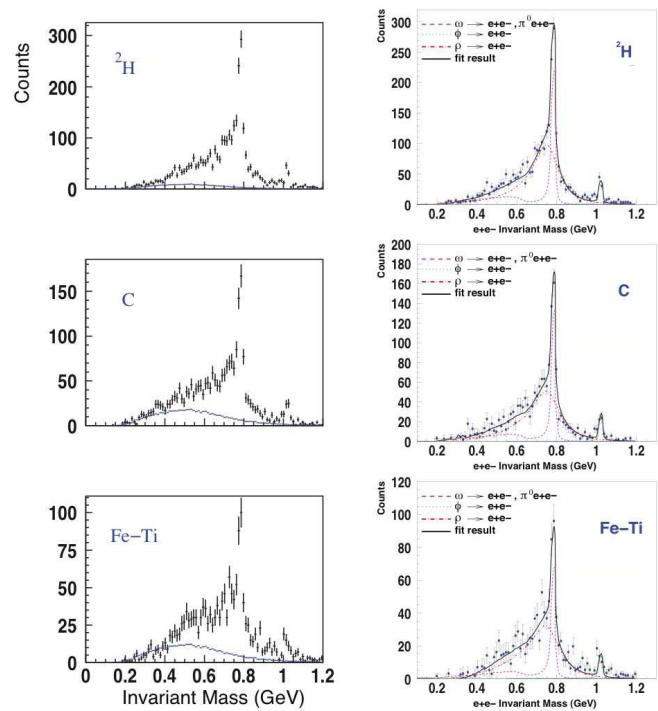}
\caption{\label{fig:clas-g7-d-c-fe-spectra} (left) Normalized combinatorial background for individual targets compared with the data. (right) Result of the fits to the $e^+e^-$ invariant mass spectrum obtained for the $^2$H (top), C (middle), and Fe--Ti (bottom) data  \cite{wood:015201}. 
The curves on the right panels are Monte-Carlo calculations by the BUU model for various vector meson 
decay channels  \cite{Effenberger:2000,Effenberger:1999}.
}
\end{figure}

The combinatorial background\footnote{In the CLAS g7 experiment, the probability of an untagged photon and a tagged photon being in the same radio-frequency timing bunch was about 25\%. This contributed to the combinatorics, in addition to the usual case of picking up a wrong lepton produced in the same event.} was approximated by an event-mixing technique, and was normalized to the number of expected opposite-charge pairs, calculated from the number of observed like-sign pairs using Eq.(\ref{eq:combinatorics}). 
The spectra shown in Fig.~\ref{fig:clas-g7-d-c-fe-spectra} (left) are the reconstructed $e^+e^-$ distributions, compared with the normalized combinatorial background \cite{wood:015201}.
The uncertainty of the normalization was estimated at $\pm 7$\%.

Monte Carlo calculations using a code based on a semi-classical BUU transport model were used to fit the background-subtracted spectra (Fig.~\ref{fig:clas-g7-d-c-fe-spectra}, right). 
In the model, the particles produced as a result of the $\gamma N$ reaction in the target nucleus were propagated through the nucleus allowing for final-state interactions \cite{Effenberger:1999}. The acceptance-corrected BUU mass shapes for the $\rho$, $\omega$ and $\phi$ mesons were scaled separately to match the experimental mass spectra. A substantial contribution from the $\rho$ meson was found (dot-dashed curves in Fig.~\ref{fig:clas-g7-d-c-fe-spectra}, right) unlike in the KEK E325 analysis.

\begin{figure}
\includegraphics[width=.5\columnwidth]{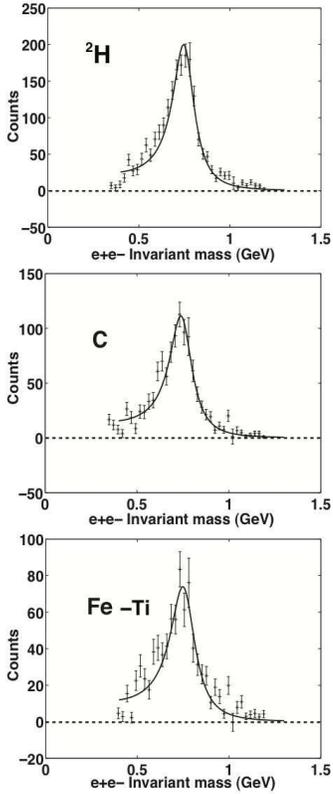}
\caption{\label{fig:clas-g7-rho-shape} Individual Breit-Wigner$/\mu^3$ fits to the $\rho$ mass spectra (background and $\omega, \phi$ contributions subtracted) \cite{wood:015201}.
}
\end{figure}

As the probabilities of the $\omega$ and $\phi$ mesons decaying inside the nucleus are low, the simulated $\omega$ and $\phi$ mass shapes were subtracted from the data, to obtain the $\rho$ mass spectra. The results are shown in Fig.~\ref{fig:clas-g7-rho-shape}. The curves therein are Breit-Wigner$/\mu^3$ fits ($\mu$ being the invariant mass). Here, the $\mu^3$ factor comes from the mass dependence of the $\Gamma_{_{e^+e^-}}(\mu)/\Gamma_{\rm tot}(\mu)$ ratio; $\Gamma_{_{e^+e^-}}(\mu) \propto 1/\mu^4$ and $\Gamma_{\rm tot}(\mu) \propto \mu$. 

\begin{table}
\caption{\label{tab:g7masswidth} Mass and width of the $\rho$ meson obtained by the CLAS-g7 collaboration from the simultaneous fits to the mass spectra for each target and the ratio to $^2$H.}
\begin{tabular}{c|cc}
\hline
Target & $M_\rho$ & $\Gamma_\rho$\\
\hline
$^2$H & $773.0\pm 3.2$ & $185.2\pm 8.6$\\
C & $726.5\pm 3.7$ & $176.4\pm 9.5$\\
Fe, Ti & $ 779.0\pm 5.7$ & $217.7\pm 14.5$\\
\hline
\end{tabular}
\end{table}

The mass and width of the $\rho$ meson in various targets were obtained by performing a simultaneous fit to the mass spectra and the ratio of each spectrum to the $^2$H data (so as to impose more constraints on the fits), and the fit results are shown in Table \ref{tab:g7masswidth}. These are consistent with collisional broadening without mass modification.

The mass shift coefficient $k$ as defined in Eq.(\ref{eq:massdropping}) was obtained by analyzing the ratio of the Fe-Ti to the $^2$H distributions to be $0.02\pm 0.02$, which corresponds to an upper limit of $k=0.053$ with a 95\% confidence level.
These results are quite different from those obtained by the KEK E325 experiment.

\subsubsection{CBELSA/TAPS experiment}
\label{sec:cbelsataps}

\begin{figure}
\includegraphics[width=.85\columnwidth]{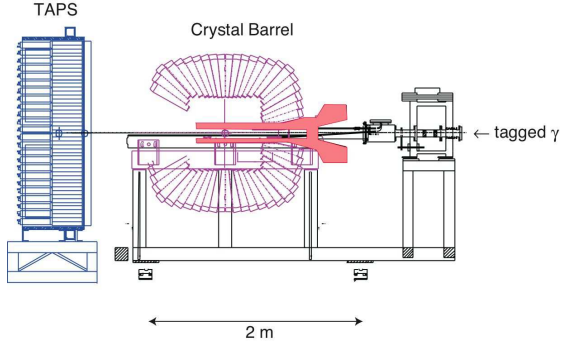}
\caption{\label{fig:cb-taps} 
Side view of the Crystal Barrel (CB) and TAPS detector combination.
}
\end{figure}

The CBELSA/TAPS collaboration at the electron stretcher accelerator (ELSA) in Bonn used the $\gamma A \rightarrow \pi^0 \gamma X$ reaction to study the $\omega$ meson in-medium behavior using the Crystal Barrel (CB) and TAPS crystal spectrometers shown in Fig.~\ref{fig:cb-taps}. 
Tagged photons in the energy range of $0.64-2.53$ GeV were incident on targets (Nb and LH$_2$) mounted in the center of the CB, a photon calorimeter consisting of 1290 CsI(Tl) crystals with an angular coverage of $30^\circ$ to $168^\circ$ in the polar angle and a complete azimuthal angle coverage. Reaction products emitted in forward direction were detected in the TAPS detector, which consisted of 528 hexagonally shaped BaF$_2$ detectors covering polar angles between $4^\circ$ and $30^\circ$ and the complete $2\pi$ azimuthal angle. The resulting geometrical solid angle coverage of the combined system was 99\% of $4\pi$. Charged particles were identified with a scintillating fiber detector placed inside the CB, and a plastic scintillator mounted in front of each TAPS crystal \cite{aker321cbs, novotny1991bps,Janssen:2000}.

\begin{figure}
\includegraphics[width=\columnwidth]{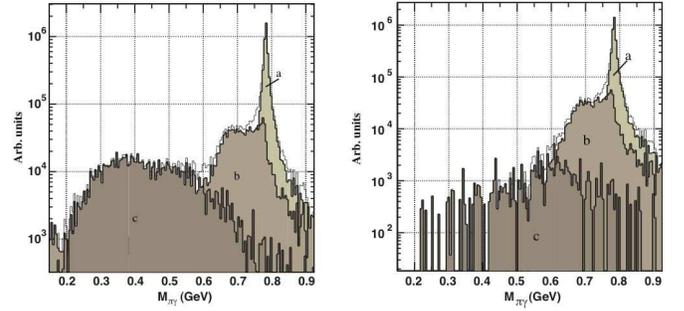}
\caption{\label{fig:pi0-gamma-fsi} 
(left) The $\pi^0\gamma$ mass distribution obtained from a Monte
Carlo simulation of the process $\gamma+{\rm Nb}\rightarrow \pi^0\gamma+X$
at $E_\gamma =1.2$ GeV. The spectrum is decomposed into different contributions corresponding to the fraction of $\omega$-mesons decaying outside the nucleus (a), the fraction of $\omega$-mesons
decaying inside  for which the $\pi^0$ does not
rescatter (b), and the fraction of $\omega$-mesons decaying inside the
nucleus for which $\pi^0$ rescatters(c). In the simulation a drop of the $\omega$ mass by 16\% at normal nuclear density was assumed.
(right) 
The same as the left panel, with the additional condition of
$T_{\pi^0} >150$ MeV
 \cite{Messchendorp:2001fc}.
}
\end{figure}

The $\omega \rightarrow \pi^0\gamma$ decay mode has a large branching ratio of 8.9\% and is a clean and exclusive mode to study the $\omega$ in-medium properties  since the $\rho \rightarrow \pi^0\gamma$ branching ratio is only $6.0\times 10^{-4}$ (see Table~\ref{tab:mesons}). 
Therefore, the study of this mode is complementary to the dilepton decays \cite{Sibirtsev:2000th}.
A serious disadvantage are possible strong final-stage interactions of the $\pi^0$ meson within the nucleus. Monte Carlo simulations \cite{Messchendorp:2001fc} show that the rescattering effect is small in the mass range of interest, and can be further reduced by removing low-energy pions ($T_{\pi}<150$ MeV), as depicted in Fig.~\ref{fig:pi0-gamma-fsi}.

\begin{figure}
\includegraphics[width=\columnwidth]{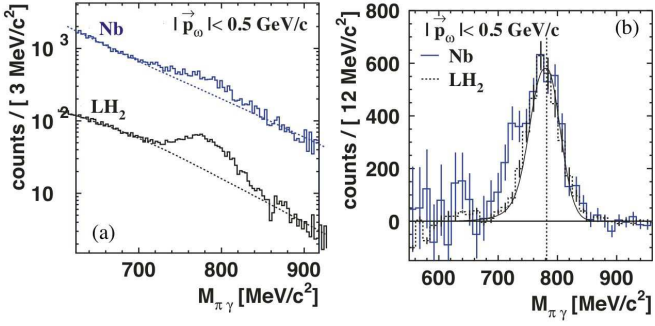}
\caption{\label{fig:trnka} 
(a) Inclusive $\pi^0 \gamma$ invariant mass spectra for momenta less than 500 MeV/$c$. 
Upper histogram: Nb data, lower histogram: LH$_2$ target reference measurement. 
The dashed lines indicate fits for the respective background.
(b) $\pi^0\gamma$ invariant mass for the Nb data (solid histogram) and LH$_2$ data
(dashed histogram) after background subtraction. The error bars show statistical uncertainties only.
The solid curve represents the simulated line shape for the LH$_2$ target
 \cite{trnka:192303}.
}
\end{figure}

The $\pi^0\gamma$ events are reconstructed from three photons, and the invariant mass spectra are  shown in the left panel of Fig.~\ref{fig:trnka}.
Here, in order to maximize the in-nucleus decay probability, slow-moving $\omega$ mesons with $\left| \mathbf{p}_\omega \right| < 0.5\, {\rm GeV} /c$ were selected.
The large continuum background 
is due to four-photon decays of $\pi^0 \pi^0$ and $\pi^0 \eta$ where one of the four photons is missed.  
A smooth polynomial background was assumed and was subtracted, and the resultant LH$_2$ and Nb data are compared in the right panel. As shown, a shoulder on the low-mass side of the $\omega$ peak was found on the Nb target. This was taken as evidence for an $\omega$ in-medium mass reduction by $60^{+10}_{-35}$ MeV at an average nuclear density of $0.6\rho_0$, or in terms of the 
 mass shift coefficient $k$ as defined in Eq.(\ref{eq:massdropping}), this gives $k\simeq 0.14$. The width was found to be $\Gamma=55{\rm MeV}/c^2$, dominated by the experimental resolution.
 
The background-subtraction procedure was criticized by \citet{Kaskulov:2007fy}, who pointed out {\em if} the same background shape is used both for LH$_2$ and Nb, the shoulder structure would disappear. The CBELSA/TAPS group pointed out that the experimental data clearly show that the background distributions are different  and hence it is not justified to {\em assume} the same background shape \cite{Metag:2008ss}. However, the fact that slightly different background assumptions lead to a complete different conclusion on the $\omega$ mass shift is quite alarming. 

The CBELSA/TAPS group has therefore started to employ the event-mixing technique to generate the background distribution, instead of using a polynomial function. Preliminary results were presented in \cite{Metag:2008ss}, but these were later found to contain some problems, and are being further investigated \cite{Metag:2008priv}. Therefore, until the reanalysis is finalized by the group, the $\omega$ mass shift reported in \cite{trnka:192303}  cannot be regarded as a conclusive evidence for the in-medium $\omega$ modification. 

\subsection{Vector-meson in-medium width from transparency-ratio measurements}

Instead of obtaining the in-medium meson width from fits to the observed invariant-mass peak, an alternative method of using the transparency ratio $T$, defined as

\begin{equation}
T = \frac{\sigma_{\gamma A \rightarrow V X}}{A \sigma_{\gamma N \rightarrow V X}},
\end{equation}

\noindent
was proposed \cite{Hernandez:1992ri,Kaskulov:2007fy,Muhlich:2006ps}, and has been used to extract $\phi$ \cite{Ishikawa:2005gf} and $\omega$ \cite{kotulla:192302} in-medium widths.
Here,  $\sigma_{\gamma A \rightarrow V X}$ is the inclusive nuclear vector-meson ($V$) photo-production cross section  and $ \sigma_{\gamma N \rightarrow V X}$ is the cross section on a free nucleon. The ratio $T$ is a measure for the loss of vector-meson flux via inelastic processes in nuclei, and is related to the absorptive part of the meson-nucleus potential. 

This is conceptually a simple measurement, but extracting the in-medium meson width from the $A$ dependence of the ratio $T$ requires comparison with theory calculations.

\subsubsection*{The $\phi$ attenuation}

\begin{figure}
\includegraphics[width=.6\columnwidth]{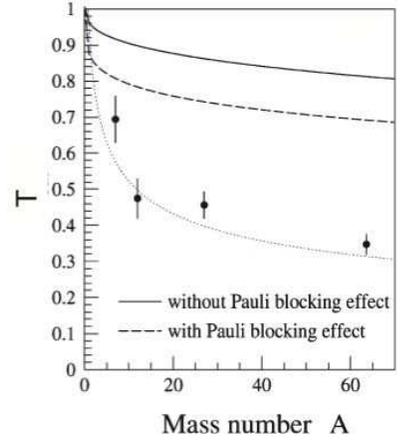}
\caption{\label{fig:phi-attenuation} 
Mass number ($A)$ dependence of the transparency ratio $T$. The dotted line corresponds to $\sigma_A \propto A^{0.72}$. The solid and dashed curves show the theoretical calculation \cite{Cabrera:2004kl} without (solid curve) and with (dashed curve) Pauli-blocking correction for the $\phi$ meson scattering angle in the laboratory frame of 0$^\circ$
\cite{Ishikawa:2005gf}. 
}
\end{figure}

The photo-produciton of $\phi$ mesons from Li, C, Al and Cu targets was measured at $E_\gamma = 1.5-2.4$ GeV, using  the laser-electron photon facility at SPring-8 (LEPS), in the $\gamma A \rightarrow K^+K^- X$ channel \cite{Ishikawa:2005gf}. The $A$ dependence of the incoherent $\phi$ photo production cross section was found to be $\sigma_{A} \propto A^{0.72\pm 0.07}$ (or $T= A^{-0.28}$, as shown in Fig.~\ref{fig:phi-attenuation}). Using a Glauber-type model calculation, the in-medium $\phi$-nucleon cross section was deduced to be $\sigma_{\phi N} = 35^{+17}_{-11}$ mb, which is much larger than the free-space value of  $\sigma_{\gamma N}^{\rm free} = 140 \mu$b used as an input to the model calculation  \cite{Amsler:2008pj}. Theoretical calculations \cite{Cabrera:2004kl} predicted much larger $T$ values (solid and dashed curves in Fig.~\ref{fig:phi-attenuation}).

Using the classical low-density relation
\begin{equation}
\Gamma_V = \hbar \rho \beta c \sigma,
\end{equation}
this would correspond to a width of $\Gamma_\phi \simeq 80$ MeV$/c^2$ at $\rho = \rho_0$ and $\beta \simeq 0.7$ (i.e., $\beta\gamma\simeq 1$), where the KEK E325 experiment reported a much smaller in-medium $\phi$ width of 15 MeV$/c^2$ (see section \ref{sec:e325})\footnote{This discrepancy may at least partly be due to the way the transparency ratio was normalized to the production cross section on the nucleon in Fig.~\ref{fig:phi-attenuation}). While the $\gamma p$ cross section is sufficiently well known, the $\gamma n$ cross section is not. This was partly avoided in \citet{Ishikawa:2005gf} by taking $^7$Li as a reference.}.

\subsubsection*{The $\omega$ attenuation}

\begin{figure}
\includegraphics[width=\columnwidth]{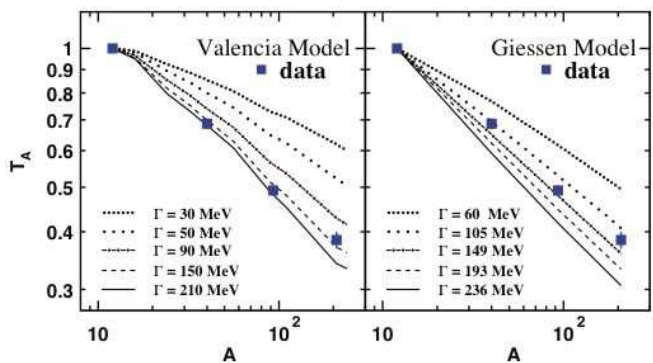}
\caption{\label{fig:omega-attenuation} 
Experimentally determined transparency 
ratio normalized to the carbon data in comparison with a theoretical 
Monte Carlo simulation \cite{Kaskulov:2007fy} (left) and a BUU calculation 
\cite{Muhlich:2004rp} (right) varying the width at 1.1 GeV/$c$ momentum, 
respectively. The width is given in the nuclear rest frame. 
Only statistical errors are shown
\cite{kotulla:192302}. 
}
\end{figure}

The CBELS/TAPS collaboration measured the $A$ dependence of the $\omega$ photoproduction cross section on the nuclei C, Ca, Nb and Pb. 
The average momenta of the mesons was 1.1 GeV$/c$, so that almost all $\omega$ mesons decay outside the nuclear target.
Since the $\omega$ photoproduction cross section on the neutron is not known, they took the transparency ratio normalized to the carbon data, as shown in Fig.~\ref{fig:omega-attenuation}.\footnote{
For this reason, the ratio shown in  Fig.~\ref{fig:omega-attenuation} is named $T_A$ instead of $T$. 
This leads to different values of the transparency ratio since the transparency loss in C is normalized away.
This normalization takes into account $\omega$ production processes on two nucleons which may be relevant in nuclei.
}

The data were then compared with three different types of models, i) a Glauber model similar to the LEPS analysis, ii) a BUU analysis \cite{Muhlich:2004rp} and iii) a calculation by the Valencia group \cite{Kaskulov:2007fy}. In all cases, the inelastic $\omega$ width was found to be $130-150$ MeV$/c^2$ at $\rho =\rho_0$ for an average $\omega$ momentum of 1.1 GeV$/c$, or in terms of $\omega N$ cross section, $\sigma_{\omega N}\simeq 70$ mb. 

\section{CONCLUDING REMARKS}

The QCD vacuum shows the dynamical breaking of chiral symmetry.
 In the hot/dense QCD medium, the chiral order parameter such as
$\left<\bar q q\right>$ (chiral condensate)
is expected to change
as a function of temperature $T$ and density $\rho$ of the medium, 
and its experimental detection is one
of the main challenges in modern hadron physics.

\subsection*{Pion}

Theoretically, all hadrons receive various spectral changes due to their strong interaction with the medium. Among those, the in-medium modification of the 
pion decay constant $f_{\pi}^t (\rho)$ and $f_{\pi}^t (T)$ is theoretically well 
under contol at low $T$ and $\rho$, and have close relation
to the in-medium change of the chiral condensate  (\ref{eq:GOR-ratio},\ref{eq:TW-ratio}).
 The predicted reduction at low density, 
 $\left<\bar q q\right>_\rho/\left< \bar q q\right>_0 \simeq 1-(0.3\sim 0.4)\rho/\rho_0$, has now been experimentally demonstrated by comparing the isovector pion-nucleon $b_1$ and pion-nucleus 
 $b_1(\rho)$ scattering lengths derived from pionic hydrogen and deeply-bound pionic Sn atoms, respectively, as well as by analyzing the differential cross sections for low-energy $\pi^\pm$ elastic scattering by several nuclei.

\subsection*{$\sigma$ meson}
One of the interesting signals associated with the in-medium chiral restoration would be the spectral enhancement on the $\sigma$ channel near the $2\pi$ threshold.
Intriguing experimental results of ``softening" 
of the $(\pi\pi)_{I=J=0}$ distribution (i.e., shift of the peak position to lower masses) 
have been obtained. These agree fairly well with
(i)  in-medium modifications of the $\pi\pi$ interaction, 
as well as with (ii) rescattering of outgoing pions with the nucleons without
 in-medium $\pi\pi$ interaction.
In fact, the results of two calculations, \citet{Roca:2002hl} and \citet{Buss:2006vh}, predict
very similar spectra, as shown in Fig.~\ref{fig:buss-roca}. However, as long as the rescattering
scenario can reproduce most of the observed ``softening'' trend,
we cannot yet extract the predicted
partial chiral restoration signature from the $\pi\pi$ spectra.
New high-statistics data taken on C, Ca, Pb with the Crystal Ball/TAPS detector, which is being analyzed, may help shed some light on this problem \cite{Metag:2008priv}.

\subsection*{Vector mesons}
\label{sec:vectormesonsummary}
Significant experimental work has been done to detect the possible in-medium
``mass shift" of vector mesons,
both using heavy-ion collisions and using elementary-particle beams.
In general, the vector spectral function receives a shift of the peak,
 broadening, new structures, etc., due to the complex interaction of the
vector current with the medium.
Also, such a spectral shift may well depend  on the spatial momentum of the current.
 Therefore,  
 it would not be appropriate to oversimplify the problem 
 to ``mass shift vs.\ width broadening".
With this caution in mind, we list
experimental results on the in-medium mass and width of 
the $\rho$, $\omega$ and $\phi$ mesons produced
 with elementary reactions, measured in different experiments in Table \ref{tab:vmsummary}.
The TAGX results are not included here for the reasons discussed in section \ref{sec:tagx}.
The $\omega$ mass shift from CBELSA/TAPS is listed in the table, but it
may change after the ongoing reanalysis, and hence we do not include this in the summary discussion.

\begin{table*}
\caption{\label{tab:vmsummary} Compilation of experimental results on the in-medium mass and width of the $\rho$, $\omega$ and $\phi$ mesons produced with elementary reactions, measured in different experiments. This is  based on and updating the table prepared  by  \citet{Metag:2008ss}}
\begin{tabular}{l||c|c|c||c|c}
\hline
& \multicolumn{3}{c||}{Invariant mass} & \multicolumn{2}{c}{Attenuation}\\
\hline
& E325 @ KEK & CLAS g7 @ Jlab& \multicolumn{2}{c|}{CBELSA/TAPS} & LEPS @ SPring-8\\
\hline
\hline
Reaction & pA 12 GeV & $\gamma A$ $0.6-3.8$ GeV & \multicolumn{2}{c|}{$\gamma A$ $0.7-2.5$ GeV}
& $\gamma A$ $1.5-2.5$ GeV\\
\hline
Momentum& $p>0.5$ GeV/$c$ & $p>0.8$ GeV$c$ & $p<0.5$ GeV/$c$ & $0.4<p<1.7$ GeV$/c$ & $ 1.1<p<2.2$ GeV/$c$ \\
\hline
$\rho$ &$\uparrow$ & $\Delta m\approx 0$ & \multicolumn{1}{c||}{--} &  \multicolumn{1}{c|}{--} &  \multicolumn{1}{c}{--} \\
&  $\Delta m(\rho_0)/m=-9\%$ & {\small some broadening} & & & \\
$\omega$ &  $\downarrow$ &  \multicolumn{1}{c|}{--} & $\Delta m(\rho_0)/m = -14\%^{\dagger}$ & $\Gamma_\omega (\rho_0)$=130-150 MeV$/c^2$&  \multicolumn{1}{c}{--}\\
&no broadening &  & &$\rightarrow \sigma_{\omega N} \approx 70$mb & \\
$\phi$ & $\Delta m(\rho_0)/m=-3.4\%$ & \multicolumn{1}{c|}{--} & \multicolumn{1}{c||}{--} & \multicolumn{1}{c|}{--} & $\sigma_{\phi N} = 35$mb\\
& $\Gamma_\phi (\rho_0) \approx 15 $MeV$/c^2$ & \multicolumn{1}{c|}{--} & \multicolumn{1}{c||}{--} & \multicolumn{1}{c|}{--} &$\rightarrow \Gamma_\phi (\rho_0) \approx 80$ MeV/$c^2$\\
\hline
\end{tabular}

\begin{flushleft}
$\dagger$ This  may change as a result of the ongoing  reanalysis \cite{Metag:2008priv}.
\end{flushleft}
\end{table*}

Upon examining this table, we realize that there are some inconsistencies, and we discuss the two most pressing issues below.

\begin{figure}
\includegraphics[width=0.5\columnwidth]{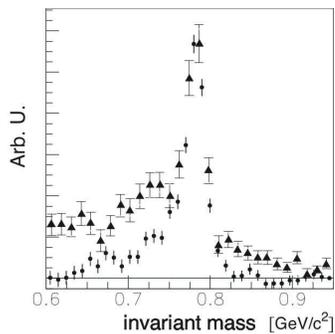}
\caption{\label{fig:e325-g7}
Comparison of the dielectron invariant-mass spectrum (carbon target) of E325 (circles) and CLAS g7 (triangles).
}
\end{figure}

(i) E325 and g7 disagree on the $\rho (\omega)$ mass shift:
The E325 result is both $\rho$ and $\omega$ masses get reduced at $\rho_0$ by 9\% (the mass shift parameter $k=0.092\pm 0.002$), while the CLAS g7 placed a 95\% confidence upper limit at $k=0.053$. The comparison of the background-subtracted dielectron distributions (carbon target) measured by the two experiments (Fig.~\ref{fig:e325-g7}) shows that the two spectra are very different\footnote{Note that the $\rho/\omega$ ratio in the $pp$ collisions at the KEK energy is about unity \cite{Blobel:1974fp}, while that in the $\gamma p$ collisions at the CLAS energy is about 3 to 1 \cite{aachen:1968,Barth:2003yq,Wu:2005kx}, which may account for the bulk of the difference found in Fig.~\ref{fig:e325-g7}.}.

\citet{wood:015201} pointed out that this difference must be due to the way the combinatorial background was subtracted in E325. In the E325 analysis,
due to the lack of a sample of same-charged
leptons by which to extract the normalization of the combinatorial background,
the background contribution was fit along with the $\omega$- and $\phi$-meson shapes.
Without an absolute determination of the combinatorial
background,  the $\rho$-meson signal was suppressed and included in the background shape.
Indeed, if the background normalization was free to vary in the CLAS-g7 fit, the g7 spectra were found to be consistent with $k\approx 0.16$ \cite{djalila:ykis06}\footnote{In reality, there is no such freedom in the g7 background normalization.  This was done just for the sake of g7-E325 comparison. }.\\


(ii) E325 and CBELSA/TAPS disagree on the $\omega$ width:
While an $\omega$ width broadening was not observed by the E325 experiment, CBELSA/TAPS found an unexpectedly large in-medium broadening. These two observations are mutually inconsistent.
Even though the extraction of the in-medium width depends on theory, the observed $A$-dependent reduction of the transparency ratio $T$ clearly shows that the $\omega$ meson is attenuated in the target nucleus. This conclusion must be robust. \\

Table \ref{tab:vmsummary} clearly shows that experimental results have not yet converged, and more work is needed to obtain consistent understanding of the in-medium behavior of vector mesons.
In view of the robustness of the method, the in-medium broadening of vector mesons deduced from the transparency-ratio measurements are hard to rule out.
On the other hand, problem(s) have been pointed out for all experiments which observed in-medium mass shifts, and hence those results must be treated with caution and further studies are needed.

\section*{ACKNOWLEDGMENTS}
This article is based on a talk given by RH at the International  Nuclear Physics Conference held in Tokyo (INPC 2007). During its preparation, the authors have beneﬁtted
from the assistance of so many of their colleagues that it
would be impossible to mention them all by name. However, we should  like to acknowledge the specific contributions of Y.~Akaishi,  C.~Djalali, H.~En'yo, D.~Gotta, D.~Jido, V.~Metag, U.~Mosel,  M.~Naruki, E.~Oset, K.~Ozawa, P.~Salabura, and S.~Schadmand.
It is a pleasure to acknowledge, too, the many insightful discussions  at various
times with S.~Hirenzaki, P.~Kienle,  T.~Kunihiro, H.~Toki, and T.~Yamazaki.

This work is supported in part by Grant-in-Aid for Specially Promoted  Research (20002003), Grant-in-Aid for Scientific Research (C)  (18540253), and the Global COE Program ``the Physical Sciences  Frontier'', MEXT, Japan.

\bibliographystyle{apsrmp}
\bibliography{rmp-hayano-hatsuda}



\end{document}